\documentclass{elsart3p}

\usepackage{natbib}

\usepackage{epsfig}
\usepackage{pgf}

\usepackage{amssymb}

\def\babar{\mbox{\slshape B\kern-0.1em{\smaller A}\kern-0.1em
    B\kern-0.1em{\smaller A\kern-0.2em R}}}
\def\Bbar    {\kern 0.18em\overline{\kern -0.18em B}{}\xspace}

\def\BB      {\ensuremath{B\Bbar}\xspace}
\def\Bz      {\ensuremath{B^0}\xspace}
\def\Bzb     {\ensuremath{\Bbar^0}\xspace}
\def\BzBzb   {\ensuremath{\Bz {\kern -0.16em \Bzb}}\xspace}
\def\Bu      {\ensuremath{B^+}\xspace}
\def\Bub     {\ensuremath{B^-}\xspace}

\def\BpBm    {\ensuremath{\Bu {\kern -0.16em \Bub}}\xspace}

\newcommand{\optbar}[1]{\shortstack{{\tiny (\rule[.4ex]{1em}{.1mm})}
  \\ [-.7ex] $#1$}}
\def\BorBbar    {\kern 0.18em\optbar{\kern -0.18em B}{}\xspace}
\def\DorDbar    {\kern 0.18em\optbar{\kern -0.18em D}{}\xspace}
\def\KorKbar    {\kern 0.18em\optbar{\kern -0.18em K}{}\xspace}
\def\CP                {\ensuremath{C\!P}\xspace}
\def\pep2{PEP-II}
\def\SLAC{Stanford Linear Accelerator Center}

\def\epem       {\ensuremath{e^+e^-}\xspace}
\def\mumu       {\ifmath{\mu^+\mu^-}}
\def\mupair     {\ifmath{\epem \to \mumu }}
\def\bhabha     {\ifmath{\epem \to \epem }}
\def\llpair     {\ifmath{\epem \to l^+l^- }}

\mathchardef\Upsilon="7107
\def\Y#1S{\ensuremath{\Upsilon{(#1S)}}\xspace}

\def\FourS {\Y4S}

\newcommand{\chisq}{\ensuremath{\chi^2}}

\def\mum  {\ensuremath{\,\mu\rm m}\xspace}

\newcommand{\tev}{\ensuremath{\mathrm{\,Te\kern -0.1em V}}\xspace}
\newcommand{\gev}{\ensuremath{\mathrm{\,Ge\kern -0.1em V}}\xspace}
\newcommand{\mev}{\ensuremath{\mathrm{\,Me\kern -0.1em V}}\xspace}
\newcommand{\kev}{\ensuremath{\mathrm{\,ke\kern -0.1em V}}\xspace}
\newcommand{\ev}{\ensuremath{\mathrm{\,e\kern -0.1em V}}\xspace}
\newcommand{\gevc}{\ensuremath{{\mathrm{\,Ge\kern -0.1em V\!/}c}}\xspace}
\newcommand{\mevc}{\ensuremath{{\mathrm{\,Me\kern -0.1em V\!/}c}}\xspace}
\newcommand{\gevcc}{\ensuremath{{\mathrm{\,Ge\kern -0.1em V\!/}c^2}}\xspace}
\newcommand{\mevcc}{\ensuremath{{\mathrm{\,Me\kern -0.1em V\!/}c^2}}\xspace}

\def\svt{SVT\,}
\def\dch{DCH\,}
\def\Svt{Silicon Vertex Tracker\,}

\def\babar{{\sl B\hspace{-0.4em} {\scriptsize\sl A}\hspace{-0.4em}
B\hspace{-0.4em} {\scriptsize\sl A\hspace{-0.1em}R}\,\,}}

\def\ifmath#1{\relax\ifmmode#1\else$#1$\fi}

\newcommand{\comment}[1]{}

\def\to{\rightarrow}

\def\CP{\ifmath{C\!P}}

\def\babar{\mbox{\sl B\hspace{-0.4em} {\scriptsize\sl A}\hspace{-0.4em} B\hspace{-0.4em} {\scriptsize\sl A\hspace{-0.1em}R}}}

\def\dirc       {{\sc DIRC}}

\def\emc 	{EMC}
\def\Emc 	{Electromagnetic Calorimeter}

\def\epem  {\ifmath{e^+e^-}}

\def\BB    {\ifmath{B\overline{B}{}}}
\def\BzBzb {\ifmath{B^0 \overline{B}{}^0}}

\def\FourS {\ifmath{\Upsilon (4S)}}

\def\Y#1S  {\ifmath{\Upsilon (#1S)}}

\def\mumu	{\ifmath{\mu^+\mu^-}}

\def\kev  {\ifmath{\mbox{\,ke\kern -0.08em V}}}
\def\mev  {\ifmath{\mbox{\,Me\kern -0.08em V}}}
\def\gev  {\ifmath{\mbox{\,Ge\kern -0.08em V}}}
\def\gevc {\ifmath{\mbox{\,Ge\kern -0.08em V$\!/c$}}}
\def\mevc {\ifmath{\mbox{\,Me\kern -0.08em V$\!/c$}}}
\def\gevcc{\ifmath{\mbox{\,Ge\kern -0.08em V$\!/c^2$}}}
\def\mevcc{\ifmath{\mbox{\,Me\kern -0.08em V$\!/c^2$}}}

\def\mum  {\ifmath{\,\mu\mbox{m}}}

\def\pf         {\ifmath{\mbox{\,pF}}}

\makeatletter
\def\@versim#1#2{\vcenter{\offinterlineskip
        \ialign{$\m@th#1\hfil##\hfil$\crcr#2\crcr\sim\crcr } }}
\def\gsim{\mathrel{\mathpalette\@versim>}}
\def\lsim{\mathrel{\mathpalette\@versim<}}
\makeatother

\def\ev  {\ifmath{\mbox{\,e\kern -0.08em V}}}

\begin{document}

\begin{frontmatter}



\title{Local Alignment of the BABAR Silicon Vertex Tracking Detector}


\vspace{0.1cm}

\author{D.\,N.\,Brown}
\address{Lawrence Berkeley National Laboratory, Berkeley, California 94720, USA }
\author{A.\,V.\,Gritsan, Z.\,J.\,Guo}
\address{Johns Hopkins University, Baltimore, Maryland 21218, USA }
\author{D.\,Roberts}
\address{University of Maryland, College Park, Maryland 20742, USA }

\begin{abstract}
The \babar \, \Svt (\svt) is a five-layer double-sided
silicon detector designed to provide precise measurements
of the position and direction of primary tracks,
and to fully reconstruct low-momentum tracks
produced in $\epem$ collisions at the \pep2 asymmetric
collider at \SLAC. This paper describes the design,
implementation, performance, and validation of the
{local alignment} procedure used to determine
the relative positions and orientations of the
340 \svt wafers.  This procedure uses a tuned mix of
in-situ experimental data and complementary 
lab-bench measurements to control systematic
distortions.  Wafer positions and orientations are
determined by minimizing a $\chi^2$ computed using
these data for each wafer individually, iterating
to account for between-wafer correlations.
A correction for aplanar distortions of the
silicon wafers is measured and applied.
The net effect of residual mis-alignments on relevant
physical variables is evaluated in special control samples.
The $\babar$ data-sample collected between November 1999
and April 2008 is used in the study of the \svt stability.
\end{abstract}




\end{frontmatter}

%

\section{Introduction}

Multi-wafer silicon (Si) tracking
and vertex detectors have become an essential part of
modern High Energy Physics experiments.
Because of the short ionization drift distances and 
the sub-\mum~feature placement accuracy of silicon 
wafer sensors, individual silicon wafers can provide 
\mum-scale position resolution over areas of  
a few tens of square centimeters.
In order to extend this precision over the square meter
areas covered by modern detectors, the relative positions 
and orientations of the constituent silicon wafers must be 
well known~\cite{LHCworkshop}.

\par

Si tracking detector construction techniques define 
the wafer positions and orientations only nominally.
Because Si tracking detectors are typically located in 
extremely confined spaces near the interaction region, the Si
wafer positions cannot be measured using conventional 
survey techniques once the detector has been installed.
Lab-bench measurements during construction using mechanical 
or optical techniques can determine wafer positions
very accurately, but because the wafers can shift due 
to mechanical and thermal stress during and after detector 
installation, and because silicon charge collection effects 
can distort the effective position of a wafer from its 
geometric value, these measurements are not sufficient.
Because of these effects, the wafer positions and orientations must be
determined primarily using signals
readout from the silicon detectors themselves when traversed 
by particles in-situ.

\par

This note describes the procedure developed and used for 
the \babar\ Silicon Vertex Tracker (\svt)\ {\em local alignment}, whereby the
positions and orientations of the wafers are determined.
Our procedure uses track data recorded during normal \babar\ running, 
filtered and prescaled to produce a fixed sample
that roughly uniformly illuminates all the wafers, 
and constrains all the local alignment degrees of freedom 
in a statistically independent and systematically complete way.  
Tracks are fit using \svt\ hits and constrained using a subset 
of Drift Chamber (DCH) and beam energy information selected to not
impose any significant systematic bias on the local alignment.
To avoid statistical bias, we select an independent subset 
of information from each track. We combine track-based 
information with direct measurements of the relative positions 
and orientations of Si wafers made during detector construction,
resulting in a statistically correct and systematically robust 
measure of the consistency ($\chi^2$) of a wafer's position 
and orientation within the detector. We use an iterative 
technique to determine the relative wafer positions that 
minimize the $\sum \chi^2$ of all wafers.  The resultant 
local alignment is then validated against several possible 
systematic effects. Each of these functions are described 
in detail in the following sections.
The related 
but simpler problem of determining the rigid-body 
position and orientation of the \svt\ within the
\babar\ detector ({\em global alignment}) 
is not covered in this note.

%
%
\section {The Silicon Vertex Tracker}
\label{sec:svt}

\begin{figure*}[htbp]
\begin{center}
\centerline{
\setlength{\epsfxsize}{1.0\linewidth}\leavevmode\epsfbox{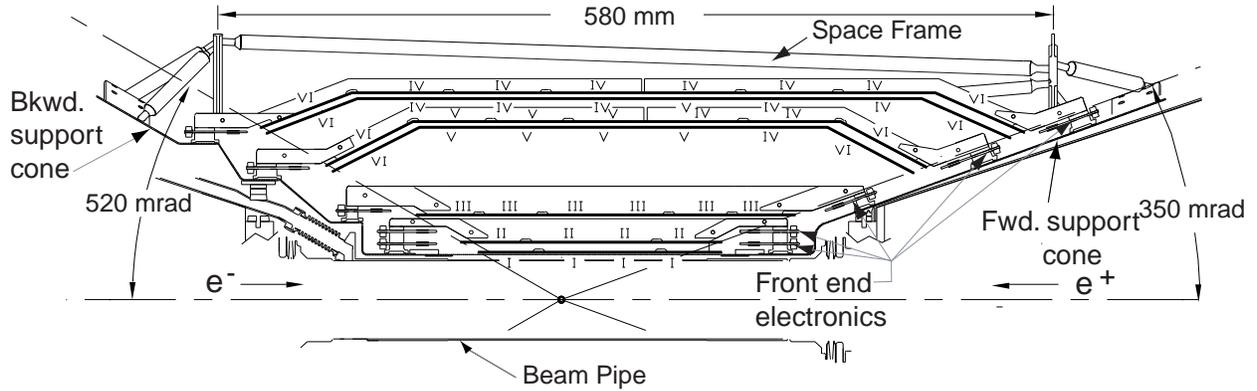}
}
\caption{ Longitudinal section of the $\babar$ \svt.
The roman numbers label the six different types of wafers.}
\label{fig:svt-1}
\end{center}
\end{figure*}

The \babar\ \svt\  
was designed primarily to provide precise 
reconstruction of charged particle trajectories and 
decay vertices near the \epem\ interaction point of \pep2 \cite{pep},
as required  by \babar's diverse physics goals.  Additionally,
the \svt\ provides the precise $\theta$ angle measurement
needed to perform charged particle identification using
\babar's Cherenkov detector (\dirc) and \Emc~(\emc)~\cite{babar}.

\par

The \svt\ is composed of 340 separate Si wafers, 
arranged in 5 co-axial roughly cylindrical layers,
see Fig.~\ref{fig:svt-1}.
Each layer is composed of between 6 and 18 
modules, arranged symmetrically around the cylinder ($z$) axis, 
held in place by a rigid carbon-fiber frame.  Each module is
in turn composed of between 4 and 8 individual Si wafers, 
which are glued to supporting kevlar ribs extending in the $z$ direction.
See Table~\ref{tab:svtgeometry} for the geometrical parameters
of each layer.
There are six different wafer shapes, including a trapezoidal
shape used to form the arch modules discussed below.
The smallest wafers are $4.2\times4.1$ cm$^2$, and the
largest (in layer 3) are $4.4\times7.1$ cm$^2$. 

\par

\begin{table}[htbp]
\caption{
Geometric parameters of five \svt layers comprised of
340 silicon wafers. The radial range for layers 4 and 
5 includes the radial extent of the arched sections.
The radius refers to the closest transverse radius. 
The length ($L$) and width ($W$) are along $z$ and $\phi$, respectively. 
}
\label{tab:svtgeometry}
\medskip
\begin{center}
\begin{tabular}{|c|c|c|c|c|c|}
\hline\hline
\vspace{-3mm} & &  &  &   &   \cr
 ~~layer~~ & ~wafers~     & ~modules~   & ~radius~  & ~$z~(L)$~   & ~$\phi~(W)$~ \cr
       & ~in module~   & ~in layer~ & ~~(mm)~~    & ~~(mm)~~~  & ~~(mm)~~  \cr
\vspace{-3mm} & &  &  &  &    \cr
\hline
\vspace{-3mm} & &  &  &     &    \cr
1 & 4 & 6  & 32       & 42  & 41    \cr
2 & 4 & 6  & 40       & 45  & 49    \cr
3 & 6 & 6  & 54       & 44  & 71    \cr
4 & 7 & 16 & 91--127  & 54-68 & 43-53    \cr
5 & 8 & 18 & 114--144 & 68 & 43-53    \cr
\hline\hline
\end{tabular}
\end{center}
\end{table}

\par

\begin{figure}[htbp]
\begin{center}
\vspace{0.4cm}
\setlength{\epsfxsize}{1.0\linewidth}\leavevmode\epsfbox{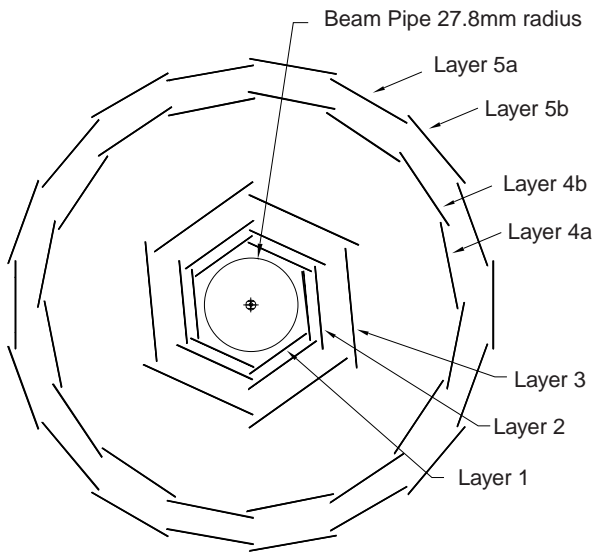}
\caption{ Transverse section of the $\babar$ \svt.}
\label{fig:svt-2}
\end{center}
\end{figure}

The modules of the inner three layers are planar,
while the modules in layers 4 and 5 are arch-shaped.
This design reduces the amount of
material and improves the point resolution for
particles originating from the interaction region compared
to a planar module.
The modules in the inner three layers are tilted by 5$^\circ$
in azimuth  ($\phi$),
allowing an overlap region between adjacent modules,
see Fig.~\ref{fig:svt-2}. This 
arrangement is advantageous for alignment and provides full
$\phi$ coverage. The outer layers cannot be tilted because
of the arch geometry. To have an overlap
and avoid coverage gaps in $\phi$, the outer two layers
are divided into two sub-layers at slightly different radii.

The \svt\ support structure is a rigid body made from
two carbon-fiber cones, connected by a space frame,
also made of carbon-fiber epoxy laminate.  While in operation
the \svt\ is mounted on the innermost magnets of the \pep2\ beamline,
supported by an assembly fixture
that allows for some relative motion with 
respect to \pep2.   Because the \svt\ is
mounted independently of the rest of the
\babar\ detector, movement between the \svt\ and the rest 
of the detector can occur. During operation
the \svt\ is cooled to remove the heat generated by the
electronics and is kept in a humidity controlled environment.

%
%
\section{Parameterizing the \svt\ Local Alignment}
\label{sec:params}

To describe the \svt\ local alignment we consider each Si wafer as 
an independent rigid body located and oriented near its nominal 
construction position and orientation. We describe the true position of a wafer
by its displacement and rotation
relative to its nominal position.  The
local alignment of the full \svt\ thus requires
2040 geometric parameters, which includes 6 redundant global degrees 
of freedom for the position and orientation of the \svt\ within \babar.
As described in detail in Sec.~\ref{sec:curvature}, 
we also model aplanar distortion of the inner 3 layers of wafers,
adding 84 more parameters. 

\par

We choose to define the local alignment parameters in the
{\em local} wafer coordinate system, 
a right-handed Cartesian system with coordinates $uvw$, defined 
uniquely for every wafer.
The $u$ axis is defined to lie in the nominal plane of the Si wafer pointing in the 
direction of increasing $\phi$.
The $v$ axis is defined to 
lie in the nominal plane of the Si wafer, orthogonal to $\hat{u}$.
The $w$ axis is defined as the normal 
direction to the nominal plane of the Si wafer, pointing roughly outwards 
from the IP.
The local wafer coordinate system origin is defined as the geometric center of the wafer.

\par

The displacement component of the local alignment is given by the vector
($\delta_u$, $\delta_v$, $\delta_w$), which describes the true position of the Si
wafer relative to its nominal position in that wafer's nominal local coordinate system.
Similarly, we describe the orientation of the wafer as the vector
($\alpha_u$, $\alpha_v$ and $\alpha_w$), which defines small right-handed 
rotations about the $\hat{u}$, $\hat{v}$ and $\hat{w}$ axes (respectively)
of the nominal local coordinate system of the given wafer,
in units of radians.  The \babar\ detector reconstruction software is written
so that this convention of local alignment can be easily and efficiently applied
to the reconstructed position of \svt\ hits \cite{detectormodel}.

\par

The readout strips on the wafers in the barrel region of the \svt\
are oriented parallel to the
local coordinates.
The strips on opposite faces of each wafer are oriented
orthogonally to each other, providing $90^\circ$ stereo coverage.
The readout strips on the wedge wafers have a pitch which varies
slightly along their length, resulting in
strip directions which change slightly with position, but which
are still roughly parallel.
Hits reconstructed in the \svt\ using the strips 
parallel to $\hat{v}$ are referred to as $u$ hits, as that is the 
dimension they constrain.  Roughly speaking, these hits  
measure the $\phi$ position of the traversing particle.
Similarly, hits reconstructed using strips parallel
to $\hat{u}$ are referred to as $v$ hits, and they
measure the $z$ position of the traversing particle.

\par

The estimated 
Lorentz shift in the position of $u$ hits induced by \babar's 
solenoid is accounted for in the \svt\ hit reconstruction.  
Any difference between this estimate and the actual Lorentz shift
is absorbed into the $\delta_u$ parameter, however this can introduce
systematic errors as described in Sec. \ref{sec:survey}.

%
%
\section{Goals and Requirements of the \svt\ Local Alignment}
\label{sec:requirements}

The goal of the \svt\ local alignment is to determine 
the local alignment parameters with sufficient accuracy
that the remaining misalignments contribute negligibly
to the final uncertainty in the physics quantities extracted using 
the tracks reconstructed in the \svt.  
For instance,
to observe \CP\ violation in $\Upsilon (4S)\to B^0\overline{B}^0$,
the \svt\ must be able to precisely measure the roughly
250\mum\ average separation between the $B$ meson decay vertices.
A full detector simulation study \cite{svt} showed this requires 
an average resolution of no worse than 10\mum\ for $u$ hits and 
20\mum\ for $v$ hits. To insure that the local alignment does not 
dilute these measurements, we require the statistical precision of
the local alignment contribute no
more than 15\% in quadrature (or  $1\%$ net) to the vertex resolution.
This implies
knowing $\delta_u$ to roughly 1.5\mum, $\delta_v$ to roughly 
3\mum, and $\alpha_w$ to roughly 2 $\mu$ radians.
To achieve this statistical precision
requires a data sample equivalent to 400 typical tracks per Si wafer.
This many tracks/wafer is recorded in less than an hour of normal
\babar\ data taking.  Thus meeting the required statistical precision
is not a challenge.  The real
challenge of the local alignment procedure is to control
the systematic errors to the required level.  To understand
the issues involved in controlling the systematics we must first
examine how the local alignment parameters are constrained
by data.

\par

A track passing through the full \svt\ and originating from the
interaction point (IP) will generally generate 
2 hits (1 $u$ and 1 $v$) in each of 5 layers. 
As a track's trajectory 
is well-described as a 5-parameter helix \cite{kalman},
a single track will constrain 5 degrees of freedom in the
local alignment.
%
%
However, because tracks scatter as they pass through material,
the most statistically powerful local alignment constraints will 
be on the relative positions of wafers in adjacent layers. 
Similarly, 
lab-bench measurements of relative wafer positions are useful only 
for nearby wafers, as mechanical and thermal stress uncertainties 
grow quickly with relative distance. 

\par

\begin{figure}
\setlength{\epsfxsize}{0.75\linewidth}\leavevmode\centerline{\epsfbox{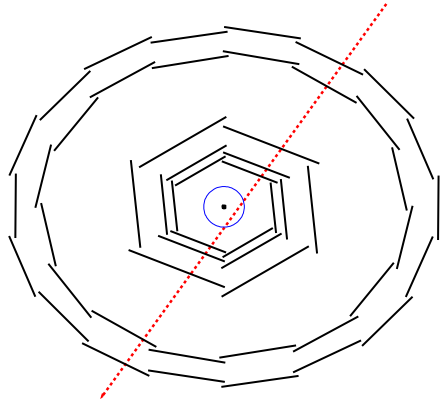}}
\caption{
Graphical illustration of one of global
distortions that affect the relative position of nearby wafers 
only to second order.  The {\em elliptical} effect shown has a
greatly exaggerated scale compared to what is used for validation studies.
}
\label{fig:distortions}
\end{figure}

\begin{figure}[hb]
\centerline{
\setlength{\epsfxsize}{0.5\linewidth}\leavevmode\epsfbox{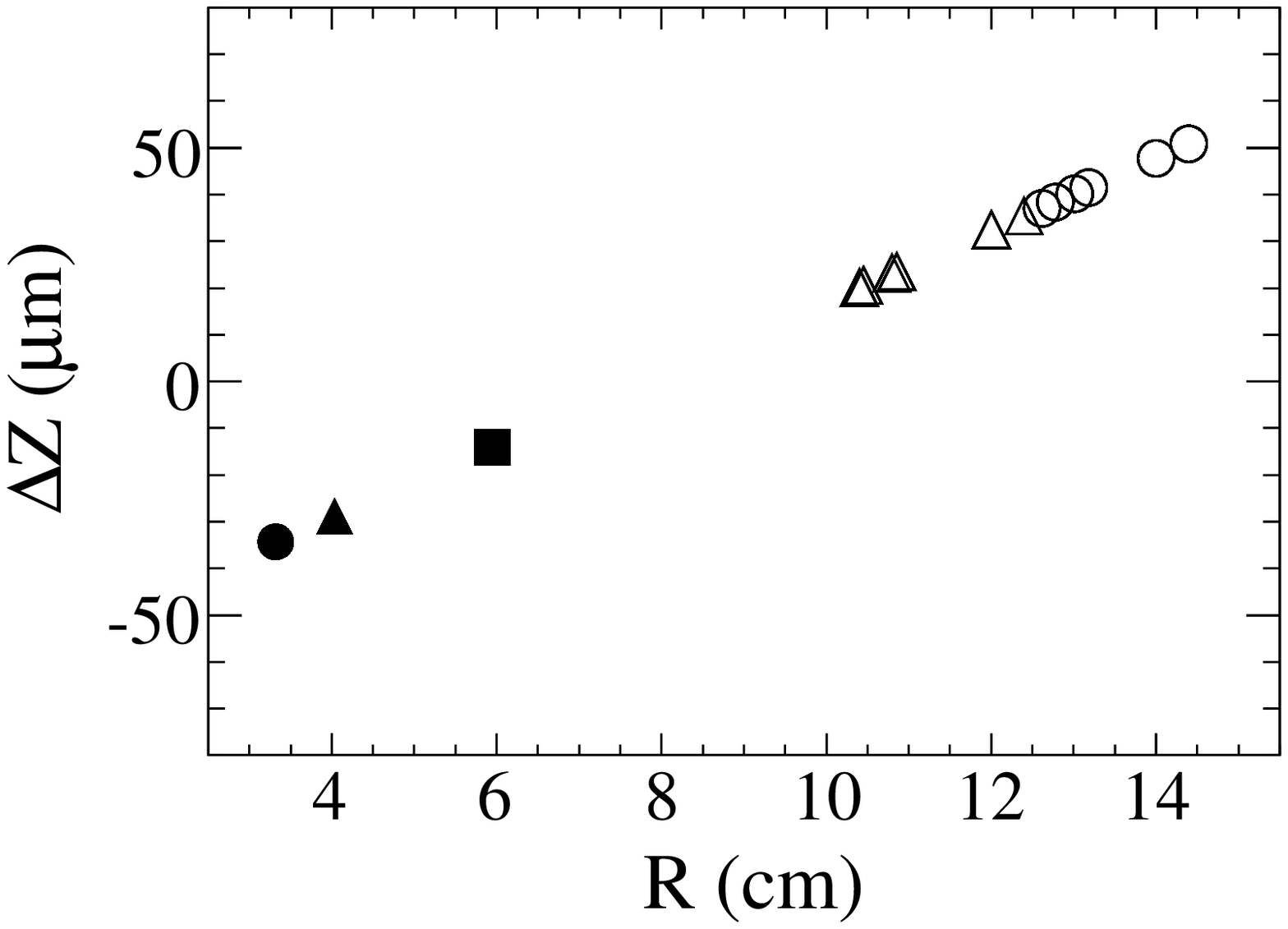}
\hspace*{-10pt}
\setlength{\epsfxsize}{0.5\linewidth}\leavevmode\epsfbox{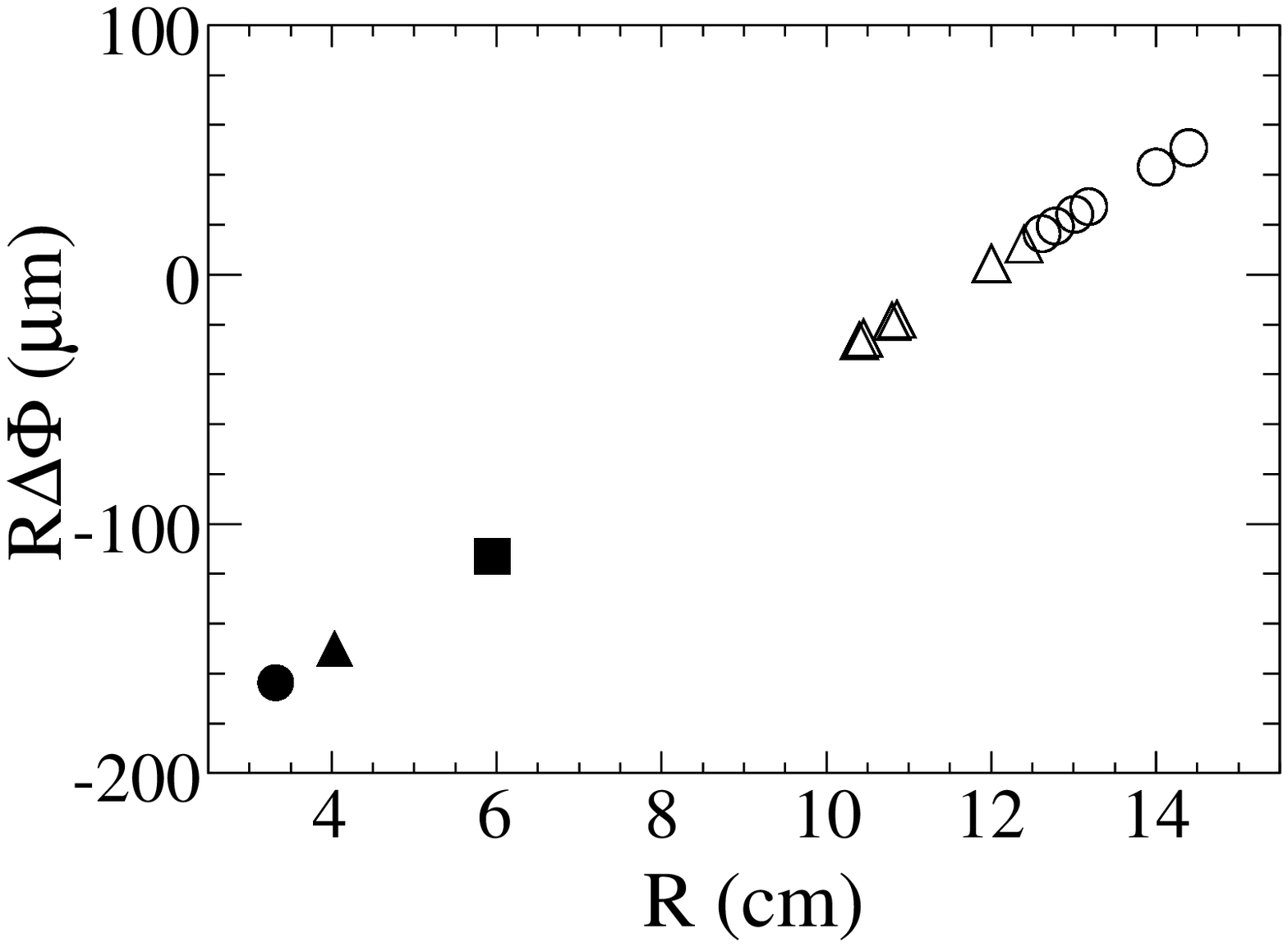}
}
\centerline{
\setlength{\epsfxsize}{0.5\linewidth}\leavevmode\epsfbox{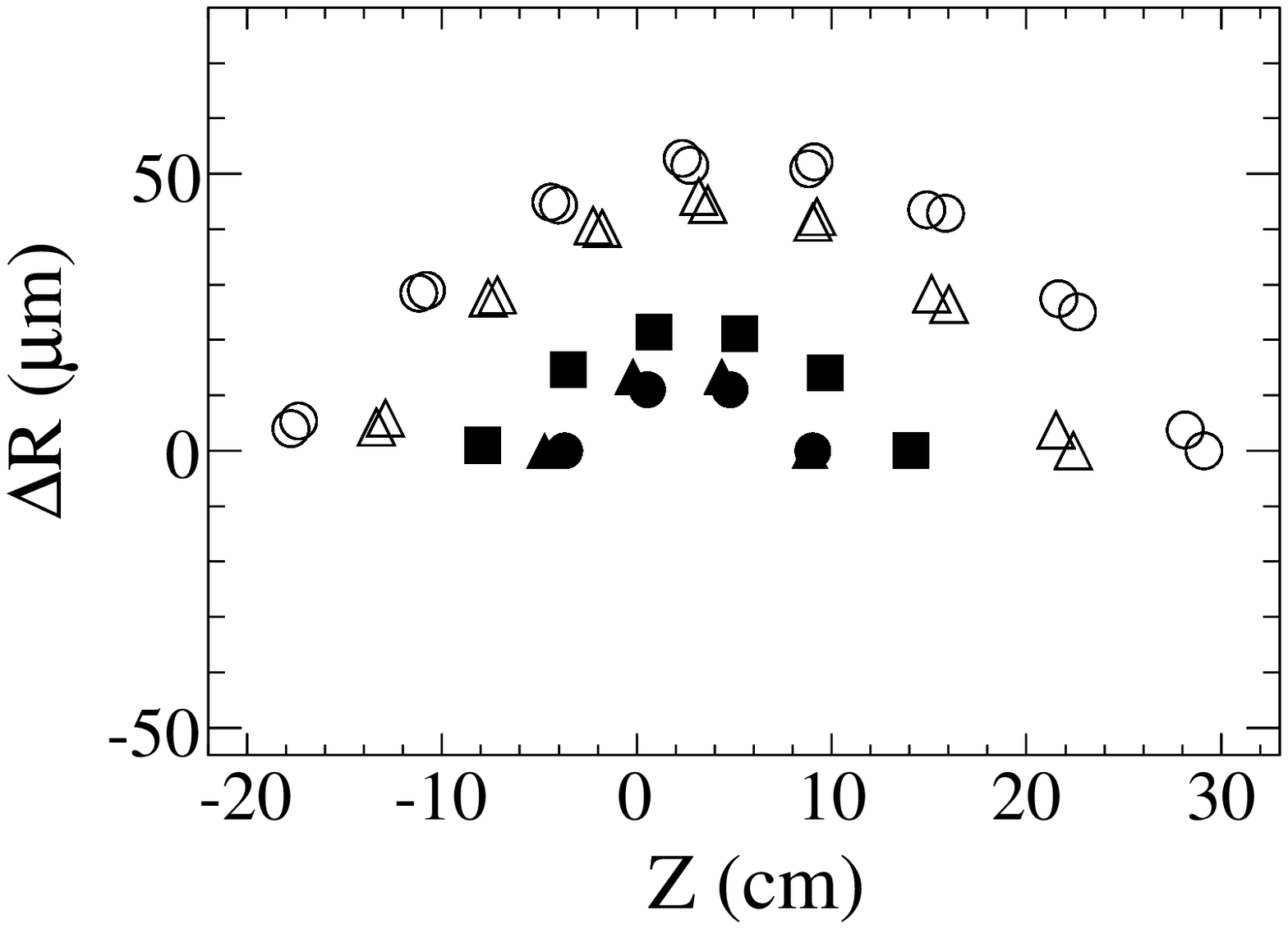}
\hspace*{-10pt}
\setlength{\epsfxsize}{0.5\linewidth}\leavevmode\epsfbox{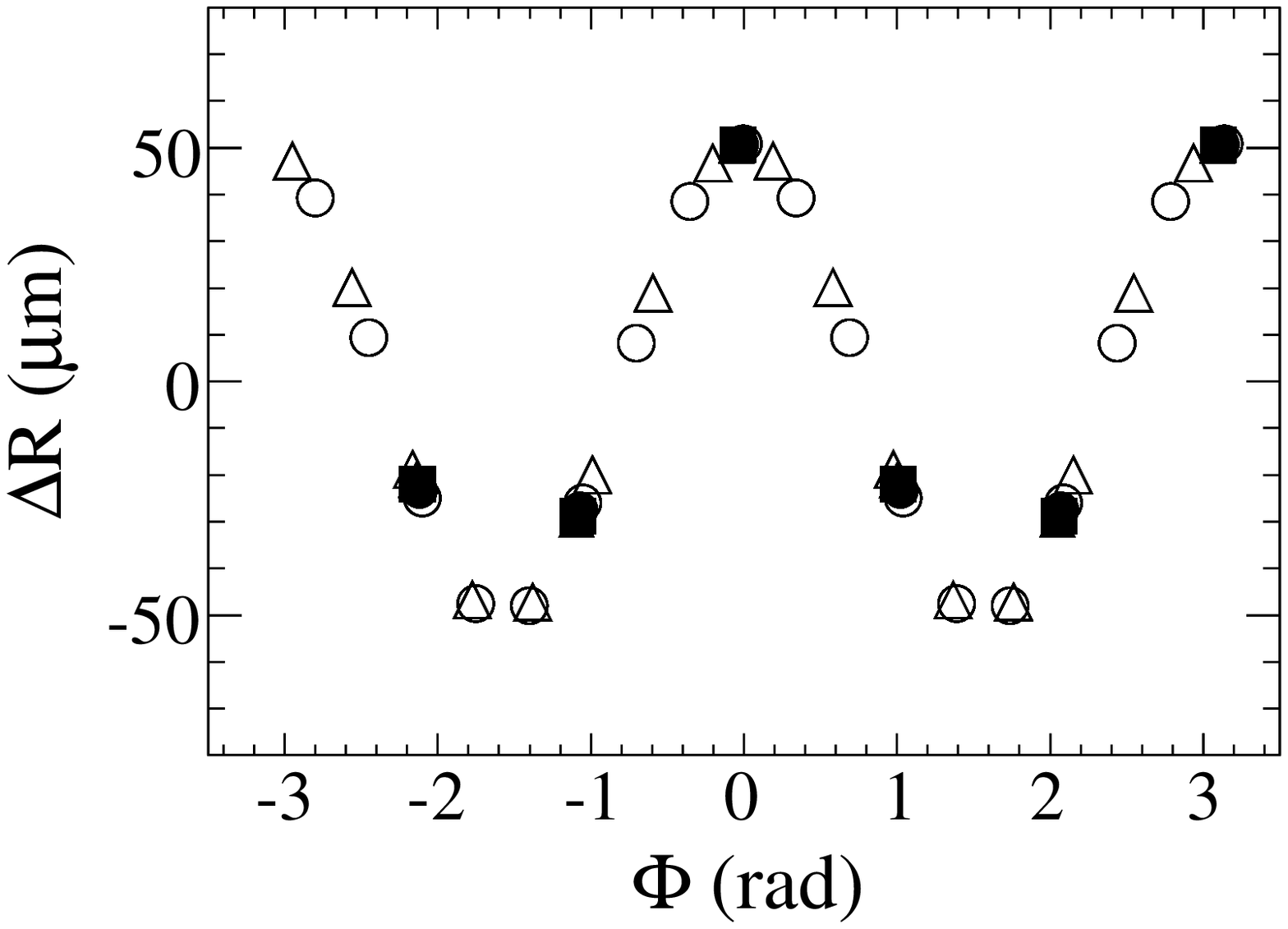}
}
\caption{
Four typical global systematic distortions out of nine 
discussed for a cylindrical
system in Table~\ref{tab:svtdistortions}:
telescope, curl, bowing, and elliptical effects.
Each point represents displacement of an \svt\ wafer shown with 
filled circles for layer one (${\bullet}$), 
filled triangles for layer two (${\blacktriangle}$),
filled squares for layer three (${\blacksquare}$),
open triangles for layer four (${\triangle}$),
and open circles for layer five (${\circ}$).
The typical scale of distortion is chosen to be 50\mum.
}
\label{fig:svt-misalign-initial}
\end{figure}

Thus, the track-based and lab-bench measurements used
by the local alignment procedure
effectively only constrain the relative positions of nearby wafers.
Many independent local constraints may of course be added together
to completely constrain the local alignment, 
but that procedure raises the risk of introducing a {\em global distortion}.
An example global distortion which correctly defines the relative position of
nearby wafers but distorts the \svt\ as a whole
is shown in Fig.~\ref{fig:distortions}. 
In Table~\ref{tab:svtdistortions} we summarize the main
global distortions in a system with cylindrical 
geometry, such as the \babar\ \svt.
In Fig.~\ref{fig:svt-misalign-initial} we illustrate
the effect of four global distortions with a natural
scale of 50\mum\ on the
position of individual wafers.

\begin{table}[ht]
\caption{
Main systematic distortions in a system with cylindrical geometry
and multiple layers. Distortions in $r$, $z$, and $\phi$ are
considered as a function these coordinates. 
}
\label{tab:svtdistortions}
\medskip
\begin{center}
\begin{tabular}{|c|c|c|c|}
\hline\hline
\vspace{-3mm} & & &    \cr
 & ~~~~$\Delta r$~~~~   &  ~~~~$\Delta z$~~~~ & $~~~r\Delta\phi$~~~   \cr
\vspace{-3mm} & & &    \cr
\hline
\vspace{-3mm} & & &    \cr
~~vs. $r$~~    & ~~radial scale~~   & ~~telescope~~  & curl   \cr
\hline
~~vs. $z$~~    & bowing  & $z$-scale    & twist      \cr
\hline
~~vs. $\phi$~~ & eliptical  & skew  & ~~~squeeze~~~  \cr
\hline\hline
\end{tabular}
\end{center}
\end{table}

If uncorrected, global distortions 
would produce unacceptable systematic biases in physics measurements.
For instance, an uncorrected {\em curl} distortion would introduce an
artificial charge-dependent momentum asymmetry to reconstructed tracks,
given the use of magnetic bending to define the charge and transverse
momentum of a track.  Likewise, a {\em radial scale} or {\em z-scale}
distortion would systematically change the measured distance scale of
the detector, distorting lifetime measurements.  To estimate specific
requirements on how well we must control the different global distortions,
we consider the implications of a subset of relevant \babar\ physics measurements.
For instance, to make competitive measurements of the $\tau$ lepton and $B$ 
meson lifetimes, the absolute distance scale
must be understood to better than 1 part in 1000.  
This implies controlling the {\em radial scale} 
and {\em z-scale} distortions to less than 1/1000, or roughly 
5\mum\ over the size of the \svt.  Similar arguments can be used to
motivate requiring that the local alignment constrain
the scale of all of the global distortions listed in Table
~\ref{tab:svtdistortions} to better than 5\mum\, or ten
times smaller than the effects shown in Fig.~\ref{fig:svt-misalign-initial}.
The verification that our local alignment procedure satisfies this requirement
is given in Secs.~\ref{sec:validation} and \ref{sec:validatesyst}.


\par

A further requirement on the the local alignment procedure is that it
be capable of following the time-dependence of actual changes in the detector.
We observed some slow relative motion of the \svt\ wafers related to humidity
changes, and due to stress changes during periods
of active access to the detector, and during changes to accelerator operation.
The timescale for observable changes was about a week, which
implies that the local alignment procedure should function
on less than a week's accumulation of \babar\ data.

\par

It is also important to be able to quickly detect when the
local alignment changes, to avoid extensive
reprocessing after the initial \babar\ reconstruction pass.  To obtain
feedback on possible alignment changes in a timely way
requires that computing the alignment constants
take no more than 24 hours.  Because it's run frequently,
the procedure must also be efficient in its use of computer resources.

%
%

\section{The Local Alignment Data Sample}
\label{sec:event}

The data used to perform the \svt\ local alignment are selected
from those collected during normal physics running of the \babar\ detector.
The \babar\ physics trigger accepts a mix of events including hadronic 
final states of $e^+e^-\to q\bar{q}$ and
$e^+e^-\to \FourS$, \llpair\ events, and cosmic rays which pass near the nominal IP.
The IP consistency requirement of a few cm was set in the trigger 
configuration during the first three years of \babar\ data-taking. 
It was relaxed for the cosmic tracks with large impact parameters for 
studies. However, in our final analysis we adopt uniform approach to all 
data periods and apply the same IP requirements discussed below. 

\begin{figure}[htbp]
\begin{center}
\setlength{\epsfxsize}{1.0\linewidth}\leavevmode\rotatebox{0}{\epsfbox{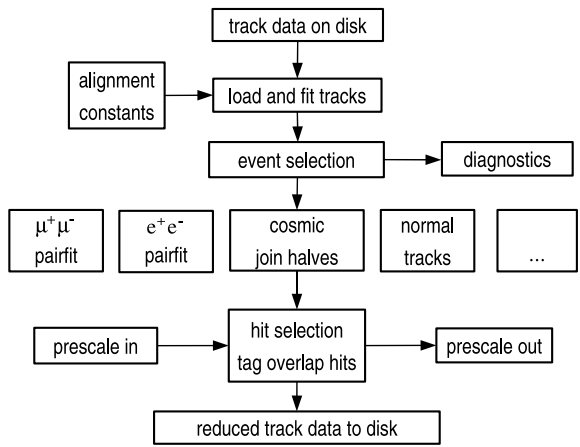}}
\caption{ Diagram of the sequence for the event, track, and hit selection,
including calculation of prescale factors. See text for details.
}
\label{fig:selection}
\end{center}
\end{figure}

Events which contain tracks useful for the
local alignment are identified, and the relevant tracks and hits are saved.
These samples are passed to the minimization procedure described in the next
section.  Details of the data selection are presented in the following subsections, and shown
graphically in Fig.~\ref{fig:selection}.

\subsection{Event Pre-selection}

A pre-selection of events of eventual interest to the local alignment procedure is integrated
into the \babar\ {\em prompt reconstruction} procedure that runs shortly after 
the events are recorded \cite{promptreco}.
All triggered events are first passed through a minimal background rejection procedure which
removes most beam-gas interactions and scales down \bhabha\ interactions.
A pseudo-random prescaling is then applied to  \bhabha, \mupair\
and cosmic ray triggers, which results in a roughly uniform illumination of the detector.  

\par

Pre-selected events are reconstructed using the standard \babar\ reconstruction program.
Tracks are found using standard pattern recognition algorithms in both the \svt\ and \dch.
Tracks found in the \dch\ (\svt) are extrapolated into the \svt\ (\dch) 
respectively, and hits consistent with the original fit are added.
Tracks are fit using a Kalman filter 
algorithm \cite{kalman} that accounts for differing hit resolutions, detector material, 
and magnetic field inhomogeneities.

\par

Prompt reconstruction uses the most recent local alignment parameters 
available at the time of processing.  If, at the end of the local alignment 
procedure, we observe local alignment change that might have affected 
the event selection, we repeat the procedure using the updated local 
alignment parameters.  Selection iteration was necessary only when
restarting the detector after a shutdown or detector opening.
A single iteration was sufficient to select an unbiased sample,
because the typical changes are not large.  The one case
when several iterations were required
was the startup of \babar\, when the starting alignment was
from the optical survey.

\par

Pre-selected events are written to a
dedicated calibration {\em stream} (file).  The calibration stream
persists the events in the standard \babar\ {\em mini-dst} event format \cite{cm2}, which records the reconstructed tracks and their associated hits.
In particular, this format records
the {\em local} centroid of all \svt\
hits associated with a track.  This allows measuring and applying
a new local alignment without first having to remove the effect of the
alignment used in prompt reconstruction.

\par 

The rarest process used in the
local alignment turns out to be cosmic rays, whose rate is independent of beam
luminosity.  Based on the cosmic ray trigger rate, and the number of tracks
required to satisfy the statistical and systematic constraints,
two days of cosmic data is sufficient to meet the alignment
goals.  We therefore define a local alignment data sample based on a fixed
calendar period of around 48 hours.

\subsection {Event Categorization and Final Selection}

A separate procedure makes a final selection of data useful for local alignment.
This reads back the calibration stream, and reconstitutes the
\svt\ and \dch\ hits using the current local alignment and calibration.
From these the full Kalman filter track fit is rebuilt, using the
recorded hit assignments.  These tracks are used to make a final event
categorization and selection.

\par

To insure a reliable momentum measurement, only tracks with at 
least 10 DCH hits, and at least 2 $v$ and 3 $u$ \svt\ hits 
(the minimum to fully constrains
all 5 track parameters) are selected for use in the local alignment.
To minimize multiple scattering effects, we also require a
transverse momentum of at least 1 \gevc.  To cut down on background from
secondary (material) interaction products,  
we accept only tracks whose point of closest approach to the
\babar\ $z$ axis is within 1.5 cm of the IP in the plane transverse
to the $z$ axis, and between $-7$ cm and $+9$ cm of the IP along the $z$ axis.
 
\par

Events are categorized and finally selected
based on the multiplicity and properties of their selected tracks.
We define three categories of events in the local alignment; 
\mupair\ or \bhabha\ (\llpair) events, cosmic ray events,
and `normal' events.  The definitions and selections of these categories is described
below.

\par

Events with exactly two selected tracks are tested as potential \llpair\ events.
Tracks in \mupair\ candidates are required to have associated \emc\ signals consistent with
a minimum-ionizing particle.  Tracks in \bhabha\ candidates are required to have associated \emc\
energy deposition consistent with the reconstructed track momentum.
All \llpair\ events are required to have a total energy (computed from track
momentum) consistent with the known combined energy of the initial \epem\ beams, and to be
back-to-back in the transverse plane.

\par

Candidate \llpair\ events that pass the above cuts are refit using 
a special form of the \babar\ Kalman filter track fit
which constrains the pair of tracks to have the same four-momentum as the initial \epem\
system, within the independently-estimated errors on the beam particle momenta.
If the $\chi^2$ of the pair fit is larger than 50, it is considered a failed fit, and
the individual tracks in these events are passed down to the `normal' track selection algorithm
described below.  The most common cause of failed pair fits is initial and/or final state
radiation.
When successful, the pair fit improves the track
momentum resolution by more than a factor of 10.
More importantly, the pair fit creates a correlated system
in which information passes from one track through the IP to the other track.
This allows the local alignment procedure to constrain the positions of wafers
relative to those on opposite sides of the detector.
We can also use \llpair\ events
to determine the beam momenta parameters simultaneously with the local alignment
parameters, without using the independent beam energy estimate.  This provides both
a systematic check on the alignment procedure, and a precise 
way to measure the beam boost.  This techinque is discussed in Sec.~\ref{sec:svtla_beam}.

\par

Because the \babar\ track finding algorithm assumes all particle originate at or near the IP,
a single cosmic ray passing through \babar\ is initially reconstructed as 2 tracks, splitting
the cosmic ray trajectory through the detector roughly
in half.  Cosmic ray event candidates are selected as having two
well-measured oppositely charged tracks which match
in angle and position at the their point of closest approach to the IP.  These tracks are
also required to have associated \emc\ signals consistent with a minimum-ionizing particle. 
If these criteria are satisfied, the
hits from the upward-going track are added to those on the downward going track, and the combined
track is refit.  

\par

Events which fail selection as  \llpair\ or cosmic ray events
are taken together with multi-track events as candidates for the `normal' category.
In this category, we select all the tracks which pass the usual selection cuts, plus
an isolation cut.  Any event with at least one selected track is classified as `normal'.

\subsection {Track Selection and Preparation}
\label{sec:tracks}

Selected tracks from events which pass final event selection in any category
are themselves labeled according to their event category.  In the case of `normal' events,
tracks are further categorized according to whether or not they have hits in
adjacent wafers of the same layer in the {\em overlap} region (see Fig.~\ref{fig:svt-2}).
As these overlap tracks have a very short extrapolation distance between
the same-layer hits, they provide a powerful constraint on the relative
position of adjacent modules, and so
are especially valuable in the local alignment procedure.  Non-overlap tracks in `normal' events
remain categorized as `normal'.

To balance the impact of the largely-independent global distortion systematic constraints afforded by the
different track categories, we perform a final track selection which roughly equalizes the statistical
power of the tracks in each category for every individual wafer.
Because the wedge wafers (see Fig.~\ref{fig:svt-1}) subtend a region of polar angle $\theta$ where the rate from
\mupair\ and other physics events changes rapidly with $\theta$, we further divide these into two roughly
equal parts.

To allow better control of the propagation of systematic misalignment effects from the DCH
into the \svt\ alignment tracks, we refit all tracks using the following technique.
First, we split the tracks into two, one with all the \svt\ hits and one with all the \dch\ hits.
Each of these associated but separate tracks are refit using the standard \babar\ Kalman
filter fit.  The parameters and covariance matrix
of the \dch-only track fit are sampled at the point where that
track enters the \svt\ detector volume, and these parameters and covariance are then used to
{\em constrain} the \svt-only track fit.
Mathematically, the parameter constraint is identical to
the effect of having left the \dch\ hits on the track.  However, by {\em masking} some of the
parameters in the constraint, the information content of the \dch-only fit can be filtered.  In particular,
by masking off all but the $\omega$ parameter (inverse curvature) of the \dch-only fit in the constraint,
we can greatly improve the momentum resolution of the constrained \svt-only track, without introducing any
dependence on possible systematic distortions in the position or orientation of the \dch.
We use the \dch-only fit $\omega$ constraint when fitting the \llpair, cosmic ray, and overlap
category tracks.

Because the sum of the local alignment parameters for all wafers include the 6 {\em global } degrees
of freedom, the local alignment procedure could introduce a global {\em drift}.  Because the $\omega$
constraint does not depend on the relative position or orientation of the \svt\ and \dch\, it cannot
constrain this global drift.
To minimize the global alignment drift, we use all 5 \dch-only fit parameters to constrain
the fit of the `normal' tracks.  As these tracks have the lowest statistical power, this introduces only
a modest and acceptable dependence on \dch\ alignment distortions.

\subsection {Hit Selection}
\label{sec:hits}

From selected tracks, we select \svt\ hits which provide information useful for
local alignment.  First, hits with questionable timing or cluster shape
are disabled, and the tracks which held them are refit.  
Then the remaining hits are filtered to make the sample
uniform over the detector, over several track categories, and over the
time window in which the data sample was accumulated.
Once the alignment procedure
is close to convergence, a final outlier removal cut is applied.  
Details of the hit selection are shown in Fig.~\ref{fig:selection} and described below.

\par

For each hit on each track in each track category in every wafer
we use a pseudo-random prescaling algorithm to select roughly 
100 (200) hits in the outer two (inner three) layers, respectively, for
use in the \chisq\ minimization.  More hits are used for the inner layers
to balance the larger number of wafers in the outer layers.
The pseudo-random prescale is seeded on the unique event time, 
and so is fully repeatable but effectively random. 
Because the data sample has a large number of tracks of each category,
only a small number of hits per track are selected,
reducing the number of correlated measurements used when
computing the alignment \chisq.
The unselected good hits are still used for the track fit.
Cosmic events have the smallest number of tracks in the sample, and thus
have the largest prescale factor, corresponding to using roughly 5 hits per track.

\par

The hit selection is done using three passes over the data, interleaved with the 
\chisq\ minimization described in section \ref{sec:minimize}.  In the first pass,
the hit prescale constants are determined by dividing
the desired number of hits/wafer/category by the number observed.
In the second pass, these prescale values
are used to pseudo-randomly select hits to be used in the alignment \chisq.  Selected
hits are persistently tagged so that the same hits are used each iteration
of the \chisq\ minimization.  In this pass we apply very loose requirements
on the hit residuals, removing only the very worst outliers, so as to avoid
biasing the alignment parameters when the initial alignment is far from optimal.
After the \chisq\ minimization  has partially converged, we repeat
the hit selection procedure using the improved alignment parameters,
applying a tighter hit residual cut to suppress outlier hits which can distort the \chisq.
Selected hits are again tagged and passed on to a final pass of \chisq\ minimization.

\par

The hit residuals used in the alignment procedure are computed as the distance of
closest approach between the given hit, defined as a line in space, and the track trajectory,
defined as a piecewise helix in space, after removing the effect of that hit from the track fit.
The residual is signed by the cross-product between the track direction and a nominal hit direction.
The error on this unbiased residual is computed as the square-root of the
quadratic sum of the projections of the hit error and the track covariance matrix onto the residual measurement.  The
hit error is estimated as a function of the hit's pulse-height and width, and the
direction of the track.  The hit position error functional form and parameters were tuned
using \babar\ data.

\par

The hit residuals and their estimated errors are combined to compute a {\em hit \chisq} for 
a particular wafer:
\begin{eqnarray}
\label{eq:chisqhits}
\chi^2_h & \equiv & \sum_{i}^{\rm hits} \epsilon_{i}^T{\bf V}^{-1}_{i}\epsilon_{i},
\end{eqnarray}
where $\epsilon_i$ is the residual for hit $i$.
This \chisq\ is relative to the
{\em test} set of alignment parameters used when the track was fit.

%
%
\section{The Optical Survey Alignment }
\label{sec:survey}

The components of the \svt\ were optically
surveyed to determine their relative positions
at several stages during construction.
First, individual modules were
surveyed on the lab bench during their construction.  Then,
each layer was surveyed after its modules were mounted on
the support structure, starting with the innermost layer and
going out.  By studying their reproducibility,
these surveys were estimated to have
a precision of roughly 5\mum\ in the wafer plane
and 20\mum\ out of the plane.

\par 

By averaging and combining
the raw survey measurements, they were converted
into a {\em survey alignment}, describing the relative positions and orientations of
all the wafers in the \svt.  The survey alignment was used as
the initial condition when the local alignment procedure described in this
paper was first performed.  The survey alignment is also used
as an additional constraint on relative wafer positions in the
local alignment procedure itself, as described in detail below.

\par

\begin{figure}[t]
\begin{center}
\centerline{
\setlength{\epsfxsize}{0.5\linewidth}\leavevmode\epsfbox{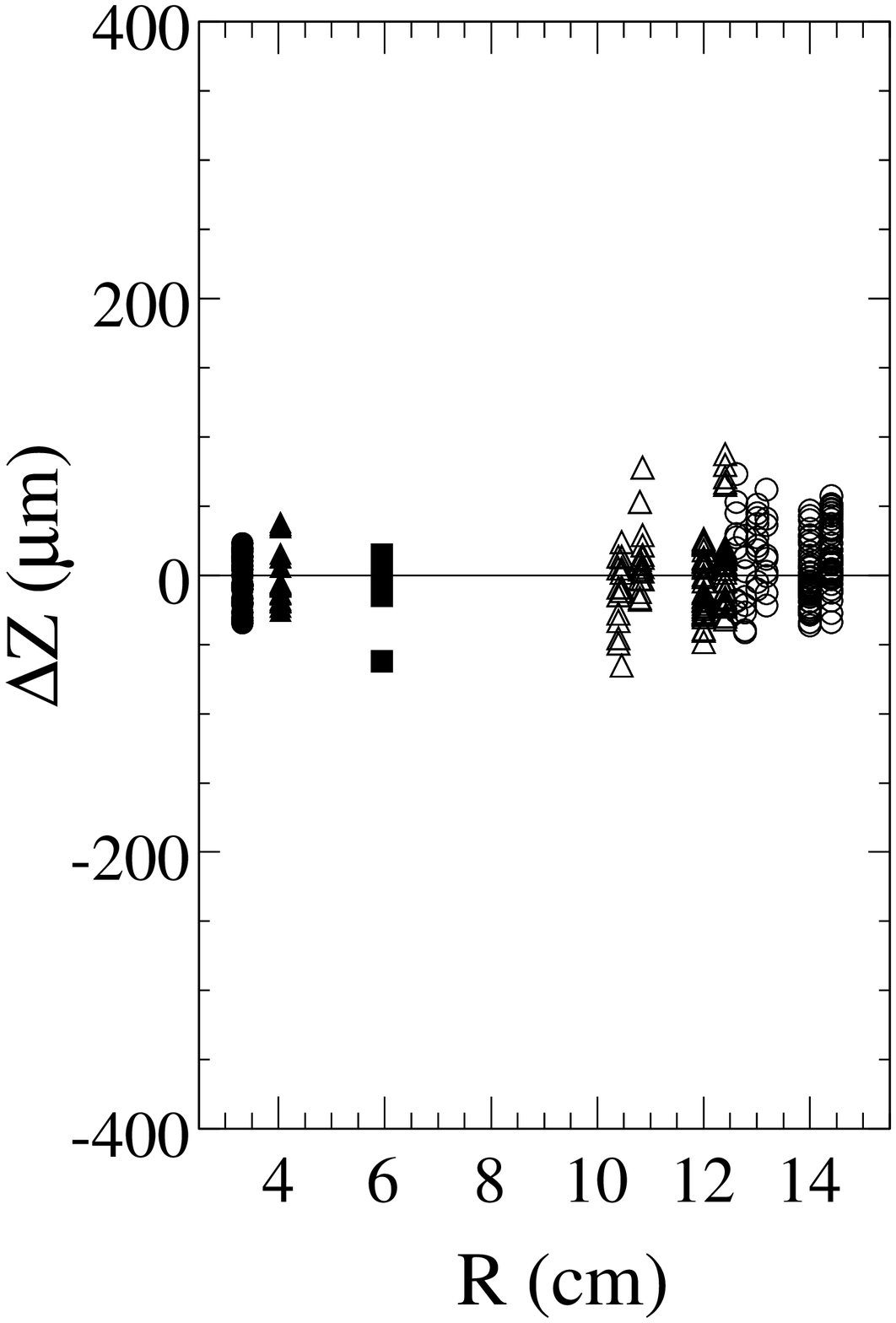}
\hspace*{-10pt}
\setlength{\epsfxsize}{0.5\linewidth}\leavevmode\epsfbox{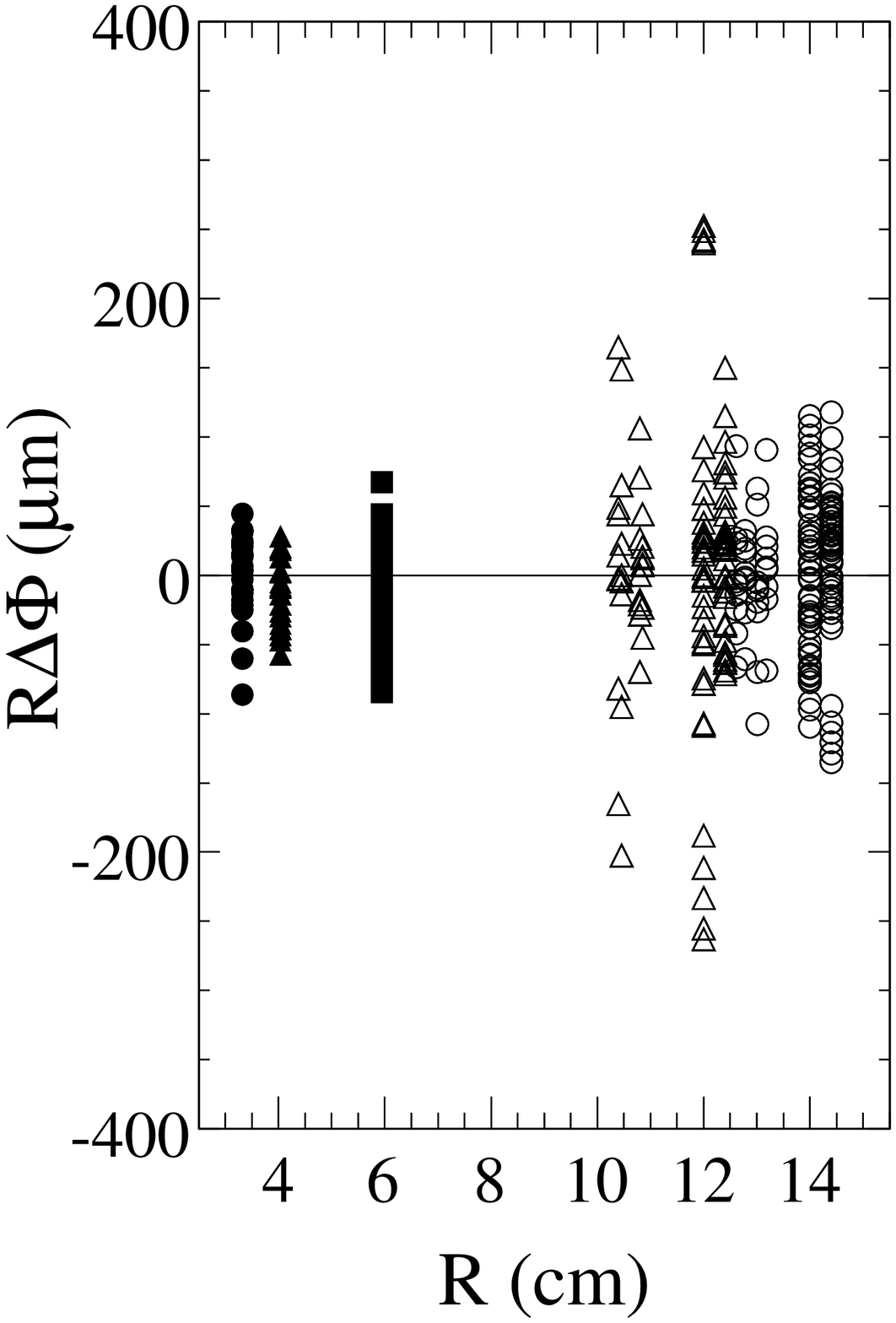}
}
\vspace*{-15pt}\centerline{
\setlength{\epsfxsize}{0.5\linewidth}\leavevmode\epsfbox{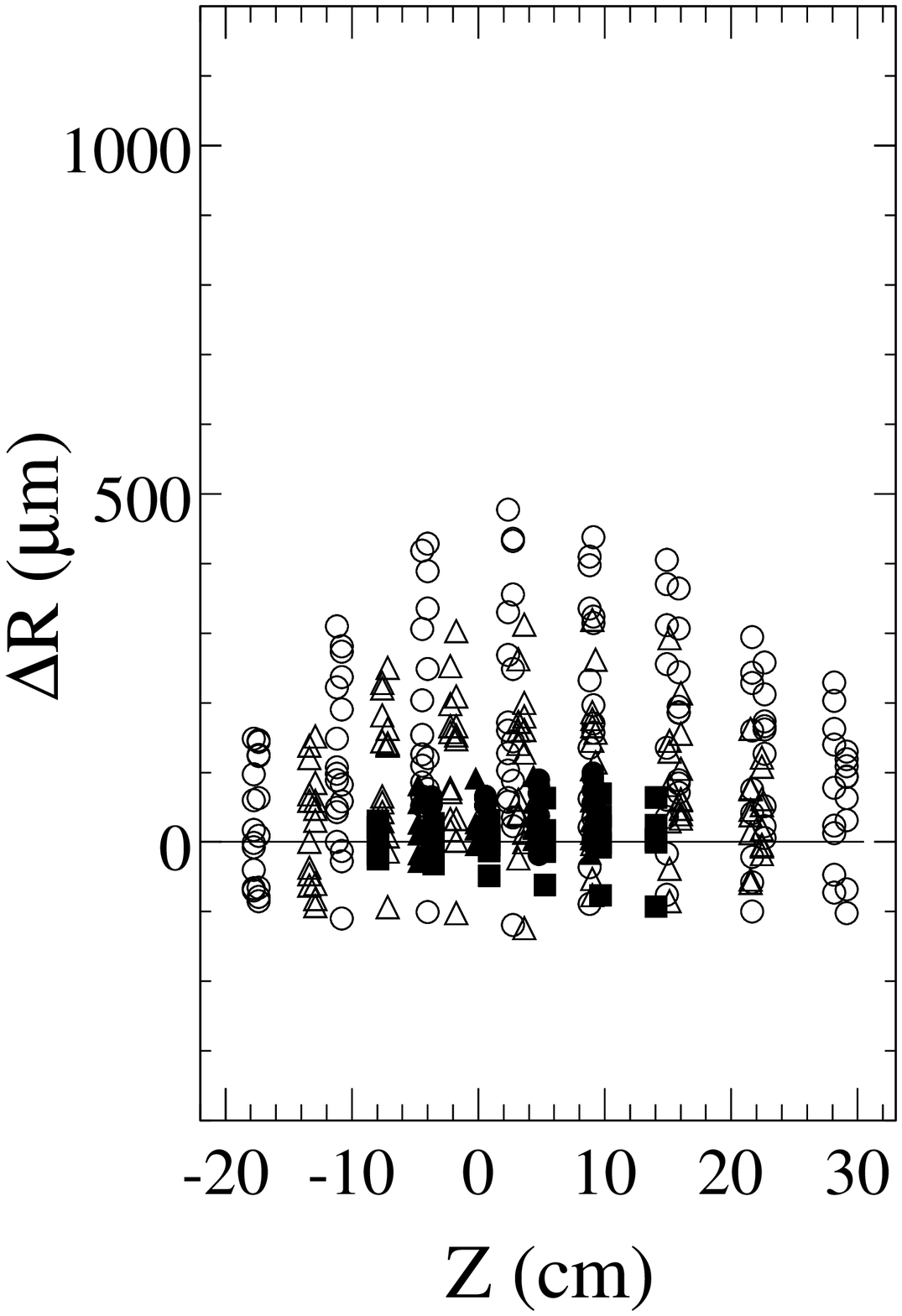}
\hspace*{-10pt}
\setlength{\epsfxsize}{0.5\linewidth}\leavevmode\epsfbox{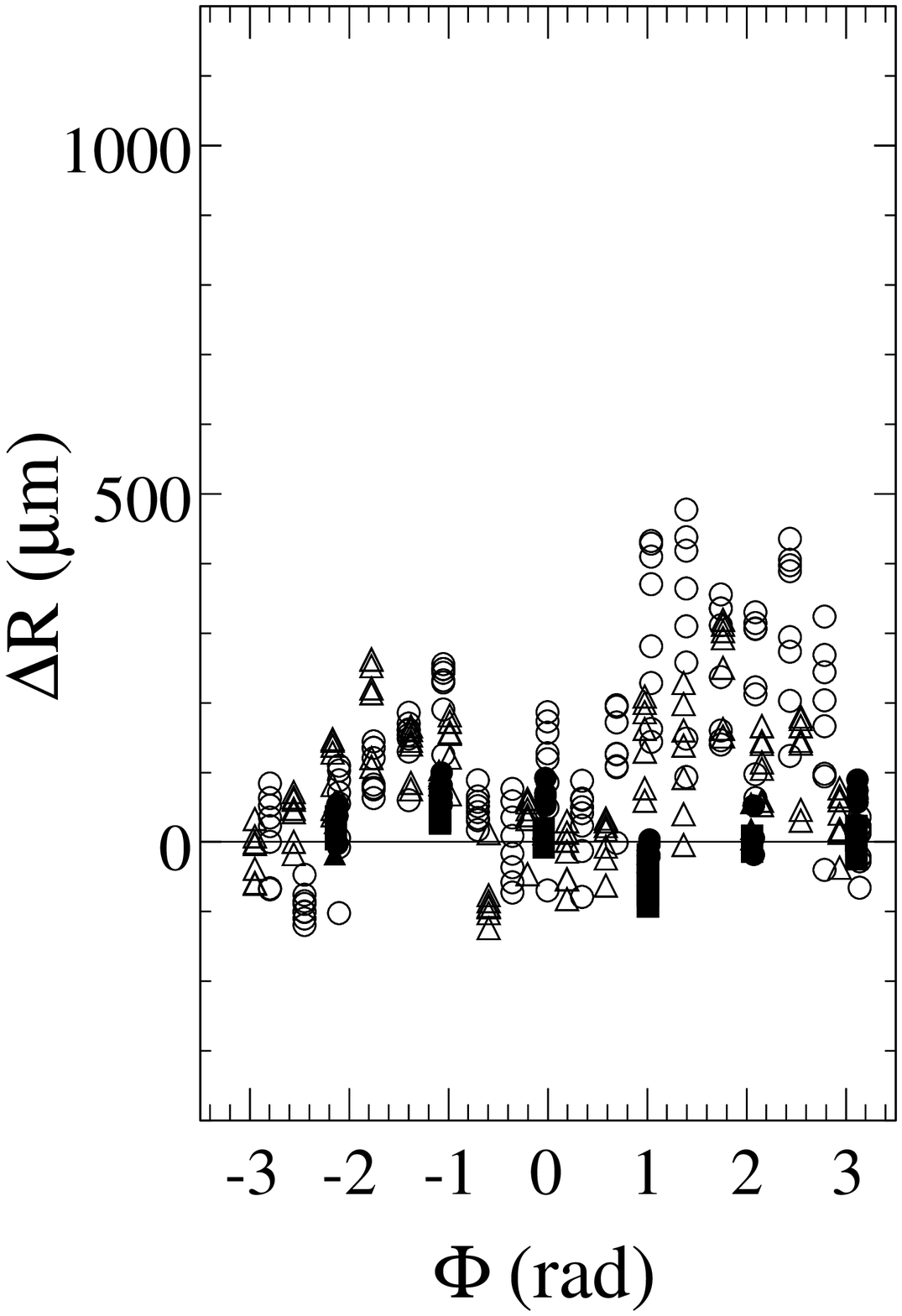}
}
\caption{
Differences in the \svt\ wafer positions between those
measured by the procedure described in this paper and the optical survey alignment,
projected in the dimensions illustrated in Fig.~\ref{fig:svt-misalign-initial}.
Each point represents a single wafer, with 
filled circles for layer one (${\bullet}$), 
filled triangles for layer two (${\blacktriangle}$),
filled squares for layer three (${\blacksquare}$),
open triangles for layer four (${\triangle}$),
and open circles for layer five (${\circ}$).
}
\label{fig:survey-1999}
\end{center}
\end{figure}
\begin{figure}[htbp]
\begin{center}
\centerline{
\setlength{\epsfxsize}{0.5\linewidth}\leavevmode\epsfbox{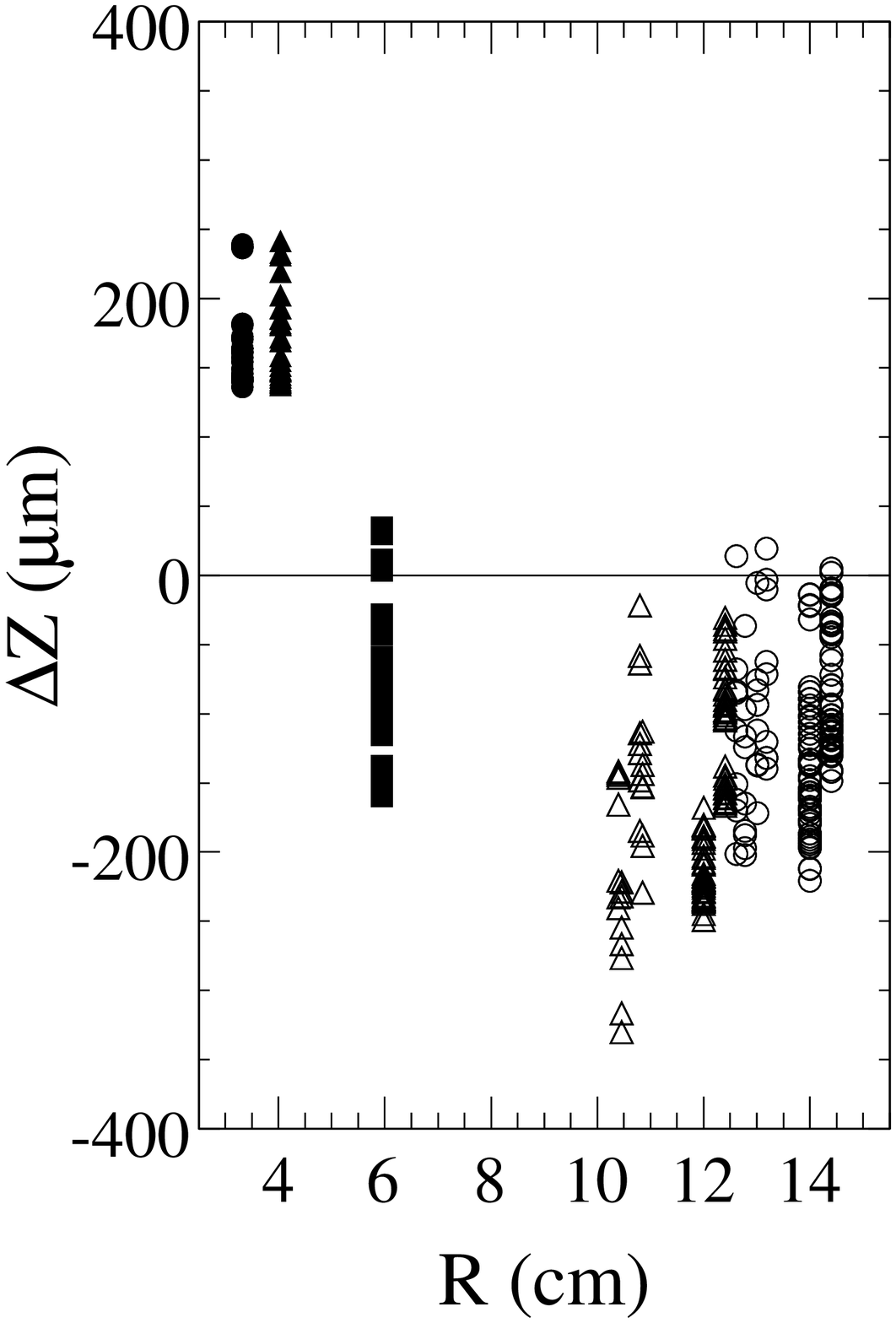}
\hspace*{-10pt}
\setlength{\epsfxsize}{0.5\linewidth}\leavevmode\epsfbox{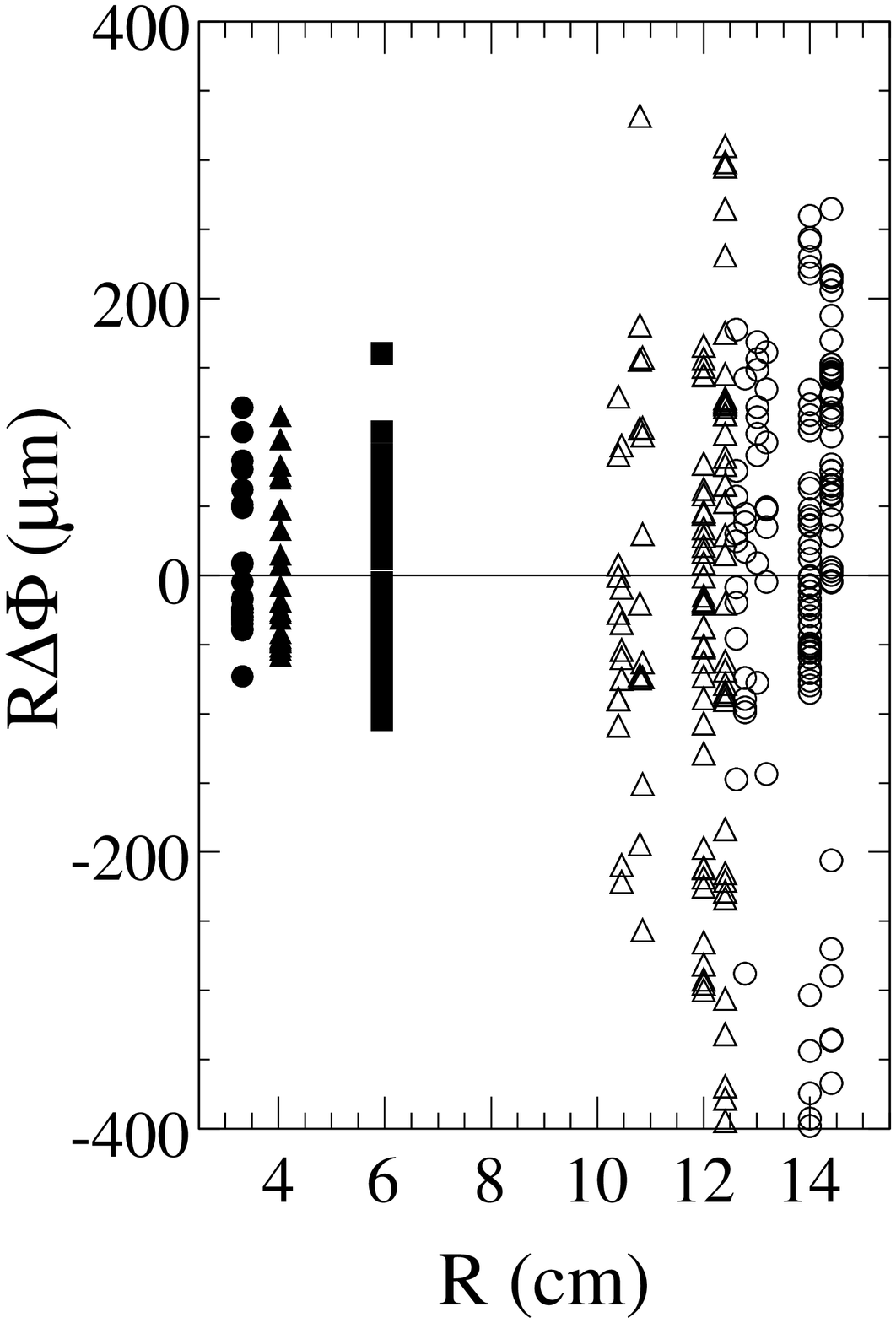}
}
\vspace*{-15pt}\centerline{
\setlength{\epsfxsize}{0.5\linewidth}\leavevmode\epsfbox{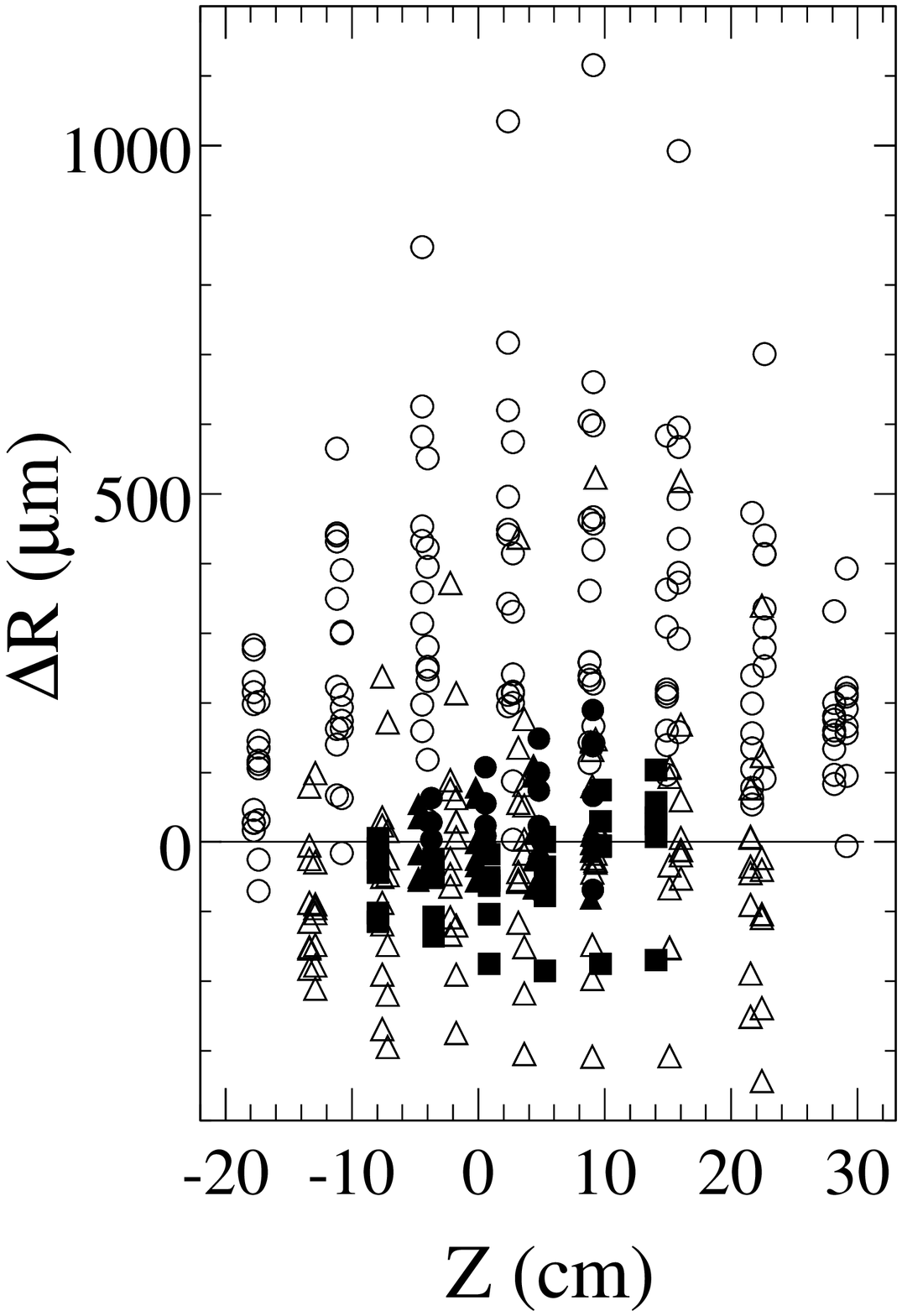}
\hspace*{-10pt}
\setlength{\epsfxsize}{0.5\linewidth}\leavevmode\epsfbox{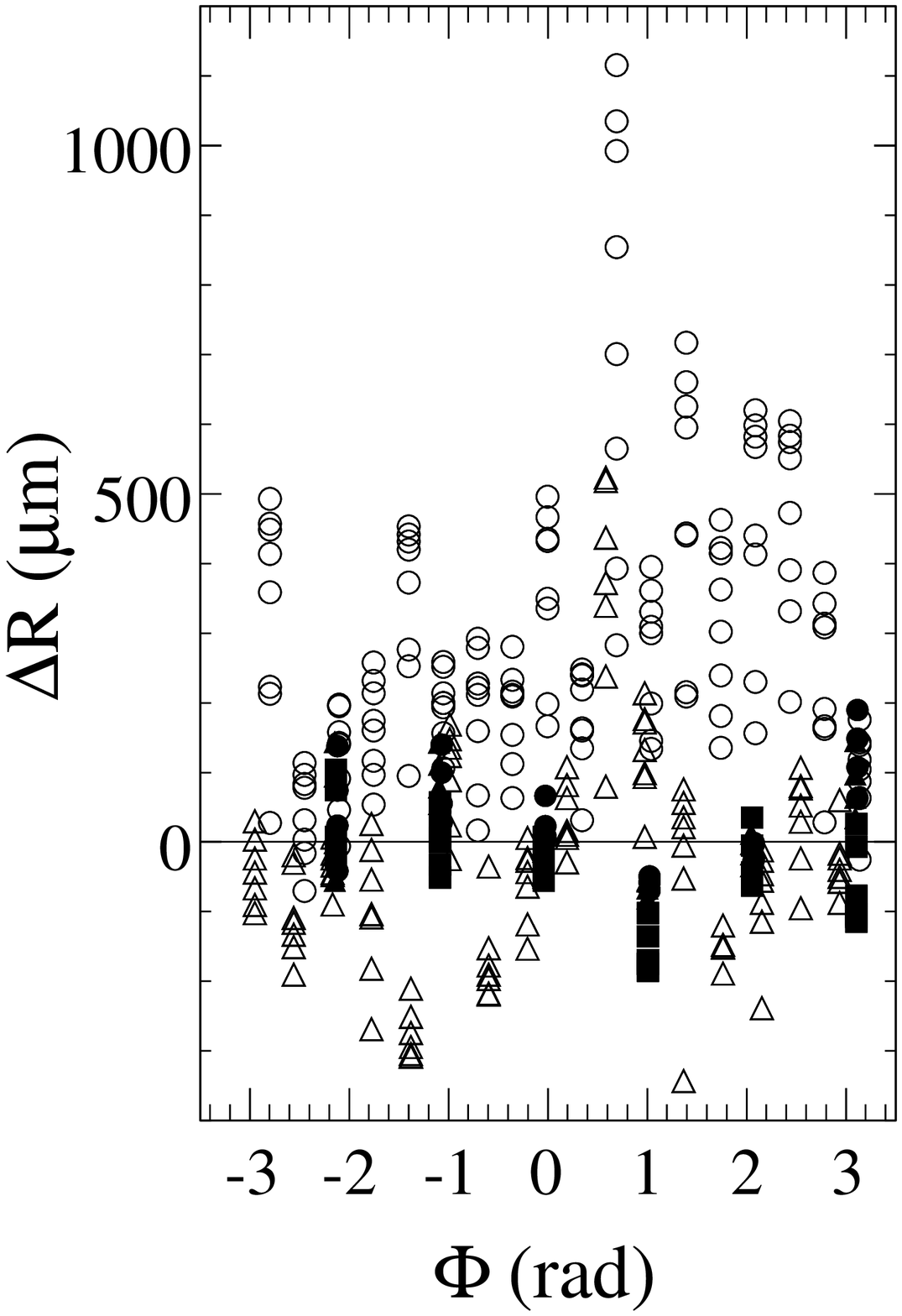}
}
\caption{
Differences in the \svt\ wafer positions between those
measured by the procedure described in this paper and the nominal geometry,
projected in the dimensions illustrated in Fig.~\ref{fig:svt-misalign-initial}.
The symbols used are defined in the caption of Fig.~\ref{fig:survey-1999}.
}
\label{fig:null-1999}
\end{center}
\end{figure}

A comparison of the wafer positions determined using the
procedure discussed in this paper with
the optical survey alignment is shown in Fig.~\ref{fig:survey-1999}.
The differences are shown after removing the overall global shift and
rotation between the two alignment descriptions.
These figures, plus tests we made in the early days of \babar\, demonstrate
that the survey alignment by itself does not meet the local alignment
requirements defined in Sec.~\ref{sec:requirements}.  This is understandable,
as mechanical stresses and other operational effects will alter the relative
wafer positions of the installed detector compared to lab-bench measurements.

\par 

The survey alignment does however contain useful information.  This is demonstrated
by comparing Fig.~\ref{fig:survey-1999} with Fig.~\ref{fig:null-1999}, which 
itself compares the measured wafer position with the \svt\ nominal geometry.
Clearly the optical survey is a better approximation to the true local alignment
than the nominal geometry.  We also expect some aspects of the survey alignment to
remain accurate even in the installed detector.  In particular, the relative positions of adjacent
wafers in a module should be well described by the survey alignment, as there is little
room for stress-induced motion between them that would not destroy the module.
Information on the relative position of wafers within a module
is orthogonal to that provided by hit residuals, which relate the
relative positions of wafers in different modules.  It is therefore
desirable to add the survey alignment information to the local alignment procedure.
However, only the reliable parts of the survey
alignment should be used, and the survey information must be appropriately
combined with the hit residuals information.

\par

We add survey information to the local alignment procedure by
constructing a {\em survey residual} for each wafer.  The survey
residual compares the wafer's position
relative to the other wafers in its same module as predicted by
the survey alignment with the relative position given by the
{\em test} alignment, analogous to the hit residual definition given in
Sec.~\ref{sec:hits}.  We assign an error to this residual
based on the survey alignment precision, corrected for systematic
effects which decrease the accuracy.  We then construct a survey \chisq\
contribution from this residual and error, which is combined
with the hits \chisq\ to form a total local alignment \chisq.

\par 

The optical survey wafer residual is computed using the
survey measurements in the module containing that wafer.
We represent the $N$ wafers of a given module
($i=1,..,N$, e.g. $N=8$ in layer five) spatially by a set of
9 points ($j=1,..,9$) lying in the wafer plane, at fixed positions
in the wafer coordinate system, located roughly at the ends
and midpoints of the sides of the rectangle defined by the wafer's active area.

The wafer points in a module can be translated into the global \babar\ coordinate
system using either the survey alignment, or the test alignment.
We use the test alignment to describe the nominal global coordinate positions
of these points ${\vec{r}_{ij}}$.  We describe the difference between
the survey alignment and the test alignment transformation of these points
into the global coordinate system as $d\vec{r}_{ij}$.

\par

We use the difference vectors to solve for the 
translation vector $\vec{R}$ and rotation vector $\vec{\Omega}$
which minimize the total distance between the points in global
coordinates as described by the two alignments (test and survey).
Together these represents the optimal transformation between the two alignments.
The optimal translation $\vec{R}$ is given by the average of the
vector differences between the measurements
\begin{eqnarray}
{ {R}}_{ l} = \left(\sum_{j, i}^{n\times N}{ d\vec{r}_{ij}}\right)_{ l} 
/ \left(\sum_{j, i}^{n\times N}1\right)       \,,
\label{eq:survey-transl}
\end{eqnarray}
where $l=1,2,3$, representing the spatial coordinates.
The optimal rotation $\vec{\Omega}$ is defined implicitly by the equation
\begin{eqnarray}
\sum_{k=1}^{3}{ \Omega_k}
\sum_{j, i  }^{n\times N}
\left(\delta_{ kl}({ \vec{r}_{ij}})^2
-({ \vec{r}_{ij}})_{ k}({ \vec{r}_{ij}})_{ l}\right)= \cr
\sum_{j, i}^{n\times N}
\left({ \vec{r}_{ij}}\times { d\vec{r}_{ij}}\right)_{ l} \,,
\label{eq:survey-rotate}
\end{eqnarray}
where $k,l$ are spatial coordinates and $\delta_{ kl}$ is the Kronecker delta.
Solving for $\vec{\Omega}$ requires inverting a $3\times3$ matrix.
The components of the rotation vector $\vec{\Omega}_k$ represent right-handed
rotations about the respective coordinate axis.  These rotations are
calculated with respect to the average nominal global position $\sum\vec{r}_{ij}$,
assuming a small angle approximation (or equivalently
small $d\vec{r}_{ij}$).  A small number of iterations 
of this procedure was found to be sufficient to solve for the optimal transformation
even when the rotation angles are large. This method of describing and solving for
coordinate transforms is based on the formalism
of rigid body rotation.
It is equivalent to minimizing a $\chi^2$ constructed from $d\vec{r}_{ij}$, if we
assign an equal error to all dimensions of all points.

\par

To compute the survey alignment residual,
we consider each wafer in the module in turn,
referred to as the {\em wafer under consideration},
with index $i=I$.
For that wafer, we first calculate the transformation
that relates the overall module position predicted 
by the survey alignment with that predicted by the test alignment.
To avoid direct bias, we exclude the
wafer under consideration when computing 
this transformation, requiring $i\ne I$ in 
Eqs.~(\ref{eq:survey-transl})~and~(\ref{eq:survey-rotate}).
To reduce the impact of potential module deformation occurring after survey, we
use only those wafer points $j,i$ which are less than 
15 cm from the center of the wafer under consideration when computing
the sums.  This cutoff was found to be sufficient to
remove systematic bias due to module deformation,
while still providing enough points to give a statistically meaningful constraint.

\par

We apply the module-level transformation 
to all the points of the survey alignment, effectively overlaying the survey
alignment on the test alignment for this module's position.
We then apply Eqs.~(\ref{eq:survey-transl}) and~(\ref{eq:survey-rotate}) to the
wafer under consideration,
taking the transformed survey points and the test alignment to compute the 
$d\vec{r}_{ij}$,
and using only the points on the wafer under consideration by requiring $i=I$.
The resulting $\Delta\vec{R}_I$ and $\Delta\vec{\Omega}_I$ effectively 
define a 6-dimensional {\em survey residual}, representing the difference between the 
position and orientation of the wafer under consideration
described by the test alignment
versus the survey alignment, relative to the rest of the module
in the region around the wafer under consideration.

\par

We define a  $6\times 6$ survey covariance matrix 
${\bf{V_s}}$ to represent the intrinsic error in the survey residual.
We use the same covariance matrix for all wafers.
We approximate the survey covariance matrix to be diagonal
in the local wafer coordinates, and set the elements according to the values found in the
survey consistency tests discussed above.  Because the module design
restricts relative wafer motion in the module plane,
we take these values literally for the in-plane errors (translation in $u$ and $v$,
and rotation about $w$).  To account for potential bowing, twisting, or other
aplanar distortions only weakly constrained by the module design, we
increase by a factor of ten the estimated errors
on the remaining degrees of freedom (translation in $w$ and rotations 
about $u$ and $v$).

\par

We construct a survey \chisq\ from the 6-dimensional survey residual
$\epsilon_s \equiv (\Delta\vec{R}_I,\Delta\vec{\Omega}_I)$ and the survey
covariance matrix as
\begin{eqnarray}
\label{eq:chisqsurvey}
\chi^2_s & \equiv & \epsilon_{s}^T{\bf V}^{-1}_{s}\epsilon_{s}.
\end{eqnarray}
The use of $\chi^2_s$ in the alignment procedure is discussed in
detail in Sec.~\ref{sec:minimize}.

\par

The use of the optical survey information in the local alignment
procedure assumes implicitly that the positions measured optically
on the surface of the wafer correspond to the hit positions
reconstructed in the data.  In particular, 
the survey constraint could introduce a bias into the alignment procedure
if the Lorentz shift is different for different wafers in a module.  
We have not studied this effect in \babar,
but we estimate it to be less than the estimated survey alignment errors,
given the similarity of wafers in a module.  Similarly, variations
in the thickness of the wafer, which can change the effective charge
integration depth, are accommodated by the large out-of-plane errors
we assign to $V_s$.

%
%
\section {Wafer Curvature}
\label{sec:curvature}

Initial tests of the local alignment algorithm with \babar\
data showed a smooth but substantial variation of 
residuals as a function of the local ($u,v$) hit position in some wafers,
especially those in the inner layers.  These variations were visible even after the
local alignment procedure had converged.  These effects were not seen
in Monte Carlo simulation of the \babar\ data or the alignment procedure.

\par

\begin{figure}[bt]
\begin{center}
\centerline{
\setlength{\epsfxsize}{1.0\linewidth}\leavevmode\epsfbox{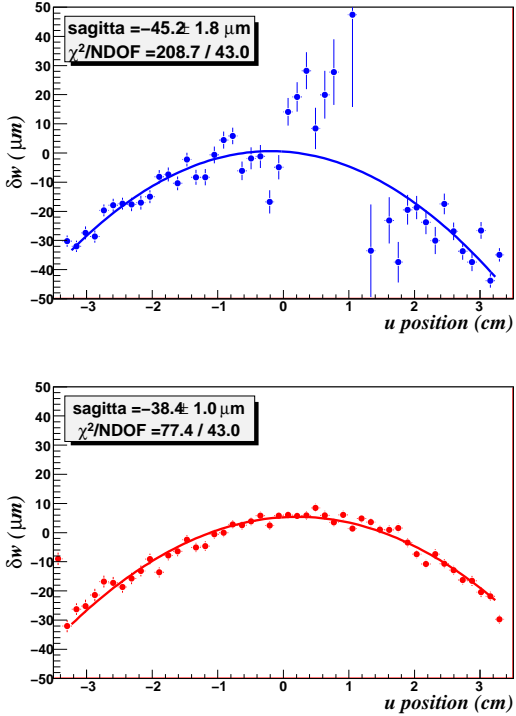}
}
\caption{Average $\delta w$ projection of the $u$ (top) and $v$ (bottom)
hit residuals in a layer 3 wafer as a function of the local
$u$ position for high-momentum tracks in the \babar\ data.}
\label{fig:sagbias}
\end{center}
\end{figure}

An example of these effects is given in Fig.~\ref{fig:sagbias}.
This plots a projection of the $u$ and $v$ hit residuals
($\epsilon_u$ and $\epsilon_v$) from a large sample of
high momentum \babar\ tracks, as a function of 
the $u$ position of the track.  The figures show the
average value of the projection, which is defined
so as to effectively interpret the residuals as a local deviation
in the $w$ position of the wafer:
\begin{eqnarray*}
	\delta w_u & \equiv & \epsilon_u / \sin(\theta_{uw}) \\
	\delta w_v & \equiv & \epsilon_v / \sin(\theta_{vw}),
\end{eqnarray*}
where $\theta_{uw}$ ($\theta_{vw}$) is the angle
between the track direction and the wafer normal in the $uw$ ($vw$) plane, respectively.
The fit to a parabola is reasonably consistent with the data, given that
the errors used are statistical only.
The points at the edges are excluded from the
fit as they are biased by hits in the overlapping wafers in the same layer.
The large uncertainty and fluctuations in the average $u$ hit residuals near
$u=1.2$ cm occurs because tracks from the IP incident at that point are nearly
normal to the wafer in the $uw$ plane, and so have very large error in $\delta w$.

\par

An incorrect local alignment
would result in a linear dependence of $\delta w$ on $u$, with an offset being an
incorrect translation and a non-zero slope being an incorrect rotation around the $v$ direction.
The clear non-linear dependence shown in Fig.~\ref{fig:sagbias}
indicates instead an {\em aplanar}
wafer distortion, not described by the standard six local alignment parameters.
This geometric interpretation is supported by the fact that
compatible effects are seen using either $u$ or $v$ residuals.
We interpret Fig.~\ref{fig:sagbias} to say that this wafer is bowed in the $uw$
plane, with a {\em sagitta} of
roughly -40\mum, or 15\% of the wafer thickness.
Bowing in the $u$ direction is possible, as the support ribs constrain against
bowing only in the $v$ direction.  The large \chisq/NDOF (number of degrees of freedom) of the parabolic fits indicates that
simple bowing may not be the only aplanar distortion present, as is discussed further in
Sec.~\ref{sec:aplanar}.

\par

We see evidence for bowing in all \svt\ wafers.  The
observed bowing is roughly proportional to the $u$ size of the wafers, with the
largest effect in layer three.
While no particular factor has been identified which causes
wafer bowing, the \svt\ detector builders agree that bowing at the
observed scale is possible \cite{privatecom}.

\par

The observed wafer curvature produced a
systematic bias on the transverse impact parameter
as a function of azimuth for high-momentum tracks.  These biases were
beyond the tolerance of the local alignment requirements.  Furthermore,
because the wafer bowing was the same direction for all of layer three,
it caused an effective bias in the average $w$ positions of
roughly 30\mum, well beyond the goal of $< 5\mum $ radial distortion given in
Sec.~\ref{sec:requirements}.
Thus we determined that the aplanar wafer distortions must be
measured and corrected for the local alignment procedure
to meet its requirements.

\par

We model the aplanar distortions as a quadratic dependence of the
wafer $w$ displacement $\delta w$ on the $u$ position of the measurement,

\begin{equation}
\delta w(u)=(u^2-u_0^2)/2R ,
\label{eq:positionbias}
\end{equation}
where $R$ is the curvature radius of the wafer, related to the 
sagitta ${\cal S}$ by $1/R=2{\cal S}/{L^2}$, $L$ being the $u$ half-width
of the wafer, and $u_0$ is a convenience parameter set to
$u_0=L/\sqrt{3}$ in order to keep the $w$ 
center of gravity of the wafer independent of the curvature radius $R$.
We do not model a first-order term as that is redundant with the
$\alpha_u$ alignment parameter.  We measure $R$ for each inner-layer
wafer by fitting the average $u$ and $v$ residuals dependence on $u$
according to this model, minimizing a residual-based \chisq\ to find
the best $R$ value.  We do not fit for curvature in the outer layers
because their smaller $u$ size makes the effect of their curvature negligible, and
other aplanar distortions are found to dominate, 
as discussed further in Sec.~\ref{sec:aplanar}.

\par

\begin{figure}[btp]
\begin{center}
\setlength{\epsfxsize}{1.0\linewidth}\leavevmode\epsfbox{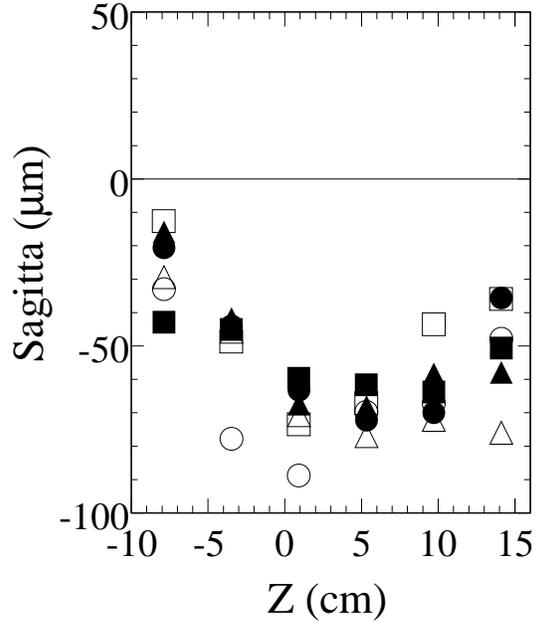}
\caption{
Results of the fits for the wafer sagitta in layer three.
Symbols represent different modules starting from $\phi=0$ in the
order of increasing azimuth: 
${\bullet}$,
${\blacksquare}$,
${\blacktriangle}$,
${\triangle}$,
${\square}$,
${\circ}$.
}
\label{fig:fitcurvature}
\end{center}
\end{figure}

Figure~\ref{fig:fitcurvature} shows the sagitta values obtained
from fits to the different wafers in layer three, as a function
of the wafer positions in global $z$.
The curvature values are generally smallest near the ends of the module, where
they are fixed to rigid hybrids, and are largest in the middle of the
detector, where there is less mechanical constraint, consistent with
expectations.

\par

Because the curvature parameter measurement depends on residuals, it
is sensitive to the local alignment.  Similarly, the local alignment
procedure depends on what value of curvature we assign to the wafers.
This correlation forces us to fit simultaneously for both the curvature
parameters and the local alignment parameters.  The organization of
the simultaneous fit for the local alignment and wafer curvature
is discussed in Sec.~\ref{sec:minimize}. 
We do not observe any time dependence to the curvature parameters,
so those parameters are normally held fixed when fitting for the local
alignment.

\par

To avoid biasing the local alignment procedure or \babar\ physics,
we must correct for wafer curvature during track reconstruction.
We correct the $u$ hits for curvature by displacing them in $w$
by the $\delta w$ amount predicted by Eq.~(\ref{eq:positionbias}), given 
the measured $u$
position of the hit.
We correct the $v$ hits by modeling them with as
a three-piece piecewise linear trajectory, where the
endpoints of the three equal-length linear segments are chosen to lie
at the $\delta w$ positions described by Eq.~(\ref{eq:positionbias}), given the $u$ coordinates of those endpoints.
This trajectory is used when computing the track-hit residual,
thereby naturally correcting for $\delta w$.


%
%
\section {Local Alignment Minimization Procedure}

To obtain the best estimate of the true alignment parameters,
the local alignment procedure combines all the available information
into a total \chisq.  We extract the optimal local alignment
parameters by minimizing this \chisq\ as a function of the
local alignment parameters.  Conceptually, this requires computing
the dependance of every residual on each of the 2040 alignment parameters
(six parameters for each wafer),
and then minimizing the total \chisq\ in this 2040-dimensional space.
However, each residual depends primarily on the alignment parameters of its
hit's wafer.  Additionally, minimizing such a large number of
dimensions is computationally challenging, raising issues of performance and accuracy.
We therefore choose to simplify the local alignment minimization procedure
by dividing the total \chisq\ into 340 separate wafer \chisq\ functions,
and minimizing each independently for that wafers alignment
parameters.  We then iterate
to account for the secondary dependence of a residual
on some other wafers alignment parameters (wafer correlation),
and stop iterating when the alignment parameters for all wafers stabilize.

\par

While our iterative procedure is less direct than a simultaneous minimization of all
parameters, we feel it offers numerous advantages over that technique.
For one, standard algorithms can be used to efficiently
invert the $6\times6$ matrices involved.
Likewise, computing the derivatives of residuals with respect to a single wafer's
alignment parameters is straightforward and fast.  Because of this,
the derivatives can be recalculated between iterations, naturally accounting for
second-order effects which are generally ignored
in a simultaneous solution.
Importantly, our iterative procedure
provides access to intermediate states of the alignment, allowing us
to monitor the convergence process directly.  This gives us confidence in the final result,
and allows us to test the sensitivity
of the procedure to physical or computational effects, as is described in Sec.
\ref{sec:validation}.  Additionally, our iterative procedure allows us to incorporate
wafer curvature (Sec. \ref{sec:curvature}) and beam boost (Sec. \ref{sec:svtla_beam})
determination during the alignment minimization, thus correctly handling correlations
between these parameters and the alignment parameters.
Finally, organizing the minimization by wafer
naturally allows for a modular software design well suited for modern Object Oriented
programming languages.

\par

\begin{figure}[thp]
\begin{center}
\setlength{\epsfxsize}{1.0\linewidth}\leavevmode\rotatebox{0}{\epsfbox{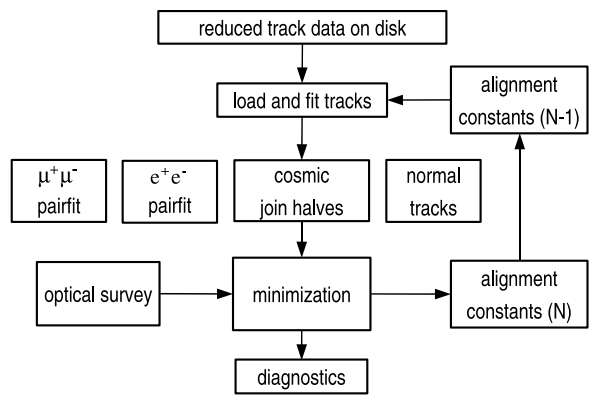}}
\caption{ Diagram of minimization sequence. 
Iterations over index $N$ is performed until diagnostics 
shows convergence.
}
\label{fig:iterations}
\end{center}
\end{figure}

The local alignment
minimization sequence is shown schematically in Fig.~\ref{fig:iterations}.
We start with the tracks and hits that were selected as described in previous
sections.  The tracks are fit using the current (test) estimate of the alignment
parameters, from which the hit \chisq\ are computed for the selected hits.  To this
is added the survey alignment \chisq, also relative to the test
alignment parameters, to form a wafer \chisq. 
We then minimize
each wafer's \chisq\ with respect to the {\em change} in that wafers
local alignment parameters, holding the parameters of every other wafer fixed.
After minimizing every wafer's \chisq, we update the alignment 
parameters for all 340 wafers by adding the computed parameter 
change to the original alignment parameters estimate.
We then use that updated local alignment to fit the tracks and
evaluate the survey information in the next iteration,
and repeat the process until the alignment converges.
After convergence, the alignment parameters are stored
in the \babar\ conditions database \cite{conddb}.

The wafer \chisq\ used in the alignment minimization is defined as the sum of the hit and survey
\chisq\ defined in Sec. \ref{sec:hits} and Sec. \ref{sec:survey} respectively.  We express
that explicitly in terms of a small change in the wafer's alignment parameters $\Delta{\bf p}$
with respect to the test alignment used in computing the hit and survey residuals:

\begin{eqnarray}
\label{eq:fullchisq}
\chi^2 & \equiv &
\sum_{i}^{\rm hits}\epsilon_{i}^T(\Delta{\bf p}){\bf V}^{-1}_{i}\epsilon_{i}({\Delta\bf p})
\cr
& + & \epsilon_{s}^T(\Delta{\bf p}){\bf V}^{-1}_{s}\epsilon_{s}(\Delta{\bf p}).
\end{eqnarray}

\par

We analytically minimize \chisq\ by finding the value of $\Delta{\bf p}$ which gives a null first
derivative with respect to all components of $\Delta{\bf p}$, evaluated to first
order in $\Delta{\bf p}$.  This requires computing the
first derivatives of the hit and survey residuals with respect to $\Delta{\bf p}$.
The hit residual derivatives are calculated analytically, given the hit direction
and the direction of the track.
The derivative formulas are presented in Appendix \ref{sec:derivs} 
expressed as the Jacobian ${\bf J}_{k}$ matrix:
\begin{eqnarray}
\label{eq:jacobian}
{\bf J}_{k} = \partial{\bf \epsilon}_{k}/ \partial{(\Delta\bf p)}.
\end{eqnarray}
Because the survey alignment residual is defined directly in terms of the
alignment parameters themselves,
its Jacobian is simply the identity
$6\times6$ matrix.
Wafers which are electronically dead or which have dead readout views naturally
have their missing parameters constrained by their survey alignment.

\par

The
matrix equation which results from setting the derivative of the wafer \chisq\
to zero can be
inverted to solve for the change in the six alignment parameters $\Delta{\bf p}$:
\begin{eqnarray}
\label{eq:chisqsolution}
\Delta{\bf p} =
\left[
\sum_{j}^{\rm all}{\bf J}_j^T{\bf V}_j^{-1}{\bf J}_j 
\right]^{-1}
\left[
\sum_{k}^{\rm all}{\bf J}_k^T{\bf V}_k^{-1}{\bf\epsilon}_k 
\right].
\end{eqnarray}
The matrix sums are computed by iterating over all the hit residuals and
the survey residuals for the wafer.
The new wafer alignment parameters are taken to be ${\bf p}={\bf p}_0+\Delta{\bf p}$,
where ${\bf p}_0$ are the test alignment parameters.  Upon the subsequent alignment
iteration, these updated parameters become the test parameters, and the procedure is
repeated.
The \chisq\ value used in convergence testing is computed directly from
Eq.~(\ref{eq:fullchisq}), and thus is one iteration behind the parameter computation.

\par

A wafer is said to have converged when its total
\chisq\ changes by less than a given threshold between iterations,
typically set to 0.01 absolute.
While this value may seem small for a \chisq\ that typically has
a few hundred degrees of freedom, we found that a low
threshold was necessary for the procedure to be sensitive to small
global distortions, as discussed in Sec.~\ref{sec:validation}.
The entire local alignment procedure is said to converge
when all but at most two wafers are converged.  This allows
for a trivial oscillation observed between dead or partially-dead
wafers constrained only by survey information.
The local alignment procedure typically converges 
after roughly 100 iterations.

\par

As discussed in Sec. \ref{sec:curvature}, the
wafer curvature parameters must be fit
simultaneously with the local alignment parameters.  This is
done in a dedicated variant of the local alignment procedure,
where we introduce a fit for the curvature parameters between
each normal local alignment parameter iteration, holding the alignment
parameters fixed.  The updated curvature parameters are used in the
subsequent alignment minimization iteration.  A similar procedure
is applied when fitting for the beam boost, as described in Sec.
\ref{sec:svtla_beam}.

\par

Because both
the derivative calculations and the \chisq\ minimization are analytically computed,
the minimization procedure is reasonably fast. 
The processing time for a single
alignment iteration is limited by the time it takes to refit the tracks.
This time is much reduced compared to the normal \babar\ reconstruction,
as the \dch\ information is applied as a single constraint instead of 40
separate hits.
A single iteration of the alignment minimization
on a standard alignment set, without the curvature fit,
takes roughly twenty minutes on a modern
multi-GHz intel-processor based Linux computer.
The entire local alignment procedure typically converges in roughly twelve hours.

\par

The local alignment minimization procedure is written within the
\babar\ software framework, using standard access to event and
conditions data.  Iterations are controlled using TCL/Tk \cite{Tk},
through either an interactive GUI or with a TCL script submitted in batch.
Bookkeeping, diagnostics, input/output, and job management are also
controlled through TCL/Tk.

\label{sec:minimize}
%
%
\section{Validation of the Alignment}
\label{sec:validation}

\begin{figure}[b]
\centerline{
\setlength{\epsfxsize}{0.50\linewidth}\leavevmode\epsfbox{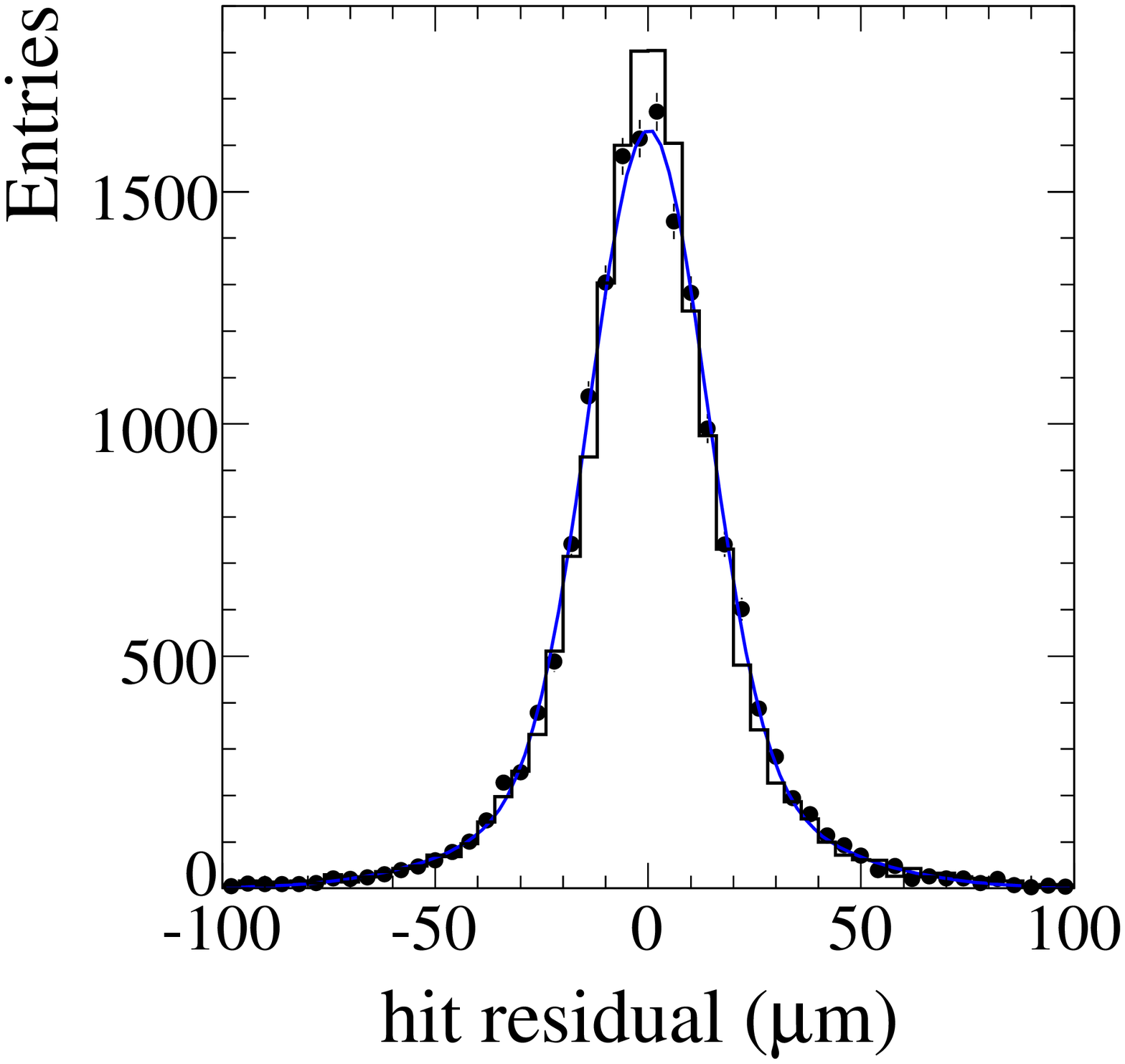}
\setlength{\epsfxsize}{0.50\linewidth}\leavevmode\epsfbox{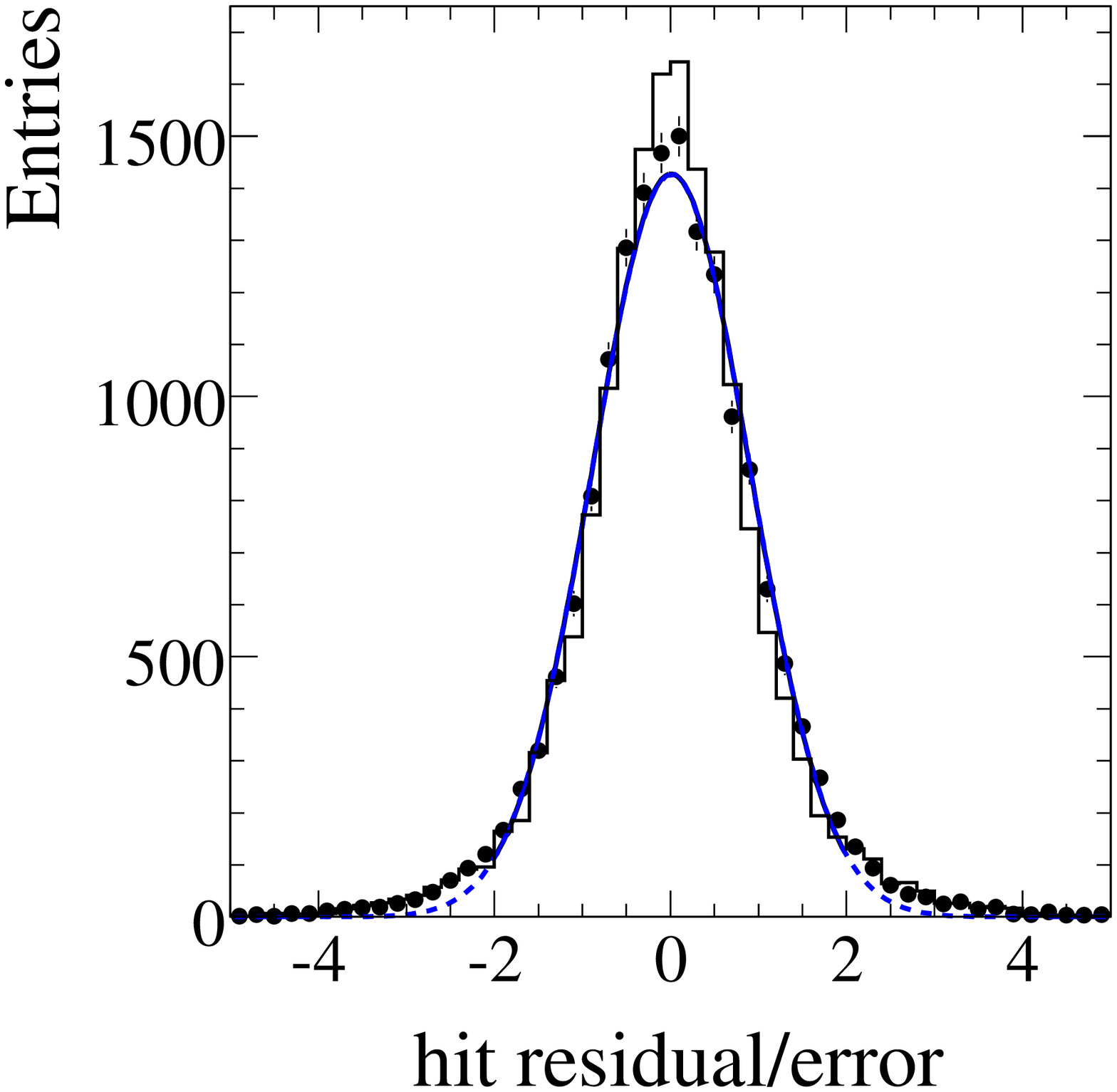}
}
\centerline{
\setlength{\epsfxsize}{0.50\linewidth}\leavevmode\epsfbox{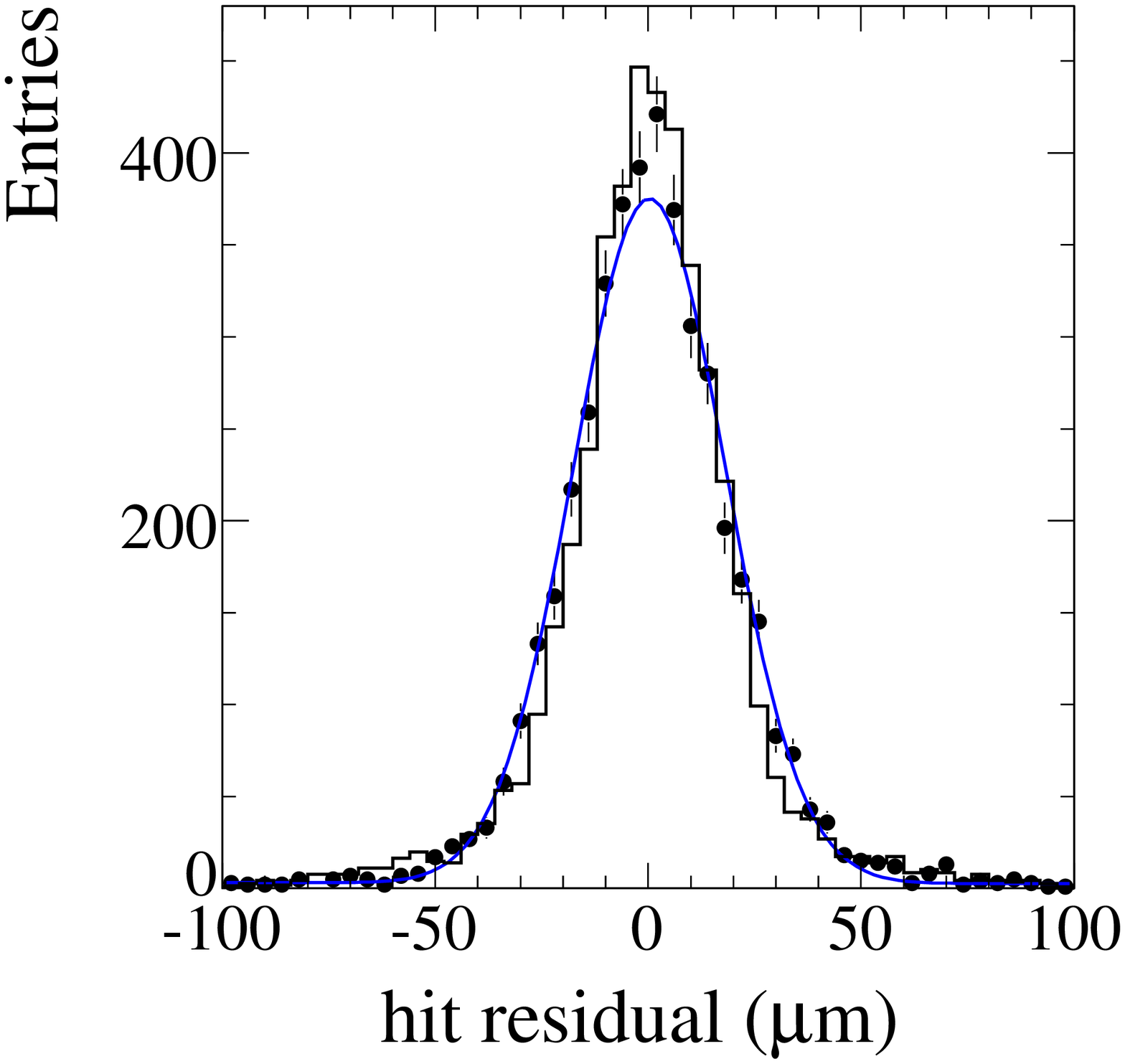}
\setlength{\epsfxsize}{0.50\linewidth}\leavevmode\epsfbox{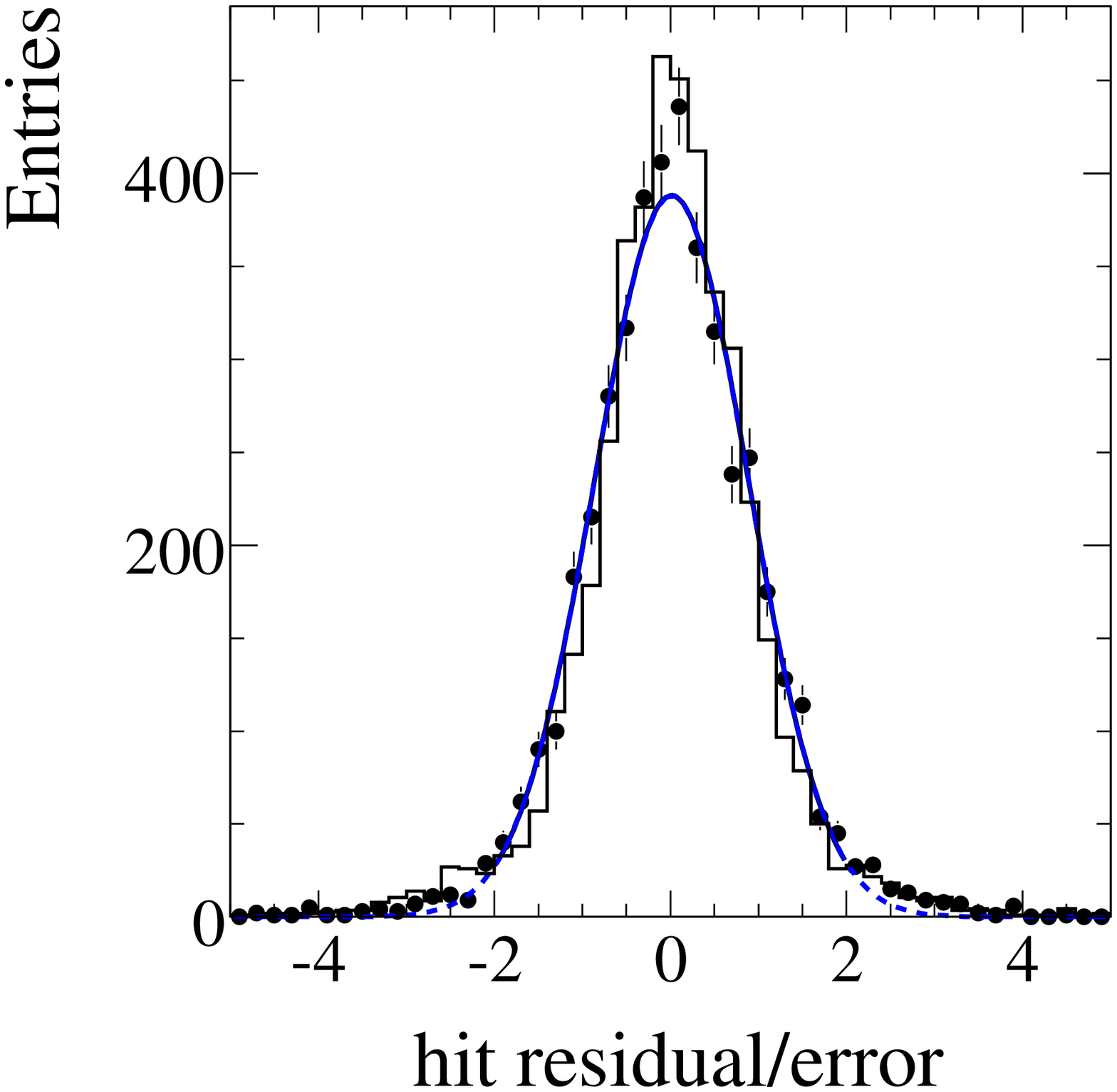}
}
\caption{
Single hit residuals (left) and normalized residuals (right) in the $u$ (top)
and $v$ (bottom) readout view.  The \babar\ data
are shown as points, Monte Carlo simulation as histograms.  The
(blue) smooth curves are the results of a Gaussian fit to the data.
}
\label{fig:align-residual}
\end{figure}

\begin{figure}[t]
\centerline{
\setlength{\epsfxsize}{0.50\linewidth}\leavevmode\epsfbox{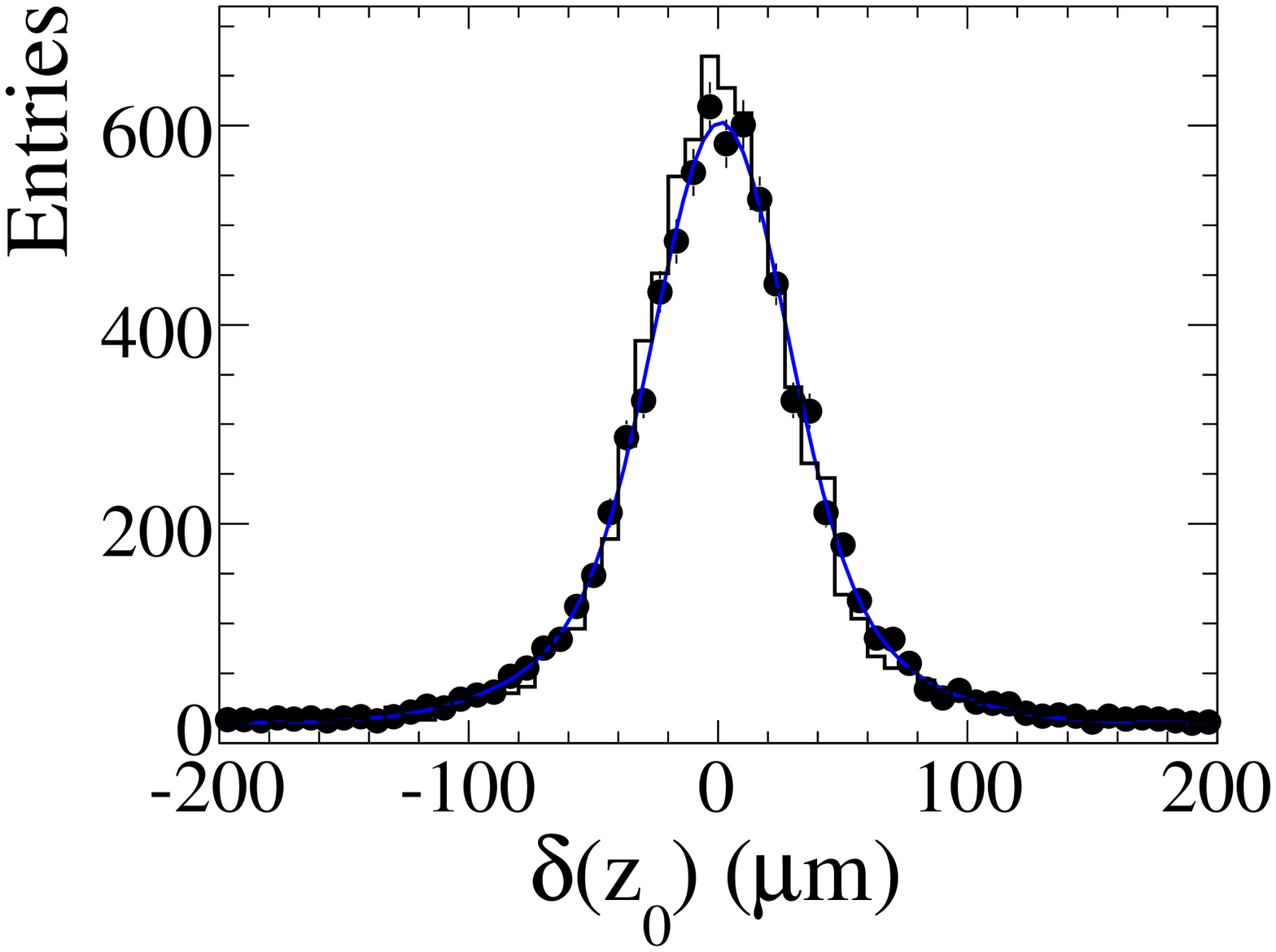}
\setlength{\epsfxsize}{0.50\linewidth}\leavevmode\epsfbox{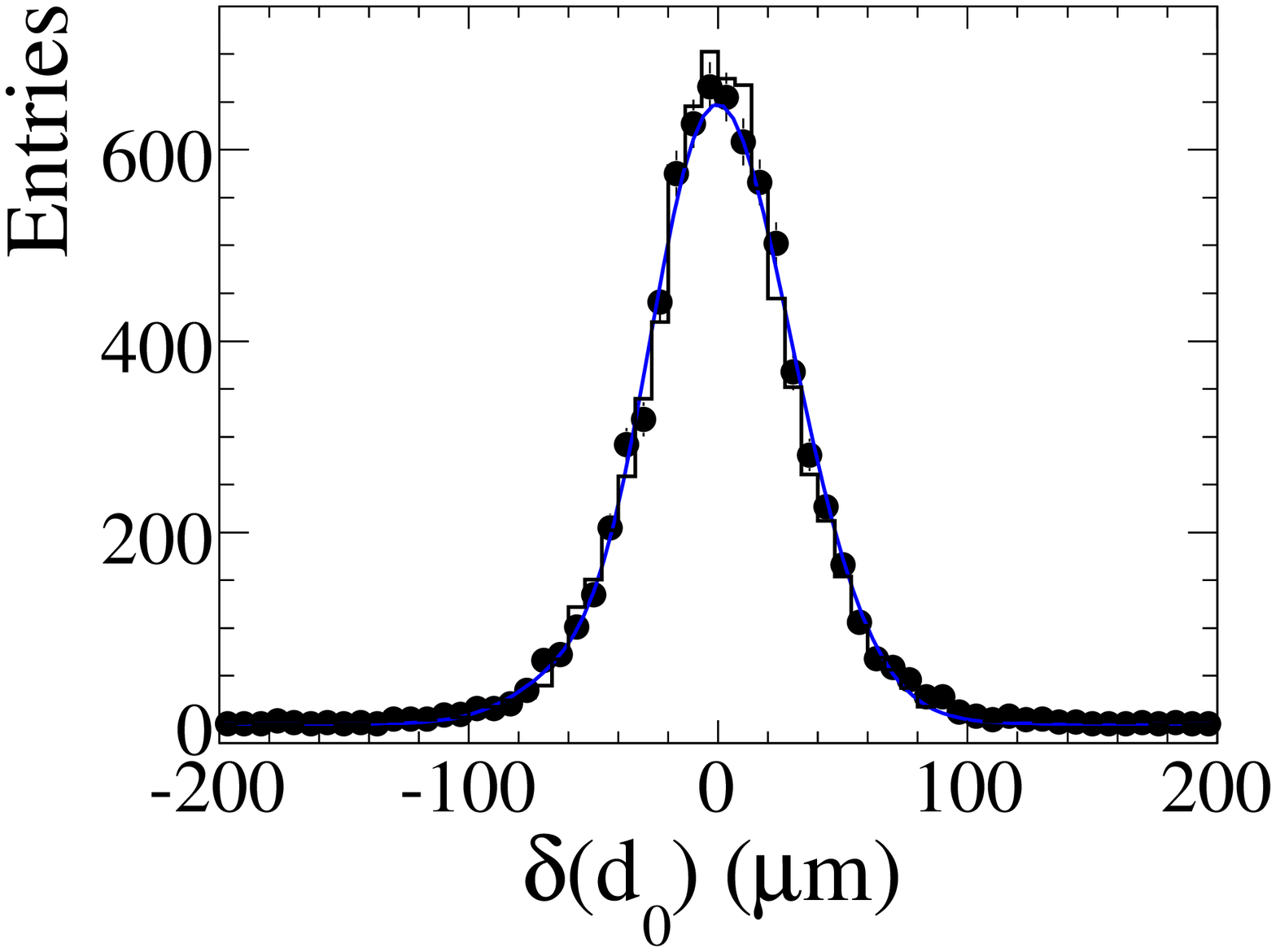}}
\centerline{
\setlength{\epsfxsize}{0.50\linewidth}\leavevmode\epsfbox{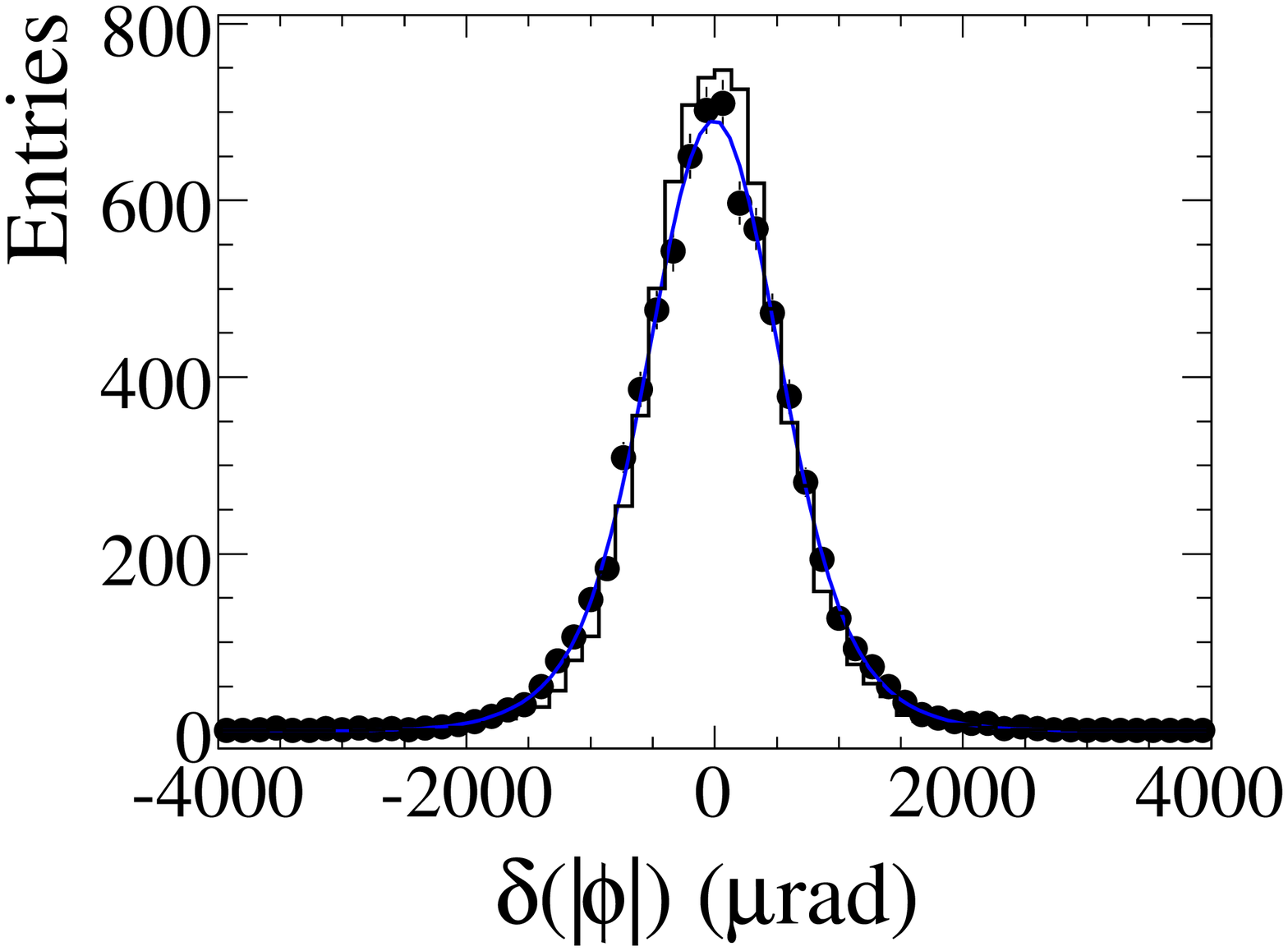}
\setlength{\epsfxsize}{0.50\linewidth}\leavevmode\epsfbox{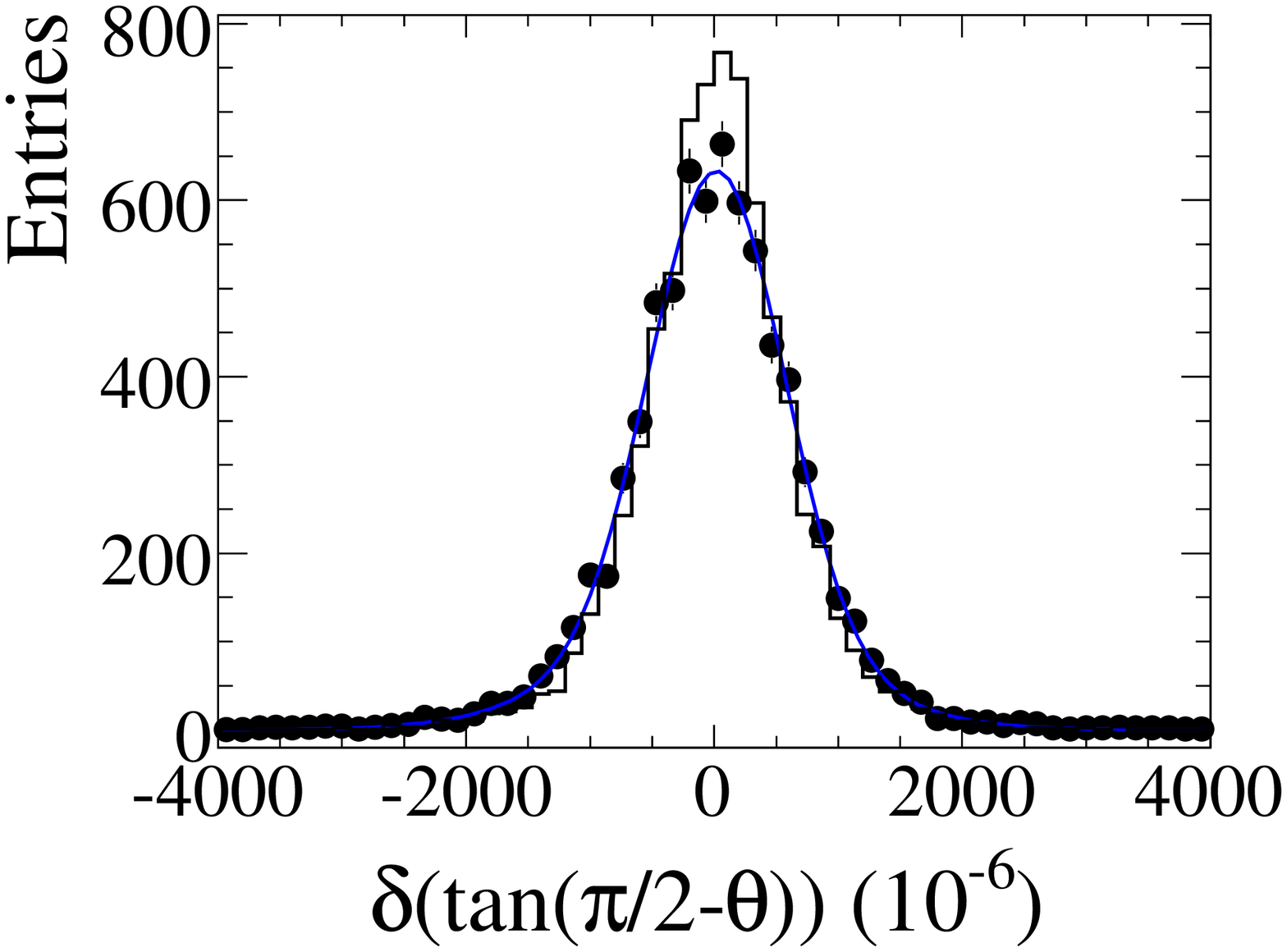}
}
\caption{
Difference between the impact parameter
along the beam axis ($z_0$), 
transverse to the beam axis ($d_0$), in
azimuthal angle ($\phi$), and in polar angle
($\tan(\pi/2-\theta)$) for split cosmic tracks. The \babar\ data
are shown as points, Monte Carlo simulation as histograms.  The
(blue) smooth curves are the results of a Gaussian fit to the data.
}
\label{fig:svt-cosmics}
\end{figure}

\begin{figure}[t]
\begin{center}
\centerline{
\setlength{\epsfxsize}{0.50\linewidth}\leavevmode\epsfbox{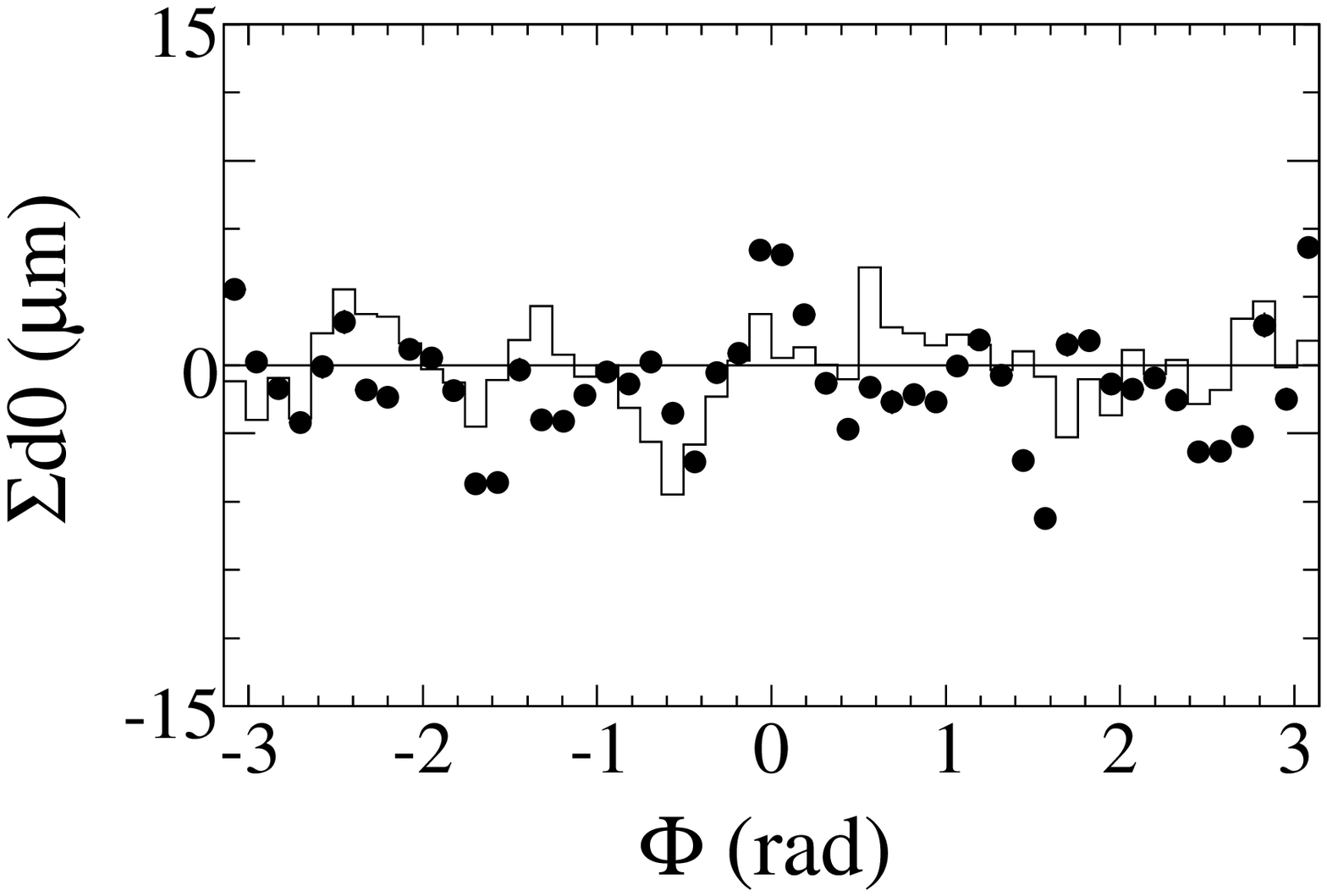}
\setlength{\epsfxsize}{0.50\linewidth}\leavevmode\epsfbox{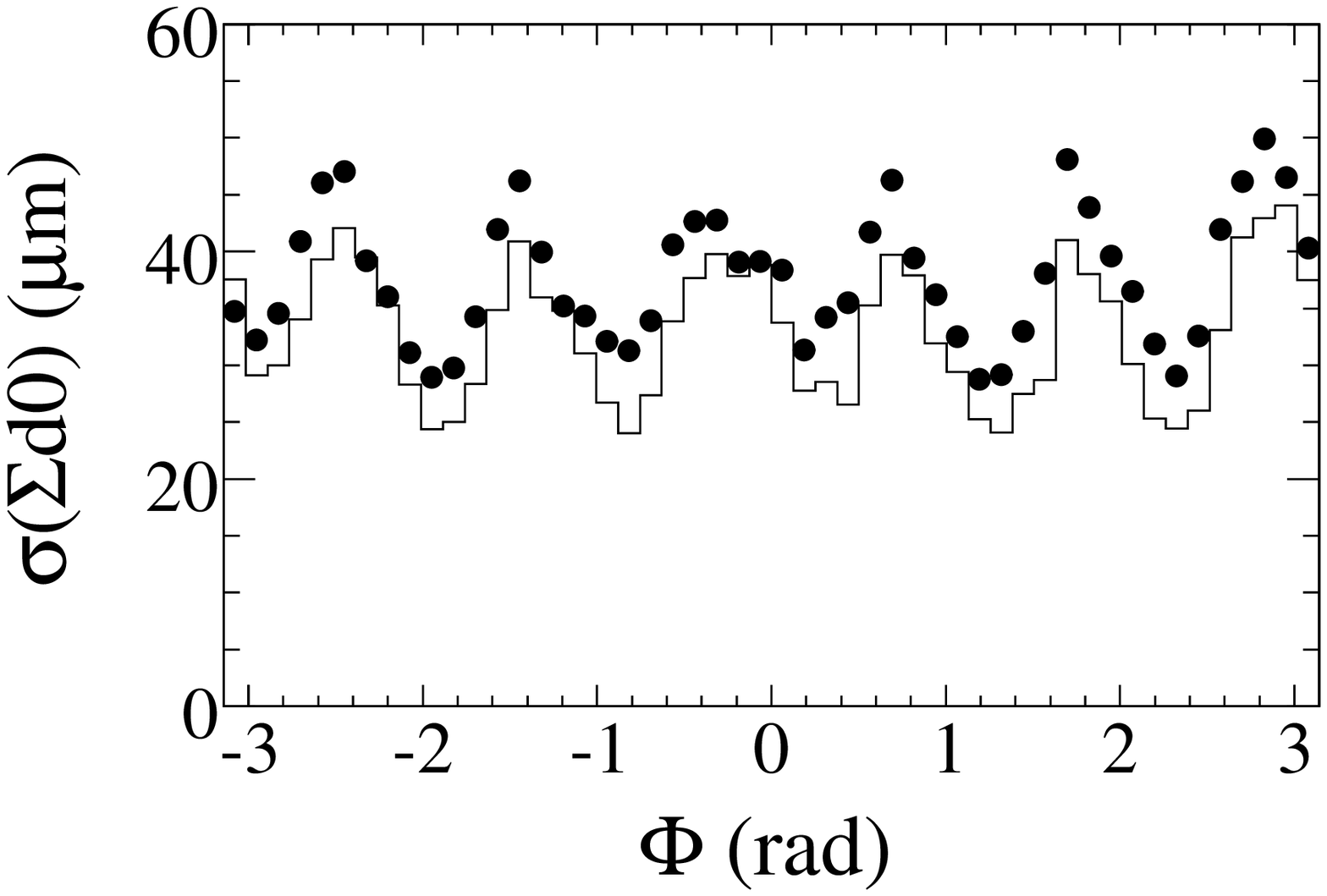}
}
\caption{
$\phi$-dependence of the mean (left) and $\sigma$ (right) of the
\mupair\ track $d_0$ mismatch.  The \babar\ data are
shown as points, Monte Carlo simulation as histograms.
}
\label{fig:svtla-dimuonphi}
\end{center}
\begin{center}
\centerline{
\setlength{\epsfxsize}{0.50\linewidth}\leavevmode\epsfbox{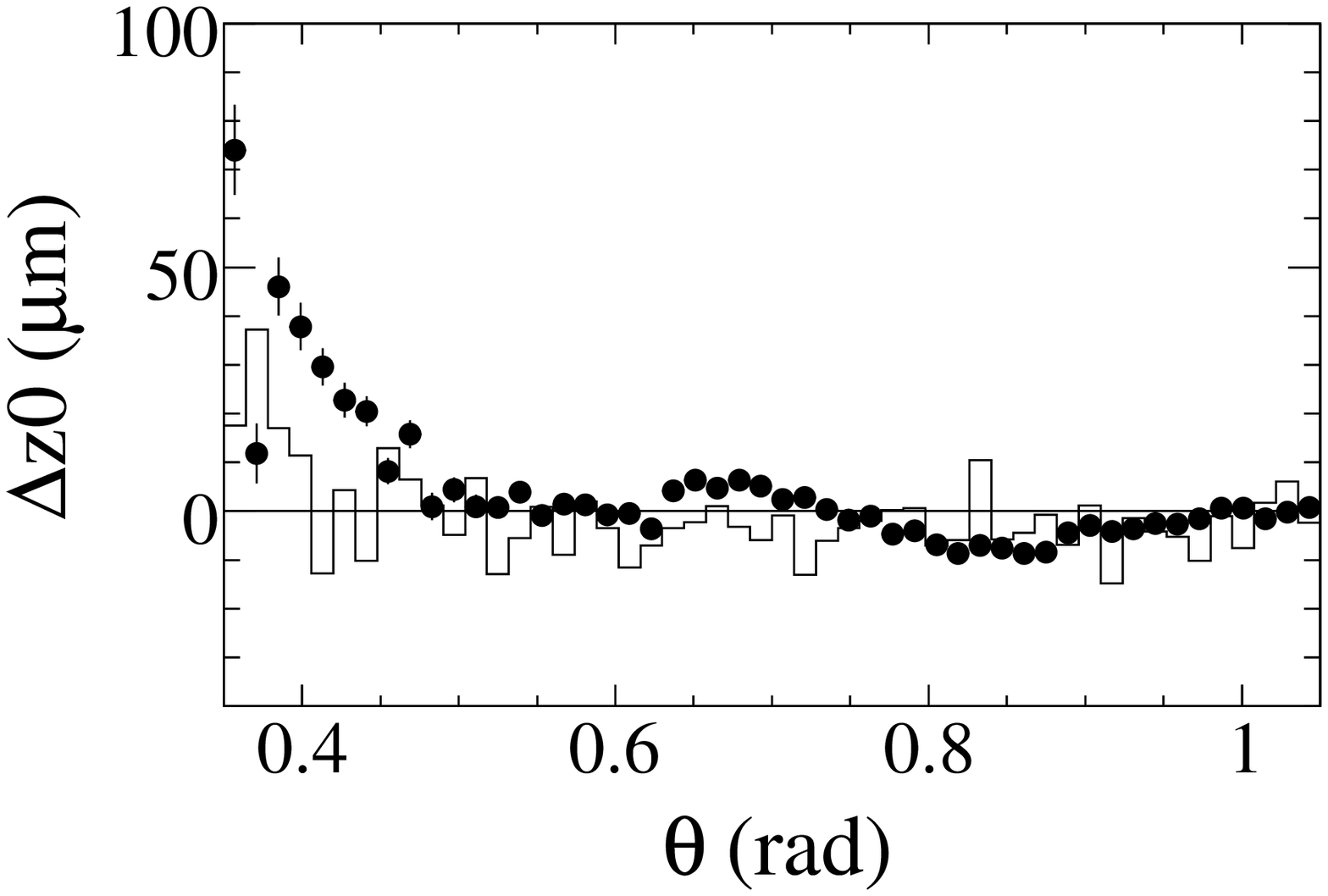}
\setlength{\epsfxsize}{0.50\linewidth}\leavevmode\epsfbox{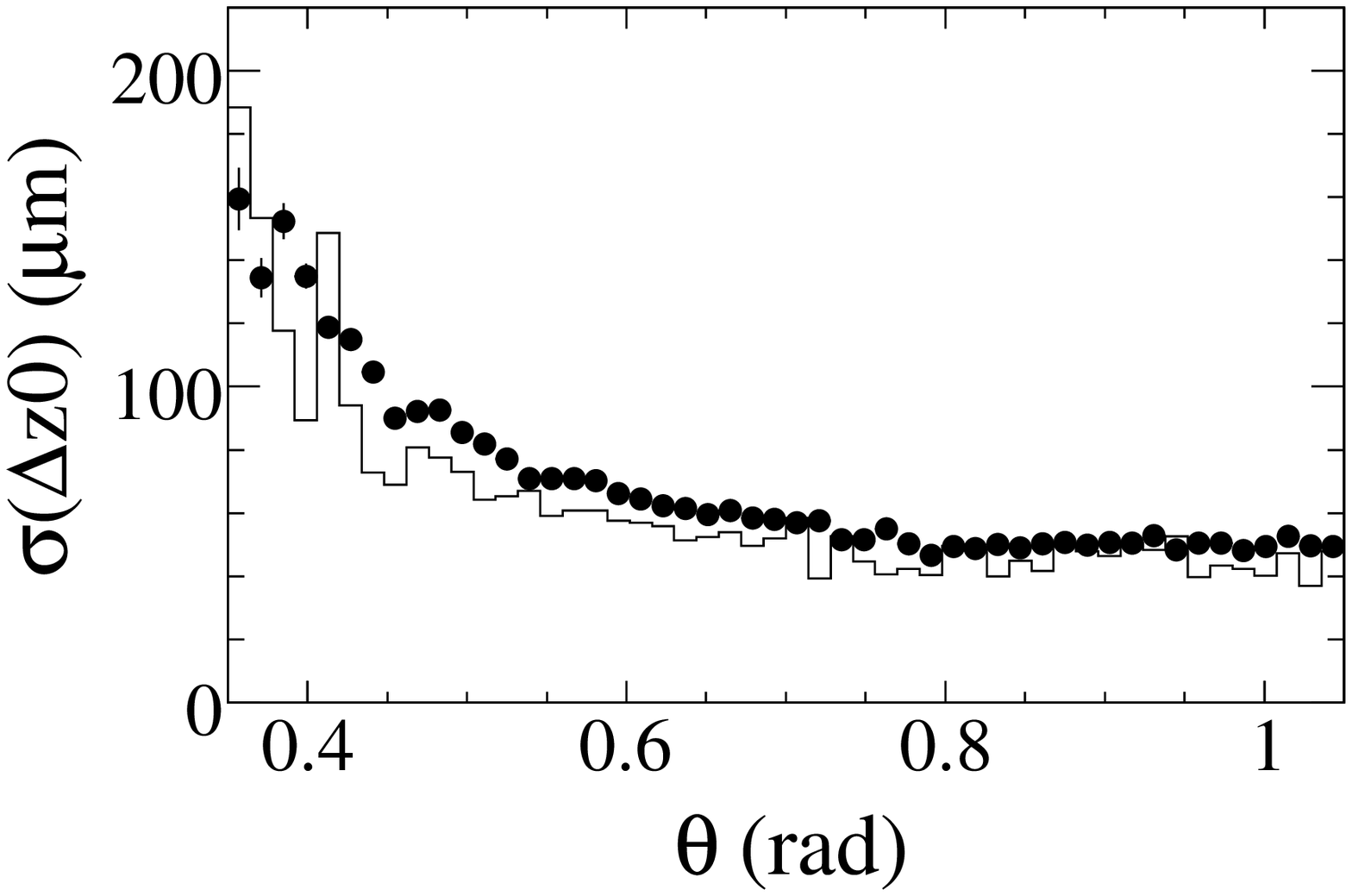}
}
\caption{
$\theta$-dependence of the mean (left) and $\sigma$ of the (right) \mupair\ track $z_0$ mismatch.
The \babar\ data are shown as points, Monte Carlo simulation as histograms.
}
\label{fig:svtla-dimuontheta}
\end{center}
\end{figure}

We validate the performance of the \svt\ local alignment procedure
through self-consistency tests using \babar\ data, where we compare
the results of the aligned detector with apriori expectations.
We also compare the performance of the aligned \babar\ detector with
perfectly-aligned Monte Carlo simulation.  The following tests were performed
using data from a typical period of \babar\ running, where the \svt\
alignment parameters were determined according to the procedure described above, applying all the
calibrations and corrections as normally done when reconstructing \babar\ data.
The validation data sample has minimal overlap with the data
used for alignment production, 
due to the pre-scaling used in the alignment procedure. 

\par

We use the track residuals from \svt
hits to perform a basic test of alignment self-consistency.
The residuals themselves
are shown in the plots on the left of Fig.~\ref{fig:align-residual}, with
the $u$ residuals on top and the $v$ on the bottom.
These show residuals from
the inner three layers of the \svt, using tracks in \mupair events selected to
be within 0.2 radians of vertical in the readout-view projection.  This selection
of hits
give the best resolution in the \svt, and so provides the most sensitivity to misalignment.
The residual distributions are centered at zero, and the
data and Monte Carlo distributions show very
similar shapes.  
Both
data and Monte Carlo distributions show some non-Gaussian tails, as expected
from scattering tails and hit resolution variation.
The RMS of the data and Monte Carlo distributions agree within
a few percent.  
A Gaussian fit to the core of the distributions gives mean values consistent
with zero, and sigma values of 14 (13) \mum\ for data (Monte Carlo)
$u$ hits, and 18 (16) \mum\ for data (Monte Carlo) $v$ hits, respectively.
The data-Monte Carlo width differences are
consistent with the predicted accuracy of the simulation of \svt\ hit resolution.

\par

The normalized residual (pull) distributions for the same tracks and hits
are shown in the plots on the right of Fig.~\ref{fig:align-residual}, with
the $u$ residuals on top and the $v$ on the bottom.  Again we see good agreement
between data and Monte Carlo.  The core of the data distribution is
well-described as a unit-width Gaussian centered at 0.  The non-Gaussian tails seen
in both data and Monte Carlo are roughly consistent with known approximations in our estimate
of the residual errors, which (for instance) do not take into account dead
electronics channels or Moliere scattering of the tracks.

\par

A higher-level self-consistency test comes from fitting the incoming and
outgoing branches of a cosmic ray as two separate tracks.  Each of these
tracks has a similar number of hits as a typical \babar\ physics track.  Both
tracks are fit independently, but because they represent the same particle,
they should have equivalent parameters at the point where they meet
if the alignment is correct.  In this study we use cosmic ray events
selected as described in Sec.~\ref{sec:event}.
Fig.~\ref{fig:svt-cosmics} compares
the impact parameters and angles of the split cosmic tracks.  These tracks were
fit using their \svt\ hits plus a 
DCH curvature constraint, which reduces
the large uncertainty in $d_0$ and $\phi$ coming from their correlation with
curvature, which is poorly measured in the \svt\ alone due to its small
lever arm.
All tracks in these plot have a momentum above 2.0 \gevc.
The plots show good agreement between \babar\ data and Monte Carlo.
Fitting the \babar\ data distributions to a Gaussian we extract single-track
resolutions of
29 $\mu$m for $z_0$,
24 $\mu$m for $d_0$,
451 $\mu$rad for $\phi$, and 
$512 \times 10^{-6}$ for $\tan(\pi/2-\theta)$,
by scaling the fitted Gaussian sigma by $1/\sqrt{2}$.

Another validation test comes from comparing the reconstructed
origin points of the two tracks produced in \mupair\ events.  Because these tracks
are known to originate at the same point, the difference in their reconstructed
parameters can be used to measure the impact parameter resolution, and to look for systematic biases left by the alignment procedure.  In this study we use \mupair events
selected approximately as described in Sec.~\ref{sec:event}.  To test the performance
of the entire \babar\ tracking system, these tracks are fit with both \svt and \dch
hits.  This brings in the possibility that misalignments inside the \dch or between
the \dch and \svt may affect our results.

Figure~\ref{fig:svtla-dimuonphi} shows the $\phi$-dependence of the \mupair\ tracks
transverse impact parameter mismatch on the left, and its resolution on the right.
The plotted points are the mean and the
$\sigma$ of a Gaussian fit to $\Sigma d_0$ in each $\phi$ bin, respectively.
The $d_0$ mean shows some structure at the level of a few microns RMS,
roughly consistent for data and Monte Carlo.
We believe this structure comes from track fit biases due to dead electronics in
the inner layers of the \svt, which are partially simulated in the Monte Carlo.

The $d_0$ resolution ($\sigma$) shows a periodic variation 
due to the six-fold symmetry of the inner layers of the \svt (see Fig.~\ref{fig:svt-2}),
which modulates the extrapolation of the hit error
according to the distance from the innermost hit
to the production point.  This periodicity is well-reproduced in Monte Carlo.
The Monte Carlo underestimates the $d_0$ resolution by roughly 10\%, consistent with the
underestimation of the individual $u$ hit residual core resolution.

Because \pep2\ produces a boosted final state, we cannot simply compare the 
\mupair\ track longitudinal
impact parameters as we did 
the transverse impact parameters, since the tracks are not back-to-back
in the lab frame in the longitudinal projection.  We can extract some information about the
longitudinal impact parameter by
constraining the production point to the event-average beamspot position, which
is well measured in the transverse plane.
However,
this couples the statistical and systematic errors of the beamspot determination
with the alignment validation.  In addition,
because the beamspot is large (roughly 100\mum) in the \pep2 bend ($x$) direction,
the comparison has meaningful precision only for vertical tracks,
where the beamspot constraint is limited by its measurement resolution
of roughly 10\mum.

Figure \ref{fig:svtla-dimuontheta} shows the polar-angle dependence of the \mupair\
longitudinal impact parameter mismatch on the left, and its resolution on the right.
The plotted points are the mean and $\sigma$ of a Gaussian fit to
$\Delta z_0$ in each $\theta$ bin, respectively.  To select vertical tracks, we
use only events with track azimuthal angles $|\phi\pm\pi/2| < 0.2$.
The agreement between data and Monte Carlo is reasonable.
The observed discrepancy of the $z_0$ mean at small $\theta$ may be due to
systematic effects in the beamspot position determination, which become
amplified at small angles.  It may also be related to remaining aplanar distortions
in the outer layer wafers at large $|z|$, as discussed in section \ref{sec:aplanar}.
The Monte Carlo underestimates the $z_0$ resolution by roughly 15\%.  This
difference is partly explained by the 10\% difference in intrinsic $v$ hit residual
core width, and by the fact that the Monte Carlo does not model the 
beamspot position measurement resolution.

\section{Validation of the Alignment Systematics}
\label{sec:validatesyst}


We test the ability of the local alignment procedure
to remove systematic distortions
by introducing a coherent misalignment of the \svt\ wafers,
and then running the alignment procedure taking that
misalignment as the initial condition.  These global
distortions are particularly difficult to remove as the residuals
used in the alignment procedure typically depend on them only to second order.
We test nine distinct distorted initial conditions, as described in 
Table~\ref{tab:svtdistortions} and Fig.~\ref{fig:svt-misalign-initial}.
We set the initial scale of these distortions to 50\mum.

\begin{figure}[t]
\centerline{
\setlength{\epsfxsize}{0.90\linewidth}\leavevmode\epsfbox{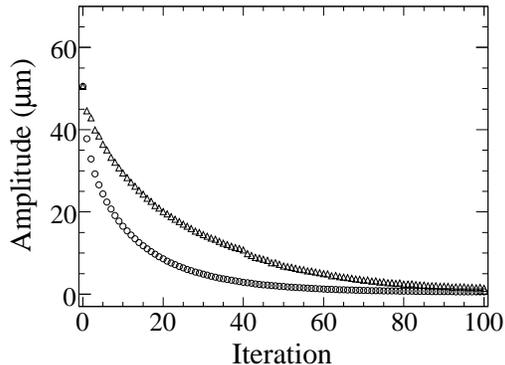}}
\caption{Amplitude of the remaining $z$-expansion distortions as a
function of iteration during the alignment procedure with data starting with
the 50\mum\ amplitude $z$-expansion distortion initial condition for data.
Circles ($\circ$) illustrate procedure with survey information, 
while triangles ($\triangle$)
illustrate procedure with this information removed.
}
\label{fig:svt-z-iteration-survey}
\end{figure}

In Fig.~\ref{fig:svt-z-iteration-survey} we show how amplitude of
the $z$-scale of the \svt\ converges back to the initial value 
after a 50$\mum$ misalignment is applied to a standard alignment set
data sample. For comparison,
convergence with and without optical survey measurements is
shown.  The survey information demonstrably provides an important
constraint on systematic distortions, which improves the
convergence of the alignment procedure, and reduces the systematic
error of the final alignment.

\begin{figure}[b]
\centerline{
\setlength{\epsfxsize}{0.50\linewidth}\leavevmode\epsfbox{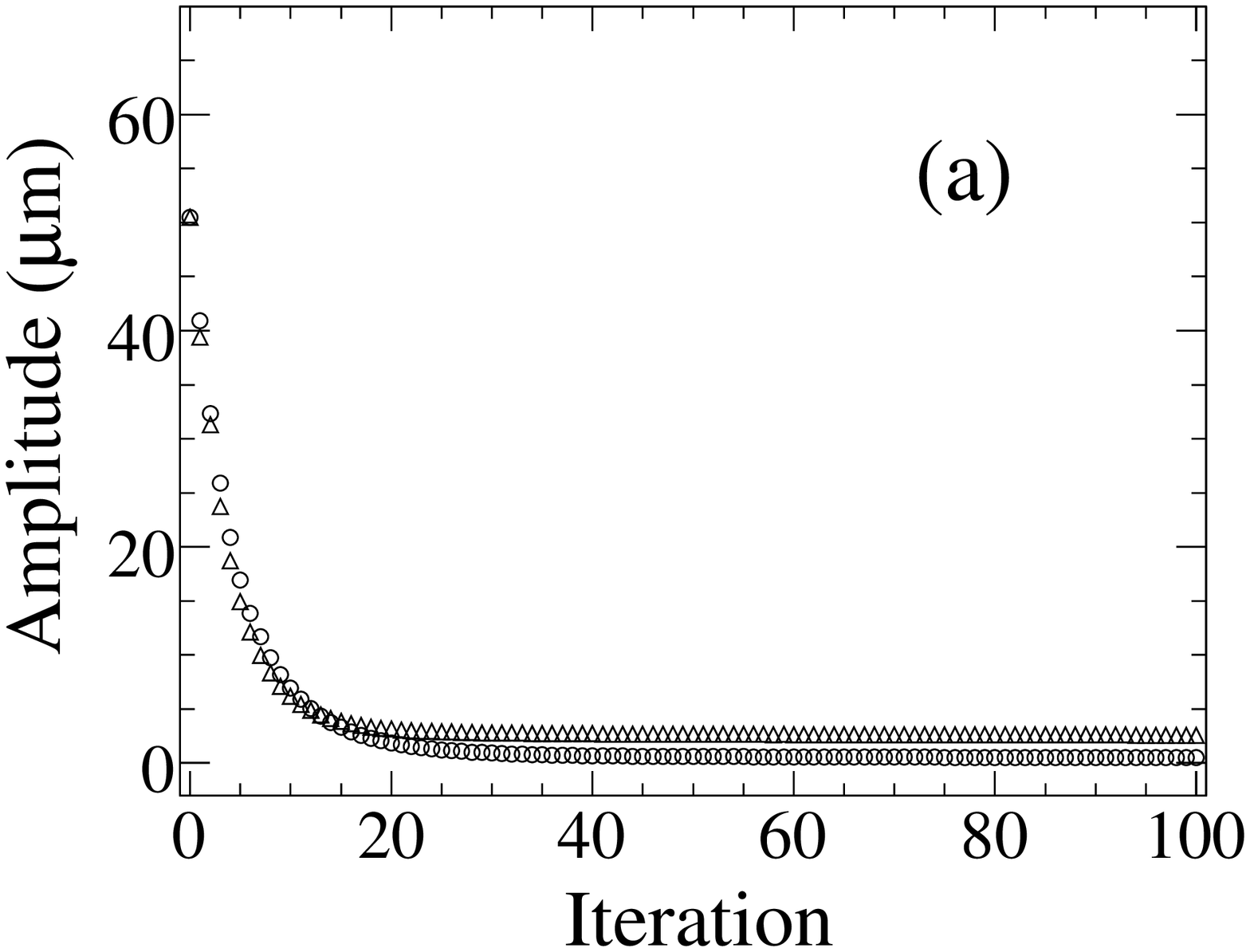}
\setlength{\epsfxsize}{0.50\linewidth}\leavevmode\epsfbox{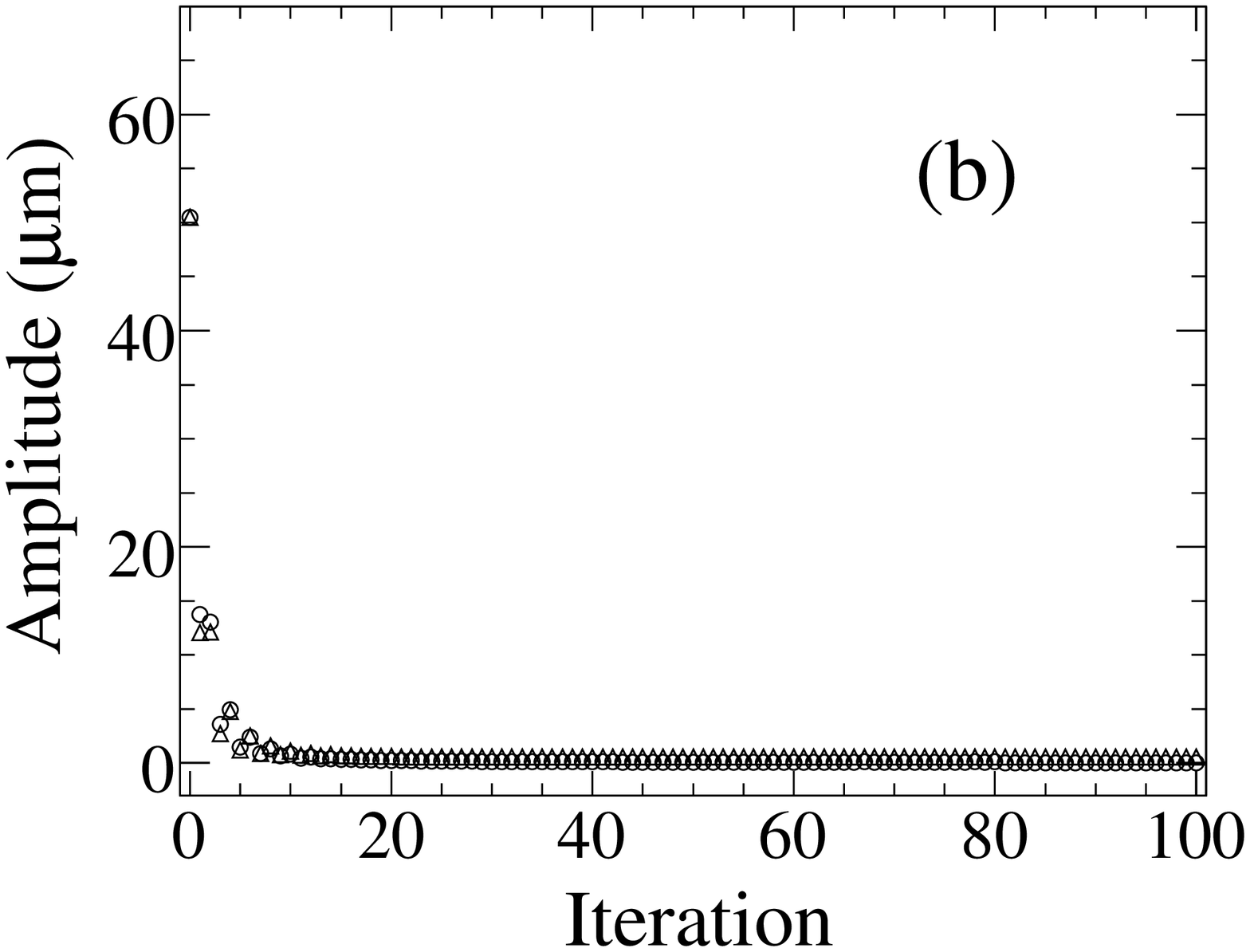}
}
\centerline{
\setlength{\epsfxsize}{0.50\linewidth}\leavevmode\epsfbox{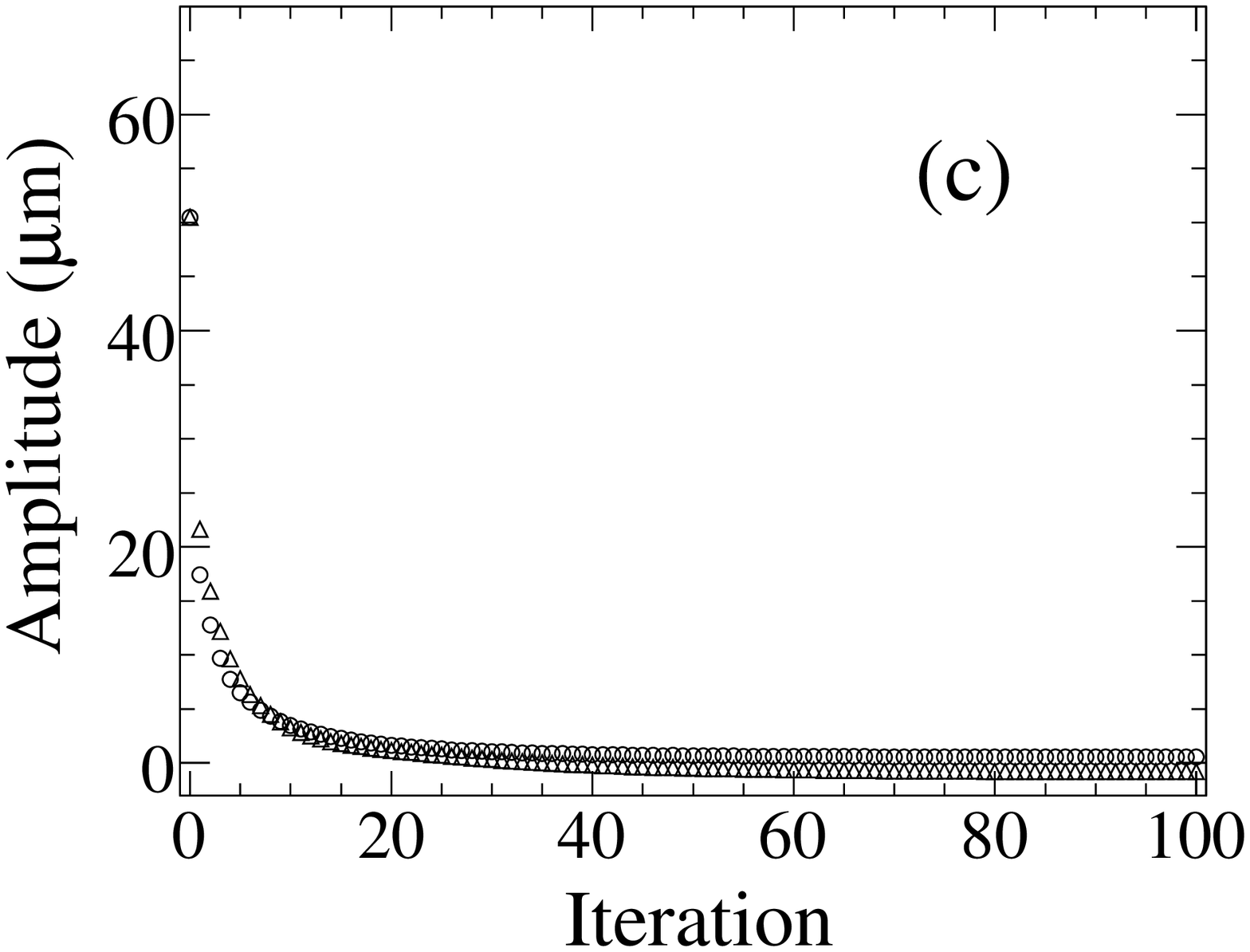}
\setlength{\epsfxsize}{0.50\linewidth}\leavevmode\epsfbox{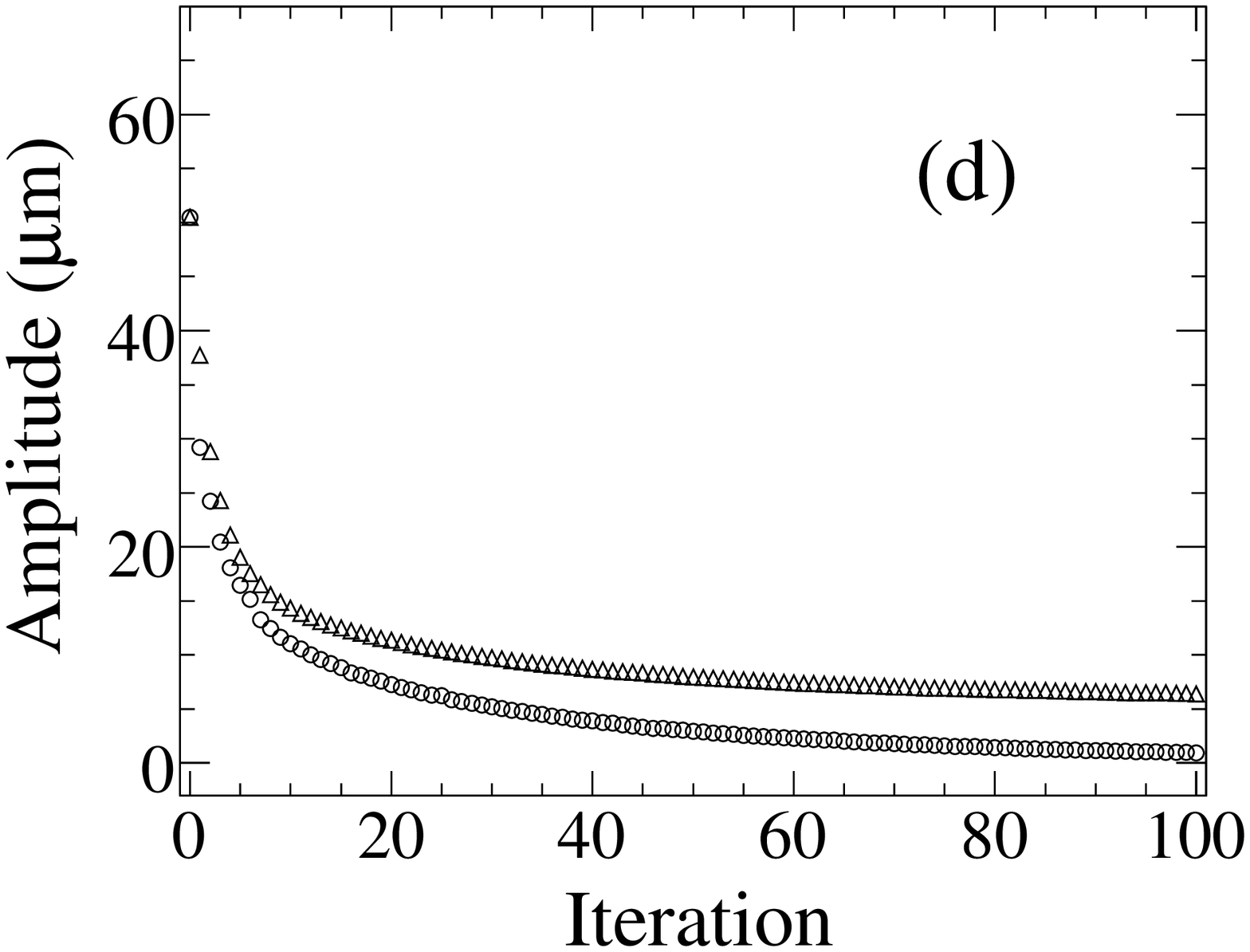}
}
\caption{Amplitude of the remaining distortions as a 
function of iteration during the alignment procedure starting with
the 50\mum\ amplitude distortion initial condition :
(a) telescope, (b) curl, (c) bowing, and (d) elliptical effects.
Circles ($\circ$) show results with data 
and triangles ($\triangle$) represent Monte Carlo.
}
\label{fig:svt-amp-iteration}
\end{figure}

\begin{table}[ht]
\caption{
Decay time (in units of iterations)
for the main systematic distortions
in the \svt\ local alignment procedure.
The initial distortion is 50\mum\ in all cases.
The remaining distortion is quoted in the table.
Distortions in $r$, $z$, and $\phi$ are
considered as a function these coordinates. 
}
\label{tab:svtdecaytime}
\medskip
\begin{center}
\begin{tabular}{|c|c|c|c|}
\hline\hline
\vspace{-3mm} & & &    \cr
 & $\Delta r$   &  $\Delta z$ & $r\Delta\phi$   \cr
\vspace{-3mm} & & &    \cr
\hline
\vspace{-3mm} & & &    \cr
~~vs. $r$~~    & ~~radial~~   & ~telescope~  & ~~~~curl~~~~   \cr
  decay (iterations)     & 5.6     & 5.1       & 1.3  \cr
  distortion ($\mu$m)         & 0.7     & 0.5       & 0.1  \cr
\hline
vs. $z$    & bowing  & $z$-scale    & twist      \cr
  decay (iterations)     &   2.6   & 11.2       & 12.0 \cr
  distortion ($\mu$m)          &   0.6   & 0.6        & 0.1 \cr
\hline
vs. $\phi$ & eliptical  & skew  & squeeze  \cr
  decay (iterations)     & 11.8    & 33.6   &  32.0 \cr
  distortion ($\mu$m)         & 0.9     & 4.9    &  4.5 \cr
\hline\hline
\end{tabular}
\end{center}
\end{table}

We fit for the amplitude of the distortion remaining
as a function of iteration for each of the nine tests.
In the data we compare wafer positions 
to the converged set of alignment
parameters prior to introducing systematic distortions.
In all nine cases we find the alignment procedure is capable 
of reducing global distortions to a negligible level.
We also perform tests with Monte Carlo where we compare wafer 
positions to the true positions known from MC generation.
In Fig.~\ref{fig:svt-amp-iteration} we show four representative
initial misalignments in 
Fig.~\ref{fig:svt-misalign-initial}. 
The rate of convergence of the nine global distortions,
defined as the decay constant of an exponential fit
to the scale of the misalignment per iteration,
is given in Table~\ref{tab:svtdecaytime}.

\begin{figure}[t]
\centerline{
\setlength{\epsfxsize}{0.9\linewidth}\leavevmode\epsfbox{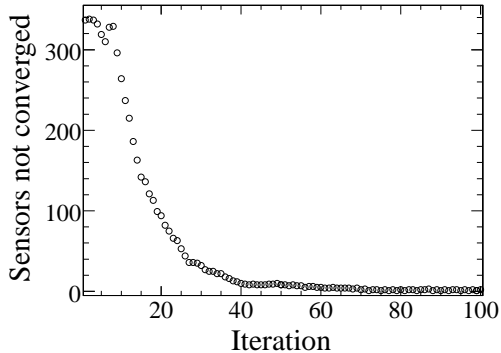}}
\caption{
Number of unconverged wafers as a function of iteration starting with
a 50\mum\ amplitude elliptical expansion distortion initial condition relative
to final alignment using $\babar$~data.
}
\label{fig:svt-nwaf-iteration}
\end{figure}

Overall, global distortions are the most weakly constrained
deformations and it was found empirically that the order
of 100 iterations were necessary to solve for these deformations.
Fig.~\ref{fig:svt-nwaf-iteration} shows the number of wafers 
which are not converged as a function of iteration.
The increase after iteration six is due to the residual
requirement applied after partial convergence of the procedure,
as discussed in Sec.~\ref{sec:minimize}.
The convergence requirement was chosen empirically to allow
convergence of the global distortions shown 
in Fig.~\ref{fig:svt-amp-iteration}.

\begin{figure}[t]
\centerline{
\setlength{\epsfxsize}{0.50\linewidth}\leavevmode\epsfbox{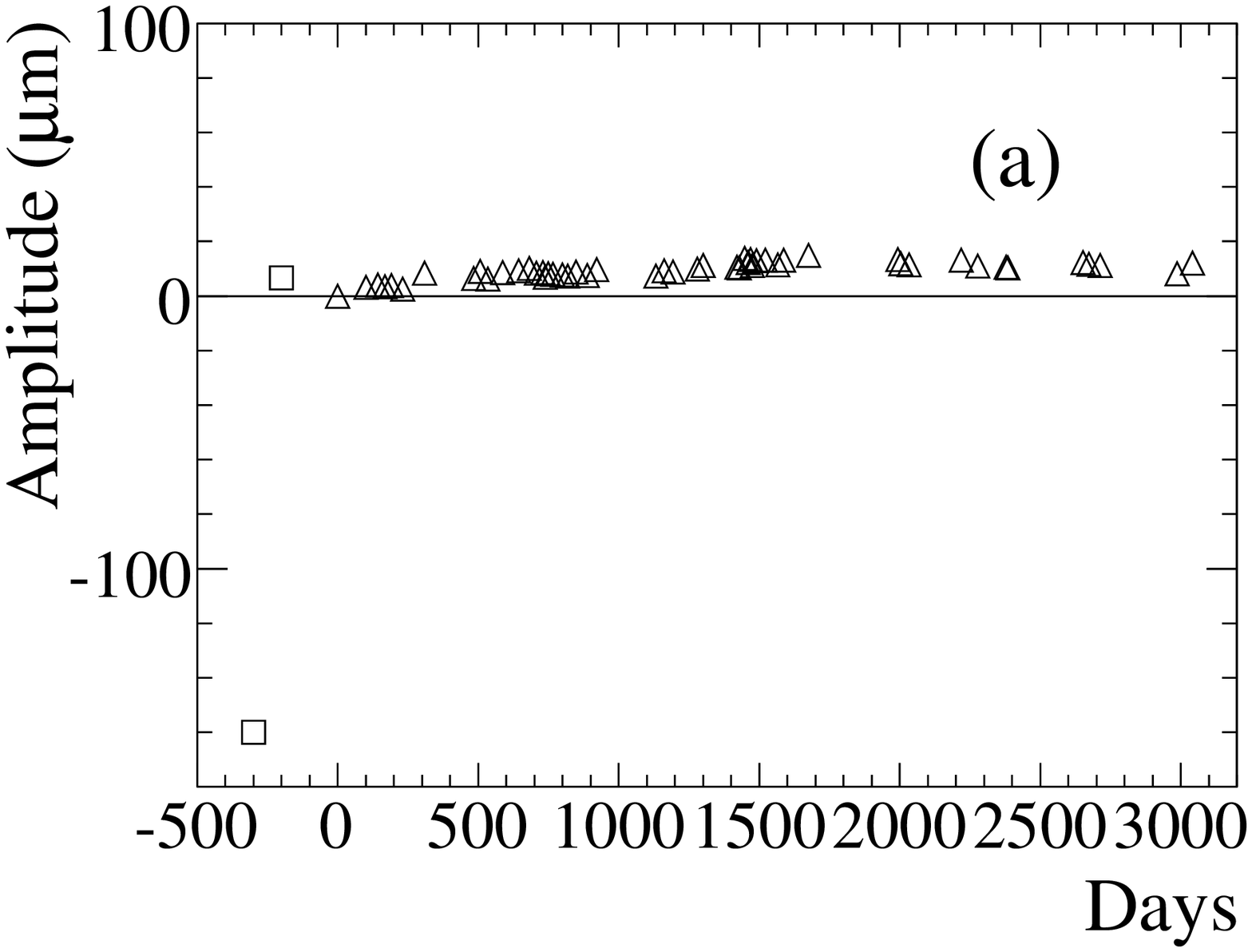}
\setlength{\epsfxsize}{0.50\linewidth}\leavevmode\epsfbox{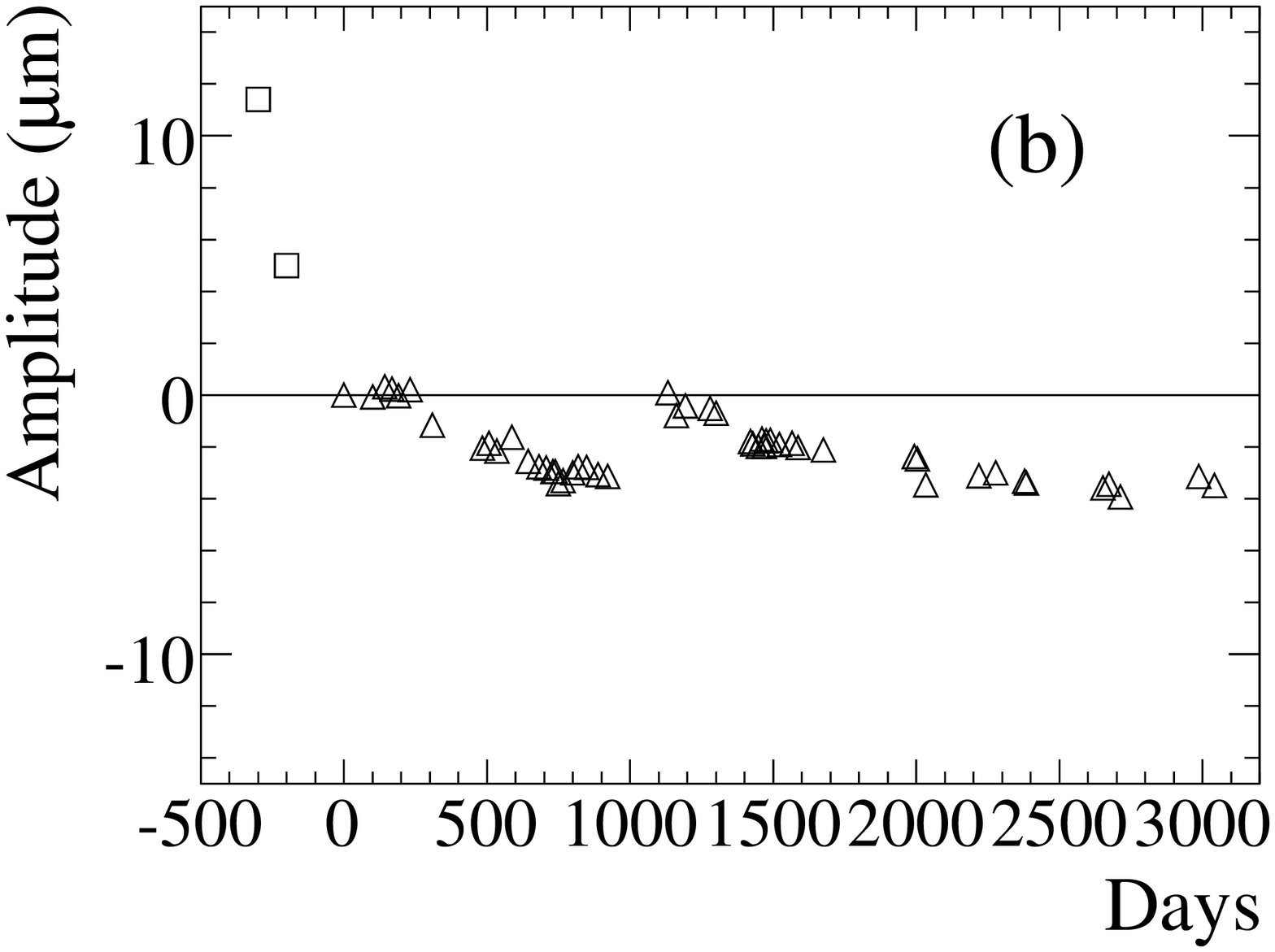}
}
\centerline{
\setlength{\epsfxsize}{0.50\linewidth}\leavevmode\epsfbox{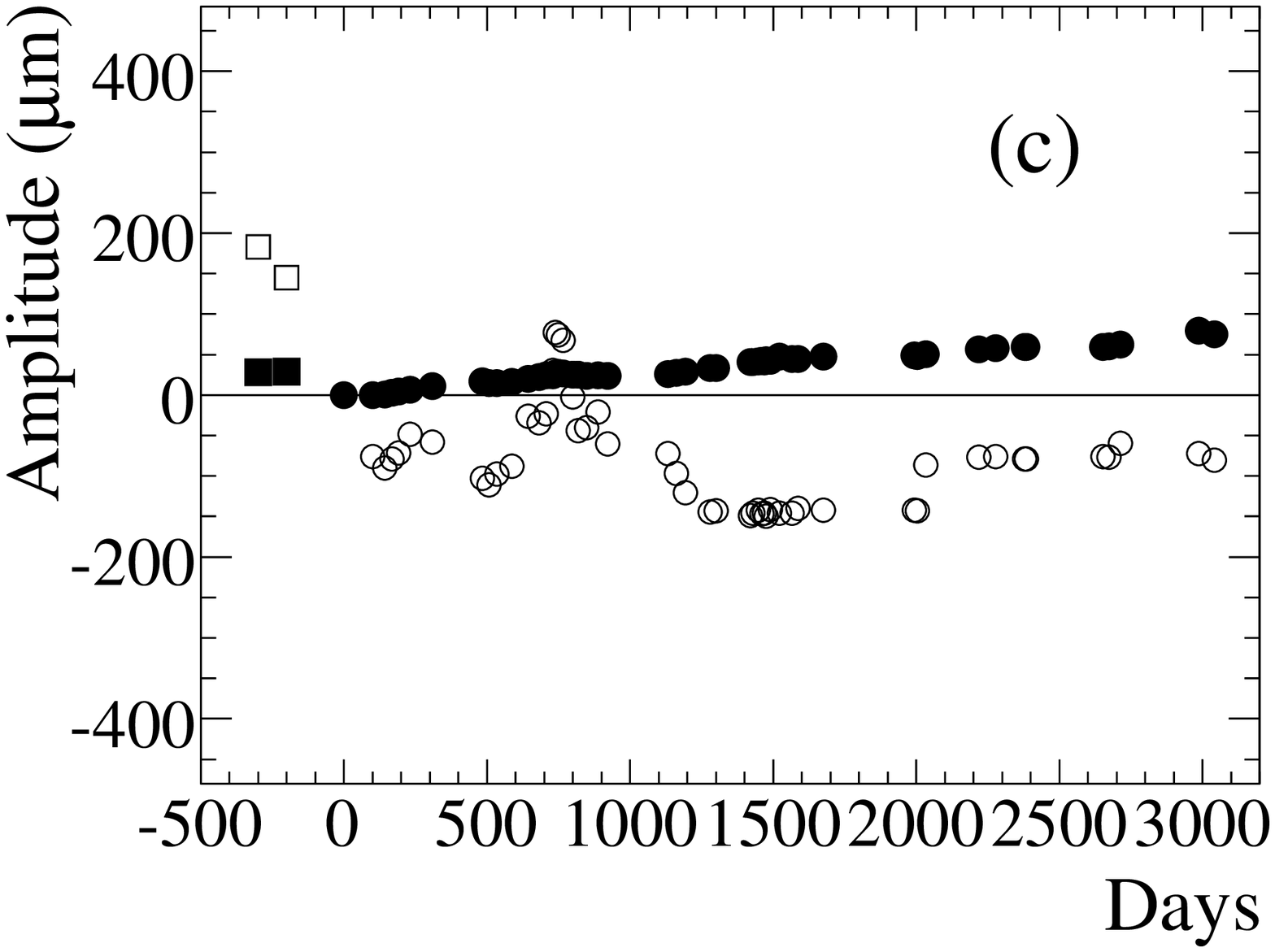}
\setlength{\epsfxsize}{0.50\linewidth}\leavevmode\epsfbox{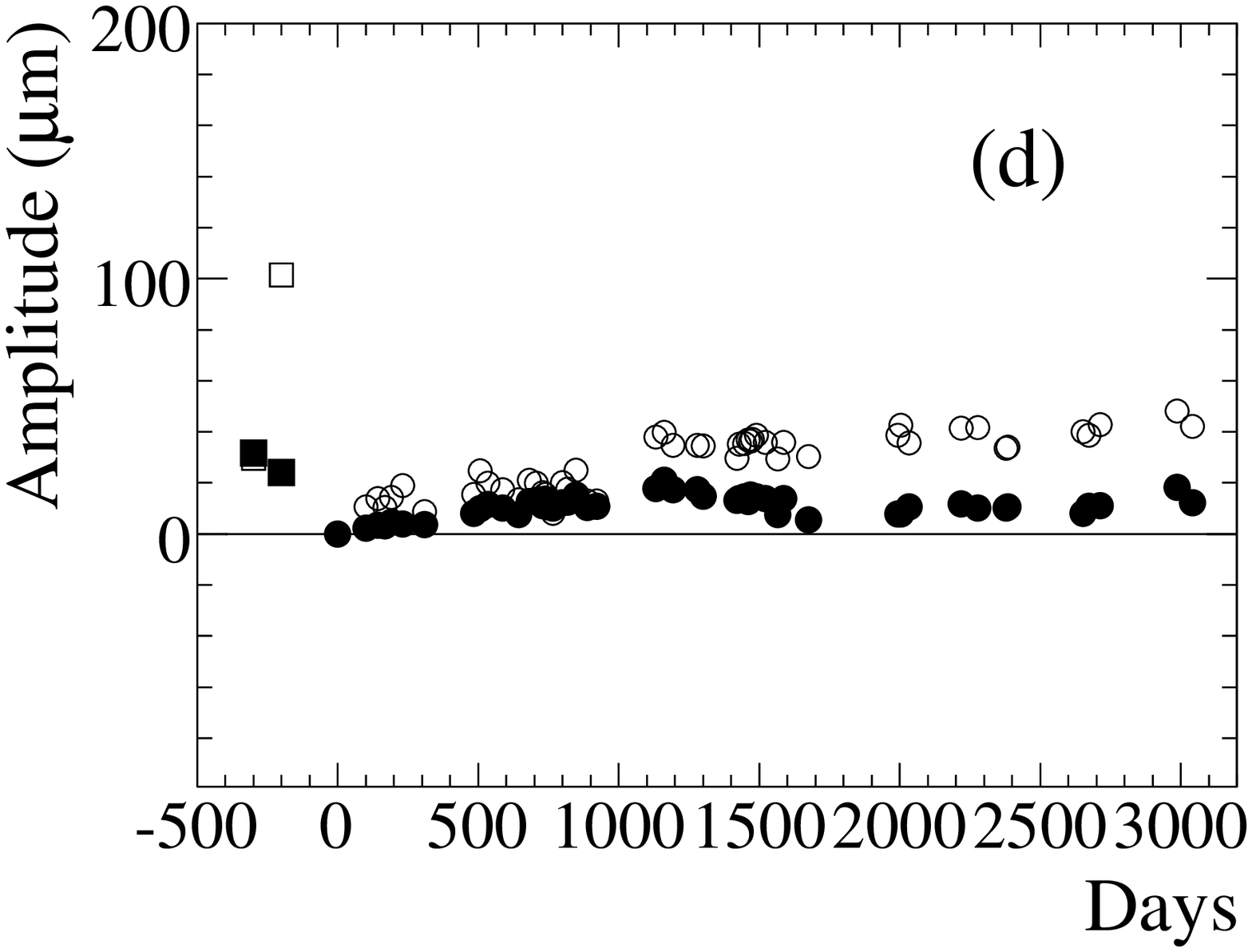}
}
\caption{Time-dependence of the global distortions of the \svt\ when compared
to day-one. Each point represents an alignment set and the day since the
first alignment set is shown on the $x$-axis, while days $-300$ and $-200$
correspond to ideal and survey geometries. Amplitudes of four distortions 
are shown: (a) telescope, (b) curl, (c) bowing, and (d) elliptical effects.
In (a) and (b) triangles ($\triangle$) represent average effects for all wafers, 
while in (c) and (d) the inner three layers and the outer two layers are shown
separately with the filled ($\bullet$) and open ($\circ$) circles, respectively.
}
\label{fig:svt-timedepen}
\end{figure}

Figure~\ref{fig:svt-timedepen}
plots the time-dependence of some
global distortions of the \svt, compared
to the initial day-one alignment.  The plot covers roughly 50 time periods when 
the internal structure of the \svt\ was suspected
to be changing due to mechanical stress during detector access
or humidity changes.
The $y$-axis plots the amplitude of the change in
a particular distortion, obtained using a method 
similar to that shown in Fig.~\ref{fig:svt-misalign-initial}.
The number of days since the
initial alignment is shown on the $x$-axis. 
The first two points at negative time
compare the initial alignment to the ideal geometry (day $-300$)
and the survey geometry (day $-200$), as shown in 
Figs.~\ref{fig:null-1999} and \ref{fig:survey-1999} respectively.
The large bowing in the outer layers 
around day 700 is due to an accidental humidity increase, which caused
the carbon-fiber support structure to expand.

Our study of potential distortions of the \svt\ wafer positions 
in the alignment procedure places limits on systematic 
uncertainties in physics measurements, such as particle 
lifetime or the Cabibbo-Kobayashi-Maskawa (CKM) matrix
parameter measurements through time-evolution studies 
of the $B$ meson decays. There are two sources of systematics: 
any misalignment due to time-variations, such as those shown 
in Fig.~\ref{fig:svt-timedepen}, and due to imperfections 
in the alignment procedure with residual misalignments 
remaining, both statistical and systematic. We minimize 
the former by having about 50 independent time periods 
which follow major changes in the detector. The systematic 
distortions are controlled in the validation plots, such 
as those shown in Figs.~\ref{fig:svt-cosmics} and 
\ref{fig:svtla-dimuonphi}. We have validated that all major
systematic distortions shown in Table~\ref{tab:svtdistortions} 
would be visible in the above validation plots. 

To facilitate studying the impact of potential remaining misalignment
on physics analysis, we create special alignment sets with intentional
misalignments, which describe remaining misalignments possible
by either of the mechanisms above.
The impact of remaining misalignment
on a physics analysis is evaluated by reconstructing the
tracks in simulated events using a misaligned parameter set,
followed by the normal analysis chain.  The difference between
that result and the one produced from analyzing the same
events with perfectly aligned track reconstruction
is taken as the misalignment systematic error.
The misalignment error is rarely 
the dominant systematic error of a \babar\ analysis.

%
%
\section {Fit for the $e^+e^-$ Beam Energies}
\label{sec:svtla_beam}

In the joint fit of the \mupair, the same four-momentum
for the initial state (from the beam energy monitoring) and the final state
is assumed. However, while beam angle measurements were found to be
precise in the PEP-II beam monitoring, measurements of the 
beam energies were not stable to better than a couple of MeV.
In addition, initial and final state radiation can systematically change
the total energy and momentum of the \llpair pair compared to the
original \epem.  Assuming the wrong momentum constraint in the pair
fit could introduce a {\em telescope} distortion,
as described in Table~\ref{tab:svtdistortions}.
This effect was indeed observed in early local alignment validation.
We use the cosmic tracks to constrain this effect.  Because these
cross the entire detector, they would detect a telescope distortion as a
kink in the track polar angle going from one side to the other.
Because of the cosmic track constraint, we can use the \svt\ local
alignment procedure to determine simultaneously the alignment and the
boost of the two muons.

\begin{figure}[tp]
\begin{center}
\centerline{
\setlength{\epsfxsize}{0.5\linewidth}\leavevmode\epsfbox{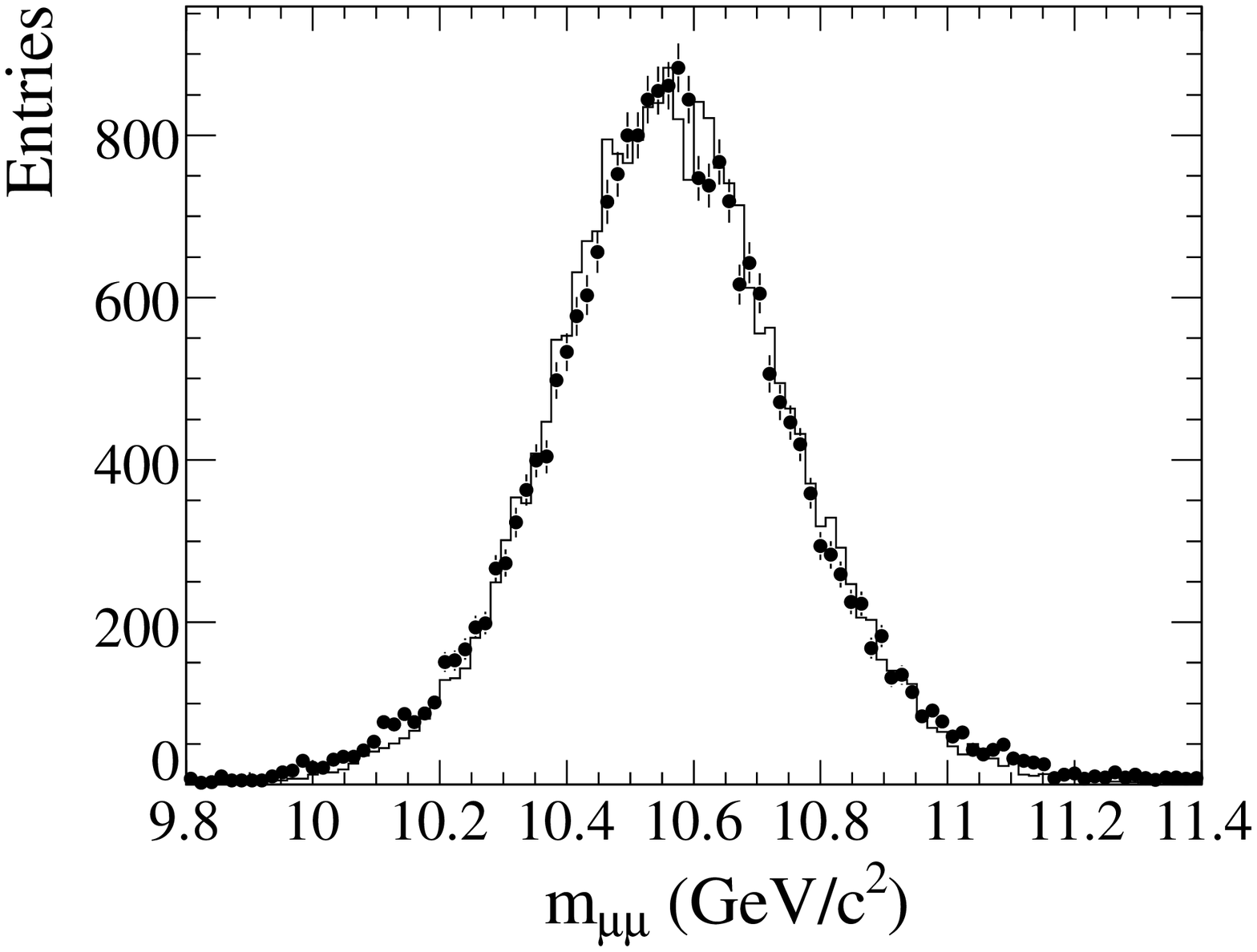}
\setlength{\epsfxsize}{0.5\linewidth}\leavevmode\epsfbox{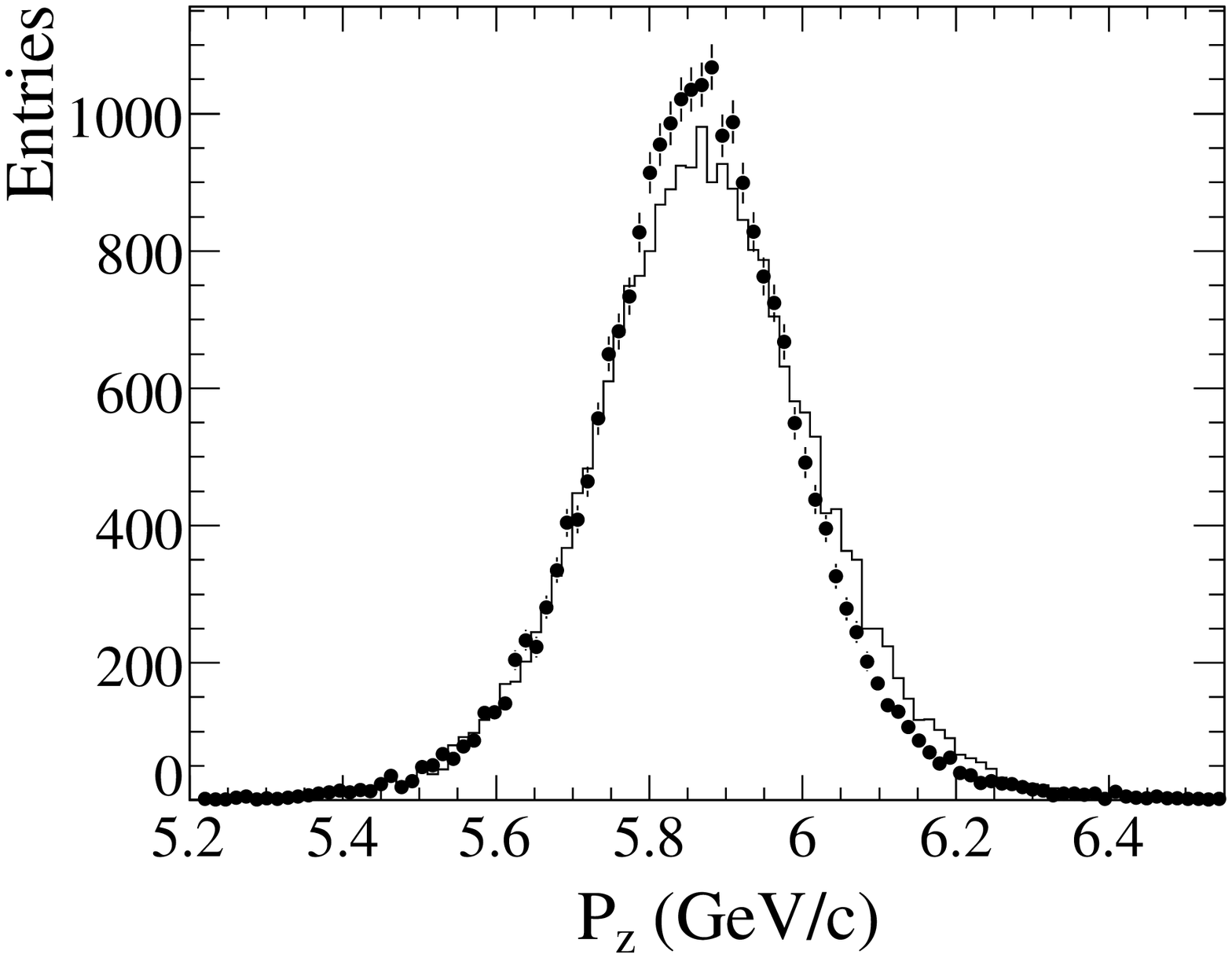}
}
\caption{
Distribution of the dimuon $\mu^+\mu^-$ invariant mass (left) and total measured momentum of the
two muons (right) in data (points with error bars) and MC (histogram).
}
\label{fig:mumumass}
\end{center}
\end{figure}

In Fig.~\ref{fig:mumumass} we show the distributions of the dimuon
invariant mass and total measured momentum of the two muons (boost).
In the alignment procedure, we fit the average $\mu^+\mu^-$ boost and use it
in the four-momentum pair constraint instead of the initial state boost.
We then iterate, allowing a simultaneous extraction of the boost value
and the geometrical constraints on the telescope
global distortion.  This simultaneous fit makes the local alignment
algorithm more stable, and allows for monitoring the $e^+e^-$ beam boost.
In principle, we can also use the \mupair\ events to measure the total
\epem\ invariant mass.  However,
because of the high momentum, and therefore poor
relative momentum resolution, of the tracks in \mupair\ events, the resolution
is poor.  A better estimate of the \epem\
invariant mass can be made using $\Upsilon(4S)\to \BB$ decays, when one $B$
meson decays fully hadronically.

%
%
\section{Residual Aplanar Distortions}
\label{sec:aplanar}


\begin{figure*}[htbp]
\begin{center}
\centerline{
\setlength{\epsfxsize}{1.0\linewidth}\leavevmode\epsfbox{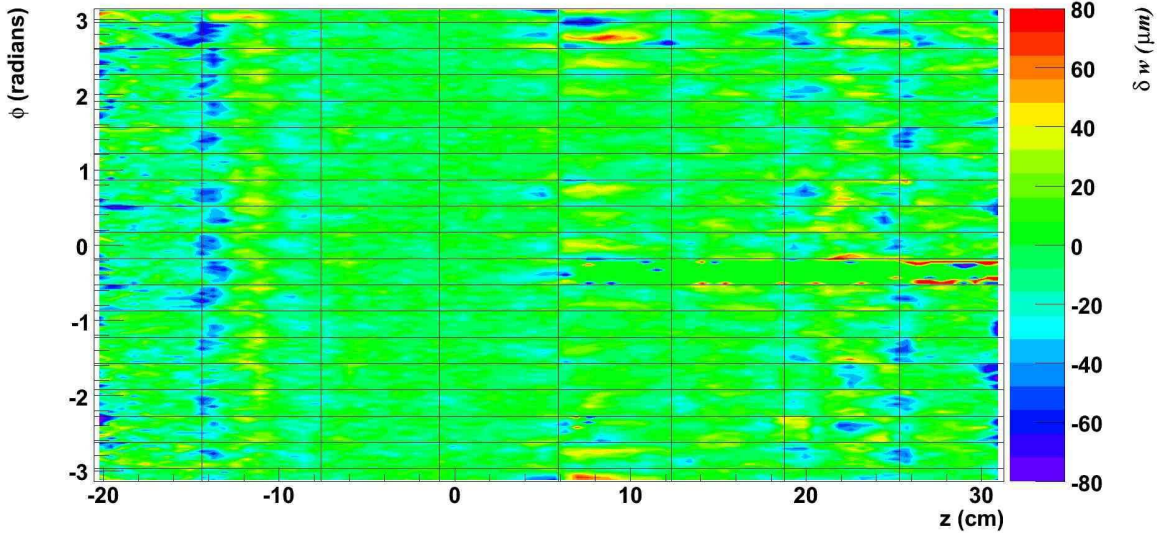}
}
\caption{Average $\delta w$ projection of the $v$ hit residuals in the
backward half of layer five as a function of the global $\phi$ and
$z$ position for high-momentum tracks in the \babar\ data.  The overlaid
lines show the approximate wafer edges.  The forward half of the
module near $\phi = -0.25$ radians
shows no data due to its failed $v$ view readout electronics.}
\label{fig:lay5_dD}
\end{center}
\end{figure*}

As described in Sec. \ref{sec:curvature}, we see clear evidence for
substantial aplanar distortions of wafers in the inner layers of the \svt.
The dominant effect can be characterized
as a bowing of the wafers along their $u$ coordinate,
which has limited external support.  We also studied more
general forms of aplanar distortions using the same techniques discussed
in Sec. \ref{sec:curvature}, where we interpret the average residuals
as due to a local $\delta w$ distortion.  From these studies we discovered
that more general forms of aplanar distortion are present in the \babar\ \svt.

\par

An example is shown in Fig.~\ref{fig:lay5_dD}.  This plots the
average $\delta w$
of layer five $v$ hit residuals for high-momentum tracks in \babar\ data
as a function of their position.  The
distortions are substantial, but they cannot be described by simple $u$ bowing of individual
wafers.
Some patterns are evident, such as the $v$ bowing of the wafers centered around $z=-10$ cm.
A similar pattern is seen at the forward end of layer five, and at both ends of layer four,
but not in similar locations in layers $1 \rightarrow 3$.
This suggests that these distortions may be related to the
bend at the end of the arch modules (see Fig.~\ref{fig:svt-1}).
Fig.~\ref{fig:lay5_dD} also shows irregular distortions in many 
layer five wafers near their edge at approximately $z=5$ cm.
This is where the forward and backward module halves, which were
constructed separately, join.
Similar distortions occur where the layer four module halves join.
No such distortions are seen in the middle of the inner layer modules,
which were built in one piece.

\par

The aplanar distortions in the outer layers
are thought to be responsible for some of the remaining irregularities
seen in the validation plots, for instance the variation of \mupair\
miss distance resolution with $\phi$ and $\theta$, 
as shown in Figs. \ref{fig:svtla-dimuonphi} and \ref{fig:svtla-dimuontheta}.
Unfortunately,
the irregularity of these distortions makes them difficult to correct, and
we do not attempt to do so in the \babar\ local alignment procedure.
A full $\delta w$ map of a vertex detector would in principle be possible
given a large and diverse collection of data.  In particular, tracks originating
at many different positions would be necessary to avoid the
lack of constraint on $\delta w$ when the projected track incident angle is normal to the wafer.

%
%
\section {Conclusions}

We have described the procedure used to determine \babar\ \svt\ local alignment.
We have shown that this procedure satisfies the requirements placed on the
\svt\ performance by the \babar\ physics goals.  We have demonstrated that this procedure
is robust against global distortions that could otherwise introduce unacceptable systematic
biases in \babar\ tracking and physics data.

%
%

\section{Acknowledgments}

The work presented in this note could not have been accomplished 
without the help of many people. The first \babar\ \svt\ alignment 
procedure had major contributions from Gerald Lynch and 
Jochen Schiek. We also wish to thank Stefan Kluth, Amir Farbin,
Vincent Lillard, Gennadiy Kukartsev, Jurgen Krosberg,
Chung Khim Lae, and Luke Winstrom for generating the alignment 
constants ultimately used in physics analysis.
We also wish to thank Eric Charles, Fred Goozen, Natalia Kuznetsova,
and Marzia Folegani for providing the optical survey measurements,
Gennadiy Kukartsev for initial studies of the boost fit algorithm,
Brandon Giles for his work on a-planar distortions,
the \svt\ group for building and operating this beautiful detector,
the tracking group for their support with validation studies,
and \babar\ and PEP-II for providing the data.

\appendix
%
%
\section{Derivative calculations for the \svt\ Local Alignment}
\label{sec:derivs}

To minimize the \chisq\ used in the \svt\ local alignment procedure, we need the
first derivatives of the track-hit residuals with respect to the six degrees of freedom of the wafer
alignment defined in Sec.~\ref{sec:params}, being specifically the
three translations along the local wafer coordinate axes ($\hat{u}$, $\hat{v}$, and $\hat{w}$)
and the three small rotations about those axes.
In \babar, residuals are defined as the distance in space between the track and the
hit trajectory at their point of closest approach (POCA), signed by the 
cross-product of the track direction ($\hat{t}$) and the hit trajectory direction ($\hat{h}$).  The hit 
trajectory is defined as a line segment in the wafer plane, with a direction
given by the strips used in this hit (generally $\hat{u}$ or $\hat{v}$ ).  Thus a
barrel-module $u$ hit
(which constrains the $u$ position of a track) has hit trajectory direction $\hat{h} = \hat{v}$,
and a $v$ hit has hit trajectory direction $\hat{h} = \hat{u}$.

The derivatives of a residual $\epsilon$ with respect to wafer translations
$d$ and rotations $\alpha$ can be expressed as:
\begin{eqnarray}
\frac{\partial \epsilon}{\partial d_i} & = &\hat{D} \cdot \hat{i} \\
\frac{\partial \epsilon}{\partial \alpha_i} & = &\hat{D} \cdot ( \hat{i} \times \vec{H}),
\end{eqnarray}
where $\hat{D} \equiv \hat{t} \times \hat{h} / | \hat{t} \times \hat{h}|$,
$\vec{H}$ is the position of the hit at POCA, relative
to the geometric center of the wafer, and
$i \in \{u,v,w\}$.
These derivatives are coded in the \babar\
local alignment procedure using the CLHEP \cite{clhep} class library geometry methods.  The
exact derivative calculations and
implementation were tested by comparison with numerically-computed derivatives.

%
%

\pagebreak

%
%
%
%





\end{document}